\input{aipcheck}

\documentclass[numberedheadings]{aipproc}
\layoutstyle{6s}

\begin{document}

\title{Electron Phonon Interaction and Strong Correlations in
High-Temperature Superconductors: One can not avoid unavoidable}
\author{Miodrag L. Kuli\'{c}}
{address={Johann Wolfgang Goethe-University, \\Institute for
Theoretical Physics, \\P.O.Box 111932, 60054 Frankfurt/Main,
Germany}}

\date{\today }

\begin{abstract}
The important role of the electron-phonon interaction (EPI) in
explaining the properties of the normal state and pairing
mechanism in high-T$_{c}$ superconductors (HTSC) is discussed. A
number of experimental results are analyzed such as: dynamical
conductivity, Raman scattering, neutron scattering, ARPES,
tunnelling measurements, isotope effect and etc. They give
convincing evidence that the EPI is strong and dominantly
contributes to pairing in HTSC oxides. It is argued that strong
electronic correlations in conjunction with the pronounced (in
relatively weakly screened materials) EPI are unavoidable
ingredients for the microscopic theory of pairing in HTSC oxides.
I present the well defined and controllable theory of strong
correlations and the EPI. It is shown that strong correlations
give rise to the pronounced \textit{forward scattering peak} in
the EPI - the FSP theory. The FSP theory explains in a consistent
way several (crucial) puzzles such as much smaller transport
coupling constant than the pairing one ($\lambda_{tr}\ll
\lambda$), which are present if one interprets the results in HTSC
oxides by the old Migdal-Eliashberg theory for the EPI. The ARPES
shift puzzle where the nodal kink at 70 meV is unshifted in the
superconducting state, while the anti-nodal one at 40 meV is
shifted can be explained at present only by the FSP theory. A
number of other interesting predictions of the FSP theory are also
discussed.
\end{abstract}

\maketitle

\contentsline {section}{\numberline {1}Introduction}{3}
\contentsline {subsection}{\numberline {1.1}Importance of strong electronic correlations and EPI}{3}
\contentsline {subsection}{\numberline {1.2}Prejudices on the EPI}{6}
\contentsline {section}{\numberline {2}Experiments related to pairing mechanism}{8}
\contentsline {subsection}{\numberline {2.1}Magnetic neutron scattering}{8}
\contentsline {subsubsection}{\numberline {2.1.1}Normal state}{8}
\contentsline {subsubsection}{\numberline {2.1.2}Superconducting state}{9}
\contentsline {subsection}{\numberline {2.2}Dynamical conductivity and resistivity $\rho (T)$}{11}
\contentsline {subsubsection}{\numberline {2.2.1}Dynamical conductivity $\sigma (\omega )$}{11}
\contentsline {subsubsection}{\numberline {2.2.2}Resistivity $\rho (T)$}{20}
\contentsline {subsection}{\numberline {2.3}Raman scattering in HTSC oxides}{23}
\contentsline {subsubsection}{\numberline {2.3.1}Electronic Raman scattering}{23}
\contentsline {subsubsection}{\numberline {2.3.2}Phonon Raman scattering}{23}
\contentsline {paragraph}{Normal state}{23}
\contentsline {paragraph}{Superconducting state}{24}
\contentsline {subsubsection}{\numberline {2.3.3}Electron-phonon coupling in Raman scattering}{27}
\contentsline {subsection}{\numberline {2.4}Tunnelling spectroscopy in HTSC oxides}{28}
\contentsline {subsubsection}{\numberline {2.4.1}$I-V$\ characteristic and $\alpha ^{2}F(\omega )$}{29}
\contentsline {subsection}{\numberline {2.5}Isotope effect in HTSC oxides}{30}
\contentsline {subsubsection}{\numberline {2.5.1}Experiments on the isotope coefficient $\alpha $}{30}
\contentsline {subsection}{\numberline {2.6}ARPES experiments in HTSC oxides}{33}
\contentsline {subsubsection}{\numberline {2.6.1}Spectral function $A(\mathaccent "017E\relax {k},\omega )$ from ARPES}{33}
\contentsline {subsubsection}{\numberline {2.6.2}Theory of the ARPES kink}{35}
\contentsline {subsubsection}{\numberline {2.6.3}ARPES and the EPI coupling constant $\lambda $}{38}
\contentsline {section}{\numberline {3}EPI in HTSC oxides}{38}
\contentsline {subsection}{\numberline {3.1}General strong coupling theory of the EPI}{40}
\contentsline {subsection}{\numberline {3.2}LDA calculations of $\lambda $ in HTSC oxides}{42}
\contentsline {subsection}{\numberline {3.3}Lattice dynamics and EPI coupling}{44}
\contentsline {section}{\numberline {4}Theory of strong electronic correlations}{44}
\contentsline {subsection}{\numberline {4.1}X-method for strongly correlated systems}{46}
\contentsline {subsection}{\numberline {4.2}Forward scattering peak in the charge vertex $\gamma _{c} $}{51}
\contentsline {section}{\numberline {5}Renormalization of the EPI by strong correlations}{53}
\contentsline {subsection}{\numberline {5.1}The forward scattering peak in the EPI}{54}
\contentsline {subsection}{\numberline {5.2}Pairing and transport EPI coupling constants}{55}
\contentsline {section}{\numberline {6}FSP theory and novel effects}{58}
\contentsline {subsection}{\numberline {6.1}Nonmagnetic impurities and robustness of d-wave pairing}{58}
\contentsline {subsection}{\numberline {6.2}Transport properties and superconductivity}{60}
\contentsline {subsection}{\numberline {6.3}Nonadiabatic corrections of $T_{c}$}{64}
\contentsline {subsection}{\numberline {6.4}Pseudogap behavior in the FSP model for the EPI}{65}
\contentsline {section}{\numberline {7}Electron-phonon interaction vs spin-fluctuations}{69}
\contentsline {subsection}{\numberline {7.1}Interaction via spin fluctuations (SFI) and pairing}{69}
\contentsline {subsection}{\numberline {7.2}Are the EPI and SFI compatible in d-wave pairing?}{72}
\contentsline {section}{\numberline {8}Is there high-temperature superconductivity in the Hubbard and t-J model?}{73}
\contentsline {subsection}{\numberline {8.1}Hubbard model}{73}
\contentsline {subsection}{\numberline {8.2}t-J model}{74}
\contentsline {section}{\numberline {9}Summary and conclusions}{75}
\contentsline {section}{\numberline {10}Appendix: Derivation of the t-J model}{77}
\contentsline {subsection}{\numberline {10.1}Hubbard model for finite U in terms of Hubbard operators}{77}
\contentsline {subsection}{\numberline {10.2}Effective Hamiltonian for $U>>t$}{78}
\contentsline {paragraph}{Slave boson (SB) method}{80}
\contentsline {paragraph}{Slave fermion (SF) method}{80}
\contentsline {paragraph}{Spin fermion method}{80}

\section{Introduction}

\subsection{Importance of strong electronic correlations and EPI}

Seventeen years after the discovery of the high-T$_{c}$ superconductors
(HTSC) \cite{BeM} there is still no consensus about the pairing mechanism in
these materials. At present two possible theories are in the focus, the
first one based on the electron-phonon interaction (EPI) and the second one
based on spin fluctuation interaction (SFI). In the meantime it was well
established that metallic compounds of HTSC oxides are obtained from
insulating parent compounds by doping with small number of carriers -
usually called holes. It turns out that the parent insulating state is far
from being conventional band insulator where usually an even number of
electrons (holes) per lattice site fill Bloch bands completely. By counting
the electron number one comes (naively) to the conclusion that the parent
compounds of copper oxides (for instance $La_{2}CuO_{4}$ and $%
YBa_{2}Cu_{3}O_{6}$) should be metallic, because in the unit cell there is
odd (nine) number of $d$-electrons per $Cu^{2+}$ ion. The way out from this
controversy is in the presence of strong electronic correlations. They are
due to the localized $d$-orbital on the $Cu^{2+}$ ion giving rise to the
strong Coulomb repulsion $U$ of two 3d$_{x^{2}-y_{2}}$ electrons (or holes)
at a given lattice site with opposite spins. This repulsion keeps electrons
apart making them to be localized on the lattice, but with localized spins ($%
S=1/2$). This type of insulating state is called the \textit{Mott-Hubbard
insulator}. Speaking in language of electronic bands, for large on-site
repulsion $U\gg W$ and for one electron per lattice site the original
conduction band (with the width $W$) is split into the lower Hubbard band
with localized spins and the empty upper band separated by $U$ from the
lower one - see more in \cite{KulicReview} and Section 4.

The relevance of strong correlations is well documented experimentally: $%
\mathbf{(i)}$ The electron-energy-loss spectroscopy \cite{Romberg} shows a
transfer of intensity (which is a measure of the number of states) from
higher to lower energies by doping. Such a property is characteristic for
the class of Hubbard models where the number of states in the upper Hubbard
band decreases by increasing the hole doping. For comparison, in typical
semiconductors the number of states in the valence band is determined by the
number of atoms, i.e. it is fixed and doping independent. $\mathbf{(ii)}$
The self-consistent band-structure calculations and the photoemission
experiments gave that the effective Hubbard interaction ($U)$ for the $Cu$
ions is of the order $U\approx 6-10$ $eV$ \cite{Schluter}, which is much
larger than the observed band width $W$ ($\sim $ $2$ $eV$) \cite{Mante}. $%
\mathbf{(iii)}$ A rather direct evidence for strong correlations comes from
the doping dependence of the dynamic conductivity $\sigma (\omega )$ in $%
La_{2-x}Sr_{x}CuO_{4}$ and $Nd_{2-x}Ce_{x}CuO_{4-y}$, particularly from the
observed shift of the spectral weight from high to low energies with doping
\cite{Uchida}. Besides the development of the Drude peak around $\omega =0$
in the underdoped systems the so called mid-infrared ($MIR$) peak is also
developed around $0.4$ $eV$.

Regarding the EPI one can put an ''old fashioned'' question: Does the EPI
makes (contributes to) the superconducting pairing in HTSC oxides?
Surprisingly, most of researchers in the field believe that the EPI is
irrelevant and that the pairing mechanism is due to spin fluctuations and
strong correlations alone- see \cite{Pines}. This belief is mainly based on
an incorrect stability criterion (which, if true, would strongly limit $%
T_{c} $ in the EPI mechanism), and also on a number of experimental results
which give evidence for strong anisotropic ($d-wave$ like) pairing with
gapless regions on the Fermi surface \cite{Scal}, etc. Moreover, the phase
sensitive $SQUID$ measurements of the Josephson effect \cite{SQUID1}, \cite
{SQUID2} in the orthorhombic material $YBa_{2}Cu_{3}O_{6+x}$ are strongly in
favor of an ''orthorhombic'' $d-wave$ superconducting order parameter, for
instance $\Delta (\mathbf{k})=\Delta _{s}+\Delta _{d}(\cos k_{x}-\cos k_{y})$%
. As experiments of Tsuei et al. \cite{SQUID1}, \cite{SQUID2} show one has $%
\Delta _{s}<0.1\Delta _{d}$ in optimally doped $YBa_{2}Cu_{3}O_{6+x}$, which
means that zeros of $\Delta (\mathbf{k})$ are near intersections of the
Fermi surface and the lines $k_{x}\approx \pm k_{y}.$ Recent experiments on
the single-layer crystals $Tl_{2}Ba_{2}CuO_{6+x}$ and on $%
Bi_{2}Sr_{2}CaCu_{2}O_{8+x}$ ($Bi2212$) done by Tsuei group \cite{SQUID3},
\cite{SQUID4}, \cite{Tsuei-recent}, prove the existence of pure $d-wave$
pairing in underdoped, optimally and overdoped systems. The recent
interference experiments on $Nd_{2-x}Ce_{x}CuO_{4-y}$ point also to d-wave
pairing in this compound \cite{Nd-Tsuei}. In that respect, we point out that
there is also an widespread (and unfounded) belief that $d-wave$ is
incompatible with the EPI pairing mechanism.

Another argument used against the EPI as an origin of superconductivity in
HTSC oxides is based on the small value of the oxygen isotope effect $\alpha
_{O}$ ($\alpha =\alpha _{O}+\alpha _{Cu}+\alpha _{Y}+\alpha _{Ba}$) in
optimally doped materials, such as $YBCO$ with highest critical temperature $%
T_{c}\approx 92K$ where $\alpha _{O}\approx 0.05$ \cite{Franck}, instead of
the canonical value $\alpha =1/2$ which would be in the case of the EPI
pairing mechanism alone and in the presence of O-vibrations only.

On the other hand, there is good experimental evidence that the EPI is
sufficiently large in order to produce superconductivity in HTSC oxides,
i.e. $\lambda >1$. Let us quote some of them: $\mathbf{(1)}$ The
superconductivity induced \textit{phonon renormalization} \cite{Thomsen},
\cite{TIR}, \cite{PhoRaman}, \cite{Hadjiev2} is much larger in HTSC oxides
than in $LTSC$ superconductors. This is partially due to the larger value of
$\Delta /E_{F}$ in HTSC than in $LTSC$; $\mathbf{(2)}$ the \textit{line-shape%
} in the phonon Raman scattering is very asymmetric (Fano line), which
points to a substantial interaction of the lattice with some quasiparticle
(electronic liquid) continuum. For instance, the recent phonon Raman
measurements \cite{PhoRaman} on $HgBa_{2}Ca_{3}Cu_{4}O_{10+x}$ at $T<T_{c}$
give very large softening (self-energy effects) of the $A_{1g}$ phonons with
frequencies $240$ and $390 $ $cm^{-1}$ by $6$ $\%$ and $18$ $\%$,
respectively. At the same time there is a dramatic increase of the
line-width immediately below T$_{c}$, while above T$_{c}$ the line-shape is
strongly asymmetric. A substantial phonon renormalization was obtained in ($%
Cu,C)Ba_{2}Ca_{3}Cu_{4}O_{10+x}$ \cite{Hadjiev2}; $\mathbf{(3)}$ the \textit{%
large isotope coefficients} ($\alpha _{O}>0.4$) in $YBCO$ away from the
optimal doping \cite{Franck} and $\alpha _{O}\approx 0.15-0.2$ in the
optimally doped $La_{1.85}Sr_{0.15}CuO_{4}$. At the same time one has $%
\alpha _{O}\approx \alpha _{Cu}$ making $\alpha \approx 0.25-0.3$. This
result tell us that other, besides O, ions participate in pairing; $\mathbf{%
(4)}$ the most important evidence that the EPI plays an important role in
pairing comes from \textit{tunnelling spectra} in HTSC oxides, where the
phonon-related features have been clearly seen in the $I-V$ characteristics
\cite{Tun1}, \cite{Tun2}, \cite{Tun3}, \cite{Tun4}, \cite{Gonnelli}; $(4)$
the \textit{penetration depth} in the a-b plane of YBCO is increased
significantly after the substitution $O^{16}\rightarrow O^{18}$, i.e. $%
(\Delta \lambda _{ab}/$ $\lambda _{ab})=(^{18}\lambda _{ab}-^{16}\lambda
_{ab})/^{16}\lambda _{ab}=2.8 $ $\%$ at $4$ $K$ \cite{lambda-isotop}. Since $%
\lambda _{ab}\sim m^{\ast }$ the latter result, if confirmed, could be due
to the nonadiabatic increase of the effective mass $m^{\ast }$.

Recent ARPES measurements on HTSC oxides \cite{Lanzara},
\cite{Cuk} show a \textit{kink} in the quasiparticle spectrum at
characteristic (oxygen) phonon frequencies in the normal and
superconducting state. This is clear evidence that the EPI is
strong and involved in pairing.

On the \textit{theoretical side} there are self-consistent $LDA$
band-structure calculations which (in spite of their shortcomings) give a
rather large bare EPI coupling constant $\lambda \sim 1.5$ in $%
La_{1.85}Sr_{0.15}CuO_{4}$ \cite{Krakauer}, \cite{Falter98}. The \textit{%
nonadiabatic effects} due to poor metallic screening along the $c$-axis may
increase $\lambda $ additionally \cite{Falter97}, \cite{Falter98}.\ All
these facts are in favor of the substantial EPI in HTSC oxides. However, if
the properties of the normal and superconducting state in HTSC oxides are
interpreted in terms of the standard EPI theory, which holds in $LTSC$
systems, some puzzles arise. One of them is related to the normal-state
conductivity (resistivity) - in optimally doped systems the width of the
Drude peak in $\sigma (\omega )$ and the temperature dependence of the
resistivity $\rho (T)$ are not incompatible with the strong-coupling theory
with $\lambda \sim 3$ and $\lambda _{tr}\sim 1$ (if $\omega _{pl}\sim 3$ $eV$%
), where $\lambda _{tr}$ is the transport EPI coupling constant \cite
{Zeyher1}. On the other side the combined resistivity and low frequency
conductivity (Drude part) measurements give $\lambda _{tr}\approx 0.3$ if
the plasma frequency takes the value $\omega _{pl}\sim 1$ $eV$ - see more
below. If one assumes that $\lambda _{tr}\approx \lambda $, which is the
case in most low temperature superconductors ($LTSC$), such a small $\lambda
$ can not give large $T_{c}(\approx 100$ $K)$.

In the past there were doubts on the ability of the EPI to explain the
linear temperature dependence of the resistivity in the underdoped system
\cite{Linear} $Bi_{2+x}Sr_{2-y}CuO_{6\pm \delta }$, which starts at low $%
T>10-20$ $K$. Because the asymptotic $T^{5}$ behavior of $\rho (T)$ (for $%
T\ll \Theta _{D}$) is absent in this sample, then it seems that this
experiment is questioning seriously the contribution of the EPI to the
resistivity. However, there are other measurements \cite{Macken} on $%
Bi_{2+x}Sr_{2-y}CuO_{6\pm \delta }$ where the linear behavior starts at
higher temperature, i.e. at $T>50$ $K$. Additionally, the resistivity
measurements \cite{Vedeneev1} on $Bi_{2}SrCuO_{x}$ samples with low $%
T_{c}\simeq 3$ $K$ show saturation to finite value at $T=0$ $K$. After
subtraction of this constant part one obtains the Bloch-Gr\"{u}neisen
behavior between $T_{c}\simeq 3$ $K$ and $300$ $K$, which is due to the EPI.

Concerning the EPI, the above results imply the following possibilities: $%
\mathbf{(a)}$ $\lambda _{tr}\ll 1<\lambda $ and the pairing is due to the $%
EPI$, or $\mathbf{(b)}$ $\lambda _{tr}\simeq \lambda \approx 0.4-0.6$ and
the EPI is ineffective (although present) in pairing; $(\mathbf{c})$ $%
\lambda _{tr}\simeq \lambda $ but the EPI is responsible for pairing on the
expense of some peculiarities of equations describing superconductivity. In
Section 5. we present a theory of the EPI renormalized by strong electronic
correlations, which is in favor of the case $\mathbf{(a)}$. It is
interesting that the similar puzzling situation ($\lambda _{tr}\ll \lambda $%
) is realized in $Ba_{x}K_{1-x}BiO_{3}$ compound (with T$_{c}\simeq 30$ $K$%
), where optical measurements give $\lambda _{tr}\approx 0.1-0.3$ \cite
{Salje}, while tunnelling measurements \cite{Jensen} give $\lambda \sim 1$.
Note, in $Ba_{x}K_{1-x}BiO_{3}$ there are no magnetic fluctuations (or
magnetic order) and no signs of strong electronic correlations. Therefore,
the EPI is favored as the pairing mechanism in $Ba_{x}K_{1-x}BiO_{3}$. It
seems that in this compound \textit{long-range forces}\textbf{,} in
conjunction with some nesting effects, may be responsible for this
discrepancy?

One can summarize, that the EPI theory, which pretends to explain the normal
metallic state and superconductivity in HTSC oxides, is confronted with the
problem of explaining why the EPI coupling is present in self-energy effects
(governed by the coupling constant $\lambda >1$) but it is suppressed in
transport properties (which depend on $\lambda _{tr}<1$), i.e. why $\lambda
_{tr}$ is (much) smaller than $\lambda .$ One of the possibilities is that
strong electronic correlations, as well as the long-range Madelung forces,
affect the EPI significantly. This will be discussed in forthcoming
sections. In light of the above discussion it is also important to know the
role of the EPI in the formation of $d-wave$ superconducting state in HTSC
oxides, i.e. why it is compatible with d-wave pairing?

In this review we discuss theoretical and experimental results in HTSC
oxides and mostly those which are related to: (i) strong quasiparticle
scattering in the normal state, (ii) the pairing mechanism \cite{Kulic1},
\cite{Kulic2}, \cite{Kulic3}, \cite{Kulic4}. \

The paper is organized as follows. In Section 2. we review important
physical properties of HTSC oxides in the normal and superconducting state,
whose understanding is a basis for the microscopic theory of
superconductivity. Only those experiments (and theoretical interpretations)
are discussed here which are in our opinion most important in getting
information on the pairing mechanism in HTSC oxides. In Section 3. we
discuss the general theory of the EPI and its low-energy version. The theory
of strong electronic correlations is studied in Section 4., where much space
is devoted to a systematic, recently elaborated, method for strongly
correlated electrons \cite{Kulic1}, \cite{Kulic2}, \cite{Kulic3}, \cite
{Kulic4} - the X-method. The latter considers strongly interacting
quasiparticles as \textit{composite objects}\textbf{,} contrary to the
slave-boson method which at some stage assumes spin and charge separation
\cite{PLee1}. A systematic theory of the renormalization of the EPI coupling
by strong electronic correlations \cite{Kulic1}, \cite{Kulic2} is exposed in
Section 5. It is shown there, that the \textit{forward scattering peak }
develops in the EPI by lowering doping, while the coupling at large transfer
momenta (the backward scattering) is suppressed .

In Section 6. we summarize the basic predictions of the theory
based on the existence of the forward scattering peak in the EPI,
impurity and Coulomb scattering, and possible relation between the
forward scattering peak in the EPI and pseudogap. The comparison
between the EPI and SFI prediction is given in Section 7. The
(im)possibility of superconductivity in the Hubbard and t-J mode
is studied in Section 8., while the obtained results are
summarized in Section 9.

\subsection{Prejudices on the EPI}

In spite of the reach experimental evidence in favor of the strong EPI in
HTSC oxides there was a disproportion in the research (especially
theoretical) activity, since the investigation of the spin fluctuations
mechanism of pairing prevailed in the literature. This was partly due to a
theoretically unfounded statement - given in \cite{Cohen}, on the upper
limit of T$_{c}$ in the phonon mechanism of pairing. It is well known that
in an electron-ion system besides the EPI there is also the repulsive
Coulomb interaction and these are not independent. In the case of an
isotropic and homogeneous system with a weak (quasi)particle interaction the
effective potential $V_{eff}(\mathbf{k},\omega )$ in the leading
approximation looks like as for two external charges ($e$) embedded in the
medium with the \textit{total macroscopic longitudinal dielectric function} $%
\varepsilon _{tot}(\mathbf{k},\omega )$ ($\mathbf{k}$ is the momentum and $%
\omega $ is the frequency) \cite{Kirzhnits}, i.e.
\begin{equation}
V_{eff}(\mathbf{k},\omega )=\frac{V_{ext}(\mathbf{k})}{\varepsilon _{tot}(%
\mathbf{k},\omega )}=\frac{4\pi e^{2}}{k^{2}\varepsilon _{tot}(\mathbf{k}%
,\omega )}.  \label{Eq1}
\end{equation}
In the case when the interaction between quasiparticles is strong, the state
of embedded quasiparticles changes substantially due to the interaction with
other quasiparticles, giving rise to $V_{eff}(\mathbf{k},\omega )\neq 4\pi
e^{2}/k^{2}\varepsilon _{tot}(\mathbf{k},\omega )$. In that case $V_{eff}$
depends on other (than $\varepsilon _{tot}(\mathbf{k},\omega )$) response
functions. However, in the case when Eq.(\ref{Eq1}) holds the weak-coupling
limit is realized where T$_{c}$ is given by $T_{c}=\bar{\omega}\exp
(-1/(\lambda -\mu ^{\ast })$ \cite{AllenMit}, \cite{Kirzhnits}$)$. Here, $%
\lambda $ is the EPI coupling constant, $\bar{\omega}$ is the average phonon
frequency and $\mu ^{\ast }$ is the Coulomb pseudo-potential, $\mu ^{\ast
}=\mu /(1+\mu \ln E_{F}/\bar{\omega})$ ($E_{F}$ is the Fermi energy). $%
\lambda $ and $\mu $ are expressed by $\varepsilon _{tot}(\mathbf{k},\omega
=0)$
\begin{equation}
\langle N(0)V_{eff}(\mathbf{k},\omega =0)\rangle \equiv \mu -\lambda
=N(0)\int_{0}^{2k_{F}}\frac{kdk}{2k_{F}^{2}}\frac{4\pi e^{2}}{%
k^{2}\varepsilon _{tot}(\mathbf{k},\omega =0)},  \label{Eq2}
\end{equation}
where $N(0)$ is the density of states at the Fermi surface and $k_{F}$ is
the Fermi momentum - see more in \cite{MaksimovReview}. In \cite{Cohen} it
was claimed that the lattice stability of the system with respect to the
charge density wave formation implies that the condition $\varepsilon _{tot}(%
\mathbf{k},\omega =0)>1$ must be fulfilled for all $\mathbf{k}$. If this
were correct then from Eq.(\ref{Eq2}) follows that $\mu >\lambda $, which
limits the maximal value of T$_{c}$ to the value $T_{c}^{\max }\approx
E_{F}\exp (-4-3/\lambda )$. In typical metals $E_{F}<(1-10)$ $eV$ and if one
accepts this (unfounded) statement that $\lambda \leq \mu \leq 0.5$ one
obtains $T_{c}\sim (1-10)$ $K$. \ The latter result, of course if it would
be true, means mean that the EPI is ineffective in producing high-T$_{c}$
superconductivity, let say not higher than $20$ $K$? However, this result is
apparently in conflict with a number of experimental results in low-T$_{c}$
superconductors (LTS), where $\mu \leq \lambda $ and $\lambda >1$. For
instance, $\lambda \approx 2.5$ is realized in $PbBi$ alloy, which is
definitely much higher than $\mu (<1)$ thus contradicting the statement made
in~Ref.\cite{Cohen}.

The statement in \cite{Cohen} that $\varepsilon _{tot}(\mathbf{k},\omega
=0)>1$ must be fulfilled for all $\mathbf{k}$ is in an apparent conflict
with the basic theory \cite{Kirzhnits}, which tells us that $\varepsilon
_{tot}(\mathbf{k},\omega )$ is not the response function. If a small
external potential $\delta V_{ext}(\mathbf{k},\omega )$ is applied to the
system it induces screening by charges of the medium and the total potential
is given by $\delta V_{tot}(\mathbf{k},\omega )=\delta V_{ext}(\mathbf{k}%
,\omega )/\varepsilon _{tot}(\mathbf{k},\omega )$ which means that $%
1/\varepsilon _{tot}(\mathbf{k},\omega )$ is the response function. The
latter obeys the Kramers-Kronig dispersion relation which implies the
following stability condition: $1/\varepsilon _{tot}(\mathbf{k},\omega =0)<1$
for $\mathbf{k}\neq 0$, i.e. either $\varepsilon _{tot}(\mathbf{k}\neq
0,\omega =0)>1$ or $\varepsilon _{tot}(\mathbf{k}\neq 0,\omega =0)<0$. This
important theorem has been first proved in the seminal article by David
Abramovich Kirzhnits \cite{Kirzhnits} and it invalidates the formula for $%
T_{c}^{\max }$ by setting aside the above\textbf{\ }restriction on the
maximal value of T$_{c}$.

Is $\varepsilon _{tot}(\mathbf{k}\neq 0,\omega =0)<0$ realized in
real systems? This question was thoroughly studied in Ref.
\cite{DKM} and in the context of HTSC in \cite{MaksimovReview},
while here we enumerate the main results. In the inhomogeneous
system, such as a crystal, the total longitudinal
dielectric function is matrix in the space of reciprocal lattice vectors ($%
\mathbf{Q}$), i.e. $\hat{\varepsilon}_{tot}(\mathbf{k+Q},\mathbf{k+Q}%
^{\prime },\omega )$, and $\varepsilon _{tot}(\mathbf{k},\omega )$ is
defined by $\varepsilon _{tot}^{-1}(\mathbf{k},\omega )=\hat{\varepsilon}%
_{tot}^{-1}(\mathbf{k+0},\mathbf{k+0},\omega )$. For instance in
the dense metallic systems with one ion per cell (such as the
metallic hydrogen ) and with the electronic dielectric function
$\varepsilon _{el}(\mathbf{k},0)$ one has \cite{DKM}
\begin{equation}
\varepsilon _{tot}(\mathbf{k},0)=\frac{\varepsilon _{el}(\mathbf{k},0)}{1-%
\frac{1}{\varepsilon _{el}(\mathbf{k},0)G_{EP}(\mathbf{k})}}.
\label{eps-toz}
\end{equation}
At the same time the frequency of the longitudinal phonon $\omega _{l}(%
\mathbf{k})$ is given by
\begin{equation}
\omega _{l}^{2}(\mathbf{k})=\frac{\Omega _{pl}^{2}}{\varepsilon _{el}(%
\mathbf{k},0)}[1-\varepsilon _{el}(\mathbf{k},0)G_{EP}(\mathbf{k})],
\label{ph-fr}
\end{equation}
where $G_{EP}$ is the local field correction $G_{EP}$ - see Ref. \cite{DKM}.
The right condition for the lattice stability requires that the phonon
frequency must be positive, $\omega _{l}^{2}(\mathbf{k})>0$, which implies
that for $\varepsilon _{el}(\mathbf{k},0)>0$ one has $\varepsilon _{el}(%
\mathbf{k},0)G_{EP}(\mathbf{k})<1$. The latter gives $\varepsilon _{tot}(%
\mathbf{k},0)<0$. The calculations \cite{DKM} show that in the metallic
hydrogen crystal $\varepsilon _{tot}(\mathbf{k},0)<0$ for all $\mathbf{k\neq
0}$. The sign of $\varepsilon _{tot}(\mathbf{k},0)$ for a number of crystals
with more ions per unit cell is thoroughly analyzed in \cite{DKM}, where it
is shown that $\varepsilon _{tot}(\mathbf{k\neq 0},0)<0$ is \textit{more a
rule than an exception}. The physical reason for $\varepsilon _{tot}(\mathbf{%
k\neq 0},0)<0$ is due to local field effects described by $G_{EP}(\mathbf{k})
$. Whenever the local electric field $\mathbf{E}_{loc}$ acting on electrons
(and ions) is different from the average electric field $\mathbf{E}$, i.e. $%
\mathbf{E}_{loc}\neq \mathbf{E}$ there are corrections to $\varepsilon
_{tot}(\mathbf{k},0)$ (or in the case of the electronic subsystem to $%
\varepsilon _{e}(\mathbf{k},0)$) which may lead to $\varepsilon _{tot}(%
\mathbf{k},0)<0$.

The above analysis tells us that in real crystals $\varepsilon _{tot}(%
\mathbf{k},0)$ can be negative in the large portion of the
Brillouin zone giving rise to $\lambda -\mu >0$, due to local
field effects. This means
that the dielectric function $\varepsilon _{tot}$ \textit{does not limit }$%
T_{c}$\textit{\ in the phonon mechanism of pairing}. The latter does not
mean that there is no limit on T$_{c}$ at all. We mention in advance that
the local field effects play important role in HTSC oxides, due to their
layered structure with ionic-metallic binding, thus giving rise to large $%
EPI $ - see more subsequent sections.

In concluding we point out, that there are no theoretical and
experimental arguments for ignoring the EPI in HTSC oxides.
However, it is necessary to answer several important questions
which are also related to experimental findings in HTSC oxides:
$\mathbf{(1)}$ If the EPI is responsible for pairing in HTSC
oxides and if superconductivity is of $d-wave$ type, how these two
facts are compatible? $\mathbf{(2)}$ Why is the transport EPI
coupling constant $\lambda _{tr}$ (entering the resistivity
formula) much smaller than the pairing EPI coupling constant
$\lambda (>1)$ (entering the formula for T$_{c}$), i.e. why one
has $\lambda _{tr}(\approx 0.4-0.9)\ll \lambda (\sim 2)$?
$\mathbf{(3)}$ Is the high T$_{c}$ value possible for a moderate
EPI coupling constant, let say for $\lambda \leq 1$?
$\mathbf{(4)}$ Finally, if the EPI interaction is ineffective for
pairing in HTSC oxides why it is so?

\section{Experiments related to pairing mechanism}

A much more extensive discussion (than here) of the experimental situation
in HTSC oxides is given in a number of papers - see reviews \cite{KulicReview}%
, \cite{MaksimovReview}. In the following we discuss briefly experimental
results, by including the most recent ones, which can give us a clue for the
pairing mechanism in the HTSC oxides.

\subsection{Magnetic neutron scattering}

\subsubsection{Normal state}

The cross-section for the \textit{inelastic neutron magnetic scattering} is
expressed via the Fourier transform of the spin-correlation function (the
spin structure factor) $S^{\alpha \alpha }(\mathbf{k},\omega )$ which is
proportional to the imaginary part of the susceptibility $Im\chi (\mathbf{k}%
,k_{z},\omega )$. In the (normal) metallic state of doped HTSC oxides
without magnetic order the inelastic scattering (in absence of the AF
magnetic order) is of interest and in most systems $Im\chi (\mathbf{k}%
,k_{z},\omega )$ is peaked around the $AF$ wave-vector $\mathbf{Q}=(\pi ,\pi
)$. The pronounced magnetic fluctuations in the underdoped metallic state is
contrary to usual metals (described by the Landau-Fermi liquid) where the
magnetic fluctuations are much weaker. In HTSC oxides $Im\chi (\mathbf{k}%
,\omega )$ depends on hole doping, and for instance, in the bilayer (two
layers per the unit cell) compound $YBa_{2}Cu_{3}O_{6+x}$ the low energy
spectra is peaked at $\mathbf{Q}$, whose width $\delta _{m}$ broadens by
increasing doping concentration - see review \cite{Bourges}. Around the
optimal doping the magnetic correlation length $\xi _{m}=(2/\delta _{m})\sim
(1-2)a$ \ is almost temperature independent. This fact contradicts the
assumption of the theory of spin fluctuation mechanism by the Pines group
\cite{Pines}, where $\xi _{m}$ is strongly T-dependent. We stress that in
the SFI theory $Im\chi (\mathbf{k},\omega )$ is important quantity since the
effective pairing potential $V_{eff}(\mathbf{k},\omega )$ and the
self-energy $\Sigma _{sf}(\mathbf{k},\omega )$ are approximately given by
(on the real frequency axis - see \cite{KulicReview})
\[
\Sigma _{sf}(\mathbf{k},\omega )\approx \sum_{\mathbf{q}}\int \frac{d\Omega}{%
\pi}G(\mathbf{q},\Omega )V_{eff}(\mathbf{k}+\mathbf{q},\omega +\Omega ),
\]
\begin{equation}
V_{SF}(\mathbf{q},\omega +i0^{+})=g_{SF}^{2}\int_{-\infty }^{\infty }\frac{%
d\Omega }{\pi }\frac{Im \chi (\mathbf{q},\Omega +i0^{+})}{\Omega -\omega }
\label{Eq3}
\end{equation}
where $G(\mathbf{q},\Omega )$ is the electron Green's function. This
approach can be theoretically justified in the weak coupling limit ($U\ll W$%
) only . Although the HTSC oxides are far from this limit this expression is
frequently used in the SFI theories of pairing, where larger $Im\chi (%
\mathbf{k},\omega )$ should give larger T$_{c}$.

\begin{figure}[tbp]
\resizebox{1.0\textwidth}{!}
{\includegraphics[width=12cm]{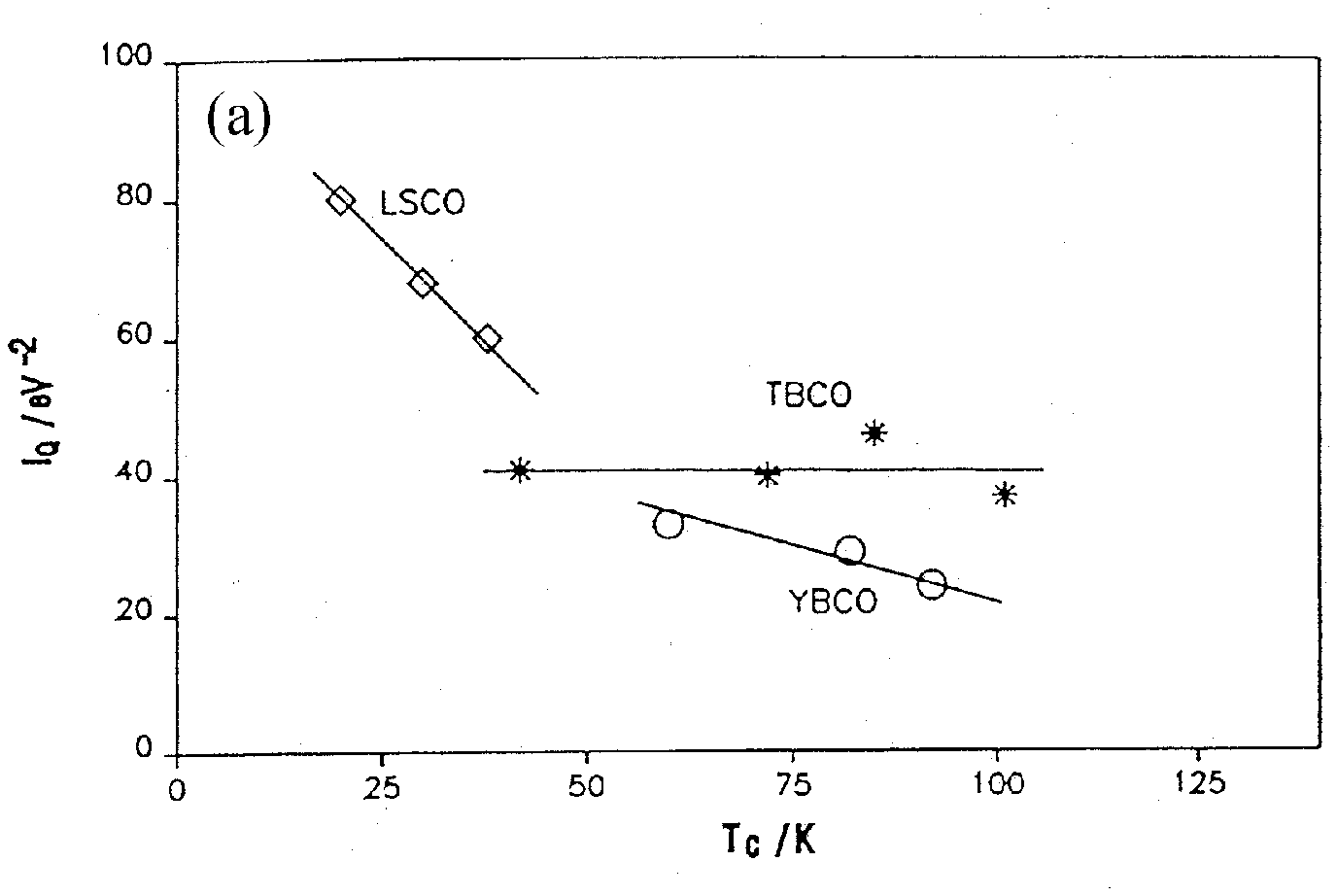}
\includegraphics[width=12cm]{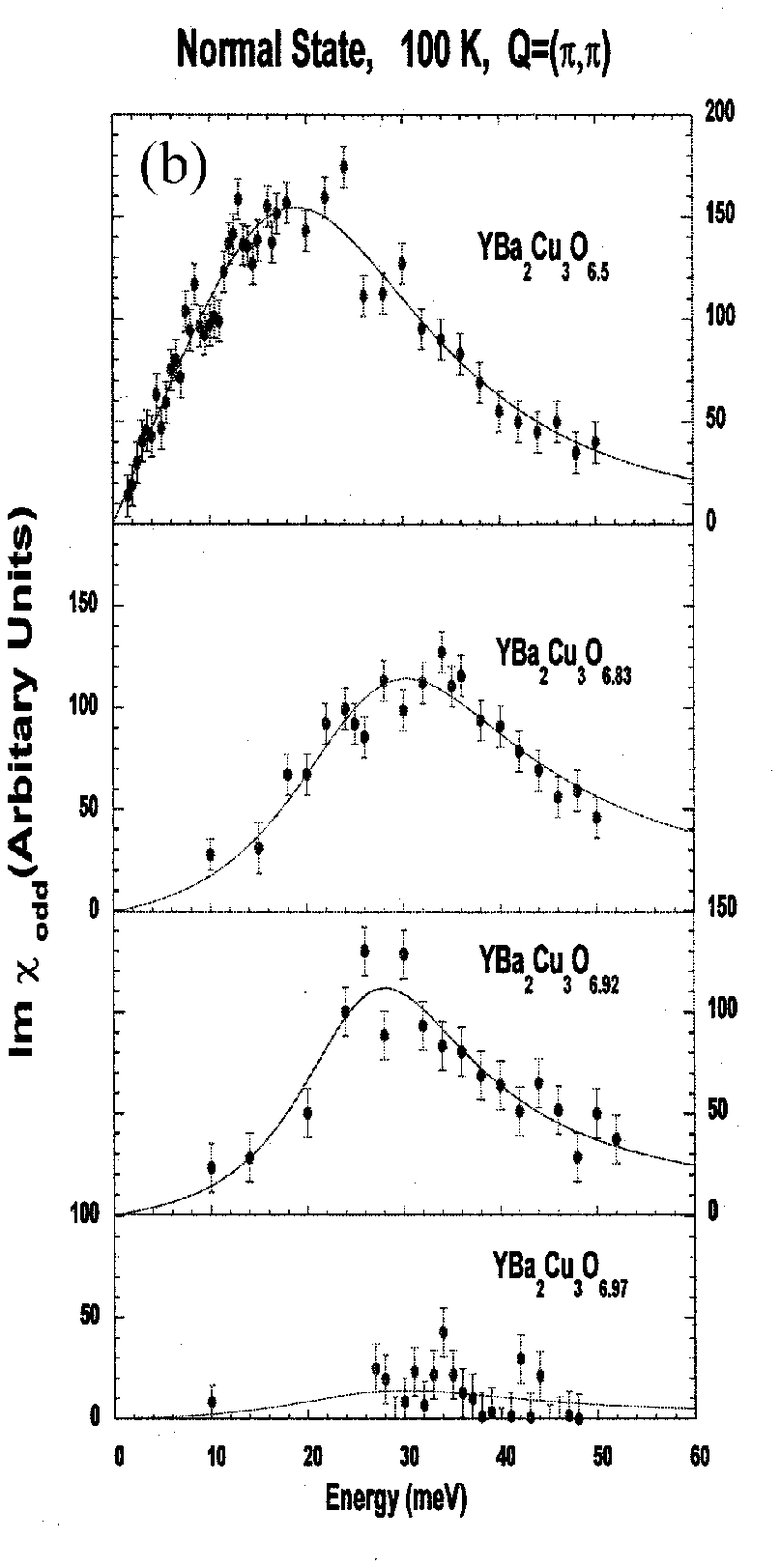}
\includegraphics[width=12cm]{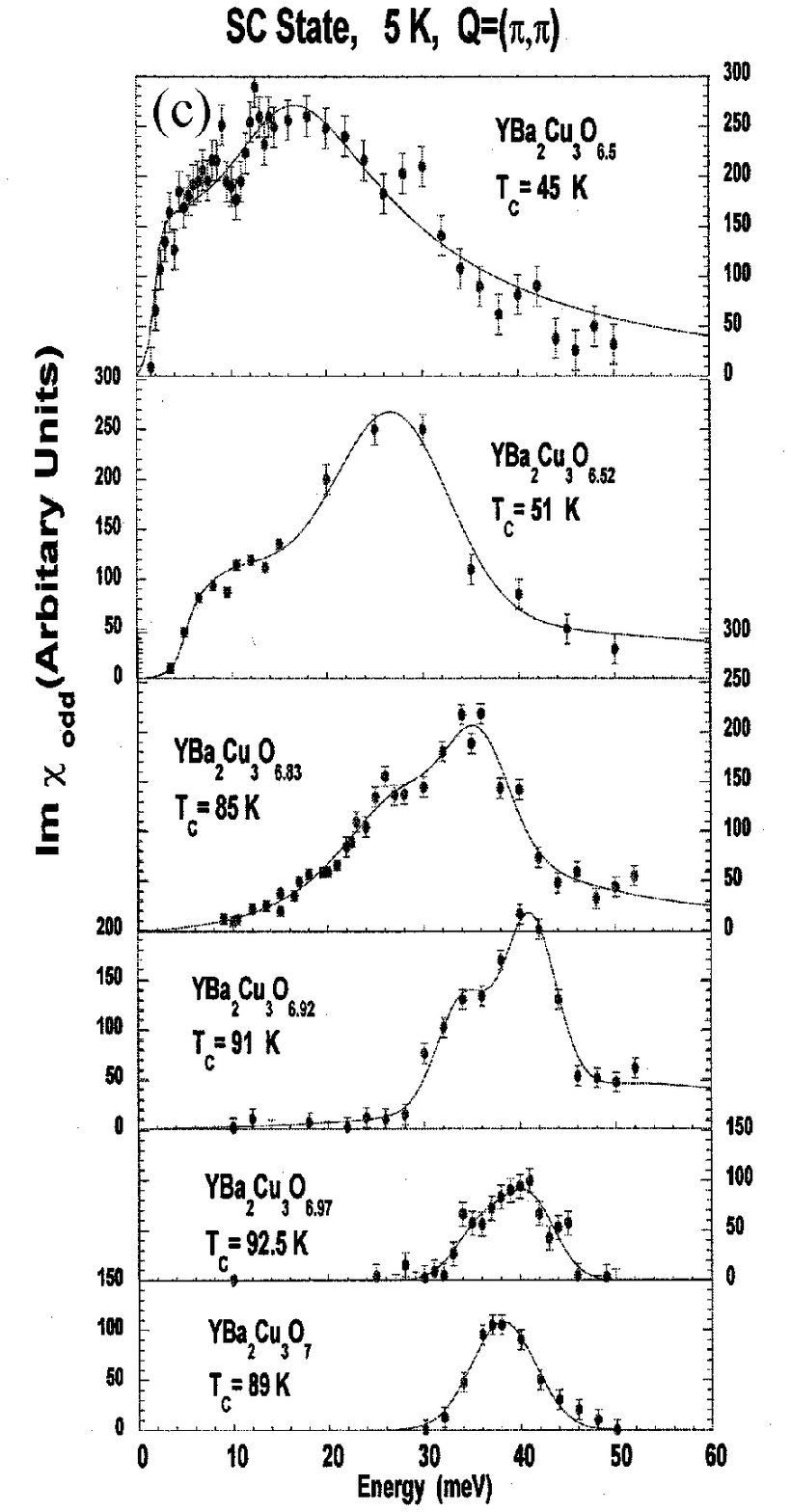}}
\caption{Magnetic spectral function $Im\protect\chi ^{(-)}(\mathbf{k},%
\protect\omega )$: (a) $I_{Q}(T_{c})$ values at $T=200$ $K$ for various HTSC
oxides: $LSCO$ - $La_{2x}Sr_{x}CuO_{4}$; $TBCO$ - $Tl_{2}Ba_{2}CuO_{6+x}$
and $Tl_{2}Ba_{2}CaCu_{2}O_{8}$; $YBCO$ - and $YBa_{2}Cu_{4}O_{8}$ - from
\protect\cite{Mehring1}; (b) for $YBa_{2}Cu_{3}O_{6+x}$ in the normal state
at $T=100$ $K$ and at $Q=(\protect\pi ,\protect\pi )$. $100$ counts in the
vertical scale corresponds to $\protect\chi _{max}^{(-)}\approx 350\protect%
\mu _{B}^{2}/eV$ - from \protect\cite{Bourges}; (c) for $%
YBa_{2}Cu_{3}O_{6+x} $ in the superconducting state at $T=5$ $K$ and at $Q=(%
\protect\pi ,\protect\pi )$ - from \protect\cite{Bourges}.}
\label{SuscFig}
\end{figure}

What is the experimental situation? The antibonding (odd) spectral function $%
Im\chi ^{(odd)}(\mathbf{k},\omega )$ of $YBa_{2}Cu_{3}O_{6+x}$ is strongly
doping dependent as it is seen in Fig.~\ref{SuscFig}. By comparing the
magnetic neutron scattering (normal state) spectra in $YBa_{2}Cu_{3}O_{6.92}$
and $YBa_{2}Cu_{3}O_{6.97}$ in Fig.~\ref{SuscFig}a the difference is
reflected in their spectral functions $Im\chi ^{(odd)}(\mathbf{k},\omega )$.
Namely, in the frequency interval which is important for superconducting
pairing $Im\chi ^{(odd)}(\mathbf{Q},\omega )$ of

$YBa_{2}Cu_{3}O_{6.92}$ is much larger than that in $YBa_{2}Cu_{3}O_{6.97}$
although the differences in their critical temperatures T$_{c}$ is very
small, i.e. $T_{c}=91$ $K$ for $YBa_{2}Cu_{3}O_{6.92}$ and $T_{c}=92.5$ $K$
for $YBa_{2}Cu_{3}O_{6.97}$. This result, in conjunction with the
anti-correlation between the NMR spectral function $I_{\mathbf{Q}%
}=\lim_{\omega \rightarrow 0}Im\chi (\mathbf{Q},\omega )/\omega $ and T$_{c}$
- shown in Fig.~\ref{SuscFig}a, is apparently against the SFI theoretical
models for pairing mechanism \cite{Pines}, \cite{Scalapino1}.

\subsubsection{Superconducting state}

In the superconducting state the magnetic fluctuations are
drastically changed, what is in fact expected for the singlet
pairing state which induces spin gap in the magnetic excitation
spectrum of s-wave superconductors. However, the spectrum in the
superconducting state of HTSC oxides is more complex due to d-wave
pairing and specificity of the band structure. For instance, at
$T<T_{c}$ the sharp peak in $Im\chi ^{(odd)}(\mathbf{k},\omega ) $
is seen at $\omega _{reson}=41$ $meV$ and at $\mathbf{k}_{2D}=(\pi
/a,\pi /a)$ of the fully oxygenated (optimally doped)
$YBa_{2}Cu_{3}O_{6+x}$ ($x\sim 1,$ $T_{c}\approx 92 $ K)
\cite{Rossat-Mignod}, \cite{Mook}. The doping dependence of the
peak position and its width \cite{Bourges} is shown in
Fig.~\ref{SuscFig}c, where it is seen that by increasing doping
the peak in the superconducting state becomes sharper and moves to
higher frequencies (scaling with T$_{c}$), while its height is
decreasing. This can be qualitatively explained by using the $RPA$
susceptibility
\begin{equation}
\chi (\mathbf{k},\omega )=\frac{\chi _{0}(\mathbf{k},\omega )}{1-U_{eff}(%
\mathbf{q})\chi _{0}(\mathbf{k},\omega )},  \label{pa-sus}
\end{equation}
where the bare susceptibility $\chi _{0}(\mathbf{k},\omega )$ contains the
coherence factor $[1-(\xi _{\mathbf{k}+\mathbf{q}}\xi _{\mathbf{q}}+\Delta _{%
\mathbf{k}+\mathbf{q}}\Delta _{\mathbf{q}})/E_{\mathbf{k}+\mathbf{q}}E_{%
\mathbf{q}}]$ - see \cite{KulicReview}. This (type II) coherence
factor reflects the (well known) fact that the magnetic scattering
is not the time reversal symmetry. In the case when $\mathbf{k}$
and $\mathbf{k}+\mathbf{q}$ are near the Fermi surface and when
$\Delta _{\mathbf{k}+\mathbf{q}}\approx -\Delta _{\mathbf{q}}$ at
$\mathbf{k}=\mathbf{Q}=(\pi /a,\pi /a)$ the
coherence factor is of the order of one at or near the Fermi surface (note $%
\xi _{\mathbf{k}+\mathbf{q}}\xi _{\mathbf{q}}\leq 0$) and therefore
contributes significantly to $\chi _{0}(\mathbf{k}=\mathbf{Q},\omega )$. The
case $\Delta _{\mathbf{q}}\Delta _{\mathbf{k}+\mathbf{q}}<0$ is realized
when the $d-wave$ order parameter, for instance $\Delta _{\mathbf{k}%
}=(\Delta _{0}/2)[\cos k_{x}-\cos k_{y}].$ So, the mechanism of the peak
formation (below T$_{c}$) is the consequence of the electron-pair creation
with an electron in the ($+$ ) lobe and a hole in the ($-$) lobe of the
superconducting order parameter. Note, that the ($\pm $) lobes of $\Delta _{%
\mathbf{k}}$ are separated approximately by the wave-vector $\mathbf{Q}=(\pi
/a,\pi /a)$. Due to the large density of states near the lobes a large peak
in $Im\chi (\mathbf{k}=\mathbf{Q},k_{z},\omega )$ is expected to be
realized, i.e. $\omega _{reson}\geq 2\Delta _{0}$. Of course the better
(than RPA) calculations of \ $\chi (\mathbf{k}=\mathbf{Q},k_{z},\omega )$ is
needed for a full quantitative analysis, where a possible resonance in $\chi (\mathbf{%
k},\omega )$ with $\omega _{reson}\leq 2\Delta _{0}$ can also
contribute. It is important to stress, that the magnetic resonance
in the superconducting state is \textit{consequence of
superconductivity }but not its cause as it was stated in some
papers. It can not be the cause for superconductivity simply
because its intensity at $T$ around T$_{c}$ is vanishing small and
not affecting T$_{c}$ at all. If the magnetic resonance would be
the origin for superconductivity (and high T$_{c}$) the phase
transition at T$_{c}$ must be first order, contrary to experiments
where it is second order.

The next very serious argument against the SFI pairing mechanism
is the \textit{smallness of the coupling constant} $g_{sf}$.
Namely, the real spin-fluctuation coupling constant is rather
small $g_{sf}\leq 0.2$ $eV$, what is in contrast to the large
value ($g_{sf}^{(MMP)}\sim 0.6$ $eV$) assumed in the SFI theory by
the
Pines group - see the MMP model in Section 7.1. The upper limit of $%
g_{sf}(\leq 0.2$ $eV)$ is extracted from : (\textbf{i}) the width of the
resonance peak \cite{Kivelson2}, and (\textbf{ii}) the small magnetic moment
($\mu <0.1$ $\mu _{B}$) in the antiferromagnetic state of LASCO and YBCO
\cite{KulicKulic}. Note, that the pairing coupling in the SFI theory is $%
\lambda _{sf}\sim g_{sf}^{2}$, and for the realistic value of $g_{sf}\leq 0.2
$ $eV$ it would produce $\lambda _{sf}\sim 0.2$ and very small $T_{c}\sim 1$
$K$. The $SFI$ model roots on its basic $t-J$ Hamiltonian. However, recently
it was shown in \cite{Pryadko} that \textit{there is no superconductivity in
the t-J model} at temperatures characteristic for HTSC oxides - see Fig.~%
\ref{NoScFig} below. If it exists T$_{c}$ must be very low.

\textit{In conclusion}, the inelastic magnetic neutron scattering
give evidence that the spin fluctuations interaction (SFI),
although pronounced in underdoped systems, is \textit{ineffective}
in the pairing mechanism of HTSC oxide. However, the SFI in
conjunction with the residual Coulomb repulsion \textit{triggers}
superconductivity from s-wave to d-wave, whose strength is
predominantly due to the EPI - see discussion in Sections 5.-7..

\subsection{Dynamical conductivity and resistivity $\rho (T)$}

Since $\sigma (\omega )$ and $\rho (T)$ give important information
on the dominant scattering mechanism, in the following we analyze
their properties in more details.

\subsubsection{Dynamical conductivity $\protect\sigma (\protect\omega )$}

$\sigma (\omega )$ is in fact \textit{derived quantity} since it
is extracted from the
\textit{measured} optic reflectivity $R(\omega )$ and absorption $A(\omega )$%
. By measuring the normal-incident (of light) reflectivity $R(\omega )$ in
the whole frequency region ($0\leq \omega <\infty $) one can determine the \
phase $\phi (\omega )$ of the complex reflectivity
\begin{equation}
r(\omega )=\sqrt{R(\omega )}e^{i\phi (\omega )}=\frac{\sqrt{\varepsilon
(\omega )}-1]}{\sqrt{\varepsilon (\omega )}+1}  \label{Refl}
\end{equation}
by the Kramers-Kronig relation, and accordingly to determine in principle
the complex dielectric function
\begin{equation}
\varepsilon (\omega )=\varepsilon _{\infty }+\varepsilon _{latt}(\omega )+%
\frac{4\pi i\sigma (\omega )}{\omega },  \label{eps}
\end{equation}
where $\varepsilon _{\infty }$ and $\sigma (\omega )$ are
electronic contributions and $\varepsilon _{latt}$ is the lattice
contribution. However, $R(\omega )$ is usually measured in a
finite $\omega $ region and extrapolations is needed, especially
at very low frequencies. This extrapolation\ of $R(\omega )$ also
contains some model assumptions on
the scattering processes in the system (on $\sigma (\omega )$), i.e. $%
1-R(\omega )\sim \sqrt{\omega }$ - the Hagen-Rubens relation for the
standard (with elastic scattering only) Drude metal, or $1-R(\omega )\sim
\omega $ for strong EPI (or for marginal Fermi liquid). So, one should be
always cautious not to overinterpret the meaning of $\sigma (\omega )$
obtained in such a way.

In HTSC oxides $R(\omega )$, $A(\omega )$ are usually measured in
a broad frequency region - up to several $eV$. At such high
frequencies the interband transitions take place and in order to
calculate $\sigma (\omega )$ the knowledge of the band structure
is needed. This problem was analyzed in the framework of the LDA
band structure calculations \cite{Maksimov2} by taking into
account the interband transitions, where a rather good agreement
with experiments for $\omega >1$ $eV$ was found. This is
surprising since the $LDA$-method does not contain the Hubbard
bands, and according to \cite{Maksimov2} there is no sign of
transitions between Hubbard sub-bands in the high energy region of
$\sigma (\omega )$. This very interesting result deserves to be
further analyzed since it contradicts the physics of the Hubbard
models.

\begin{figure}[tbp]
\resizebox{0.9 \textwidth}{!} {
\includegraphics*[width=16cm]{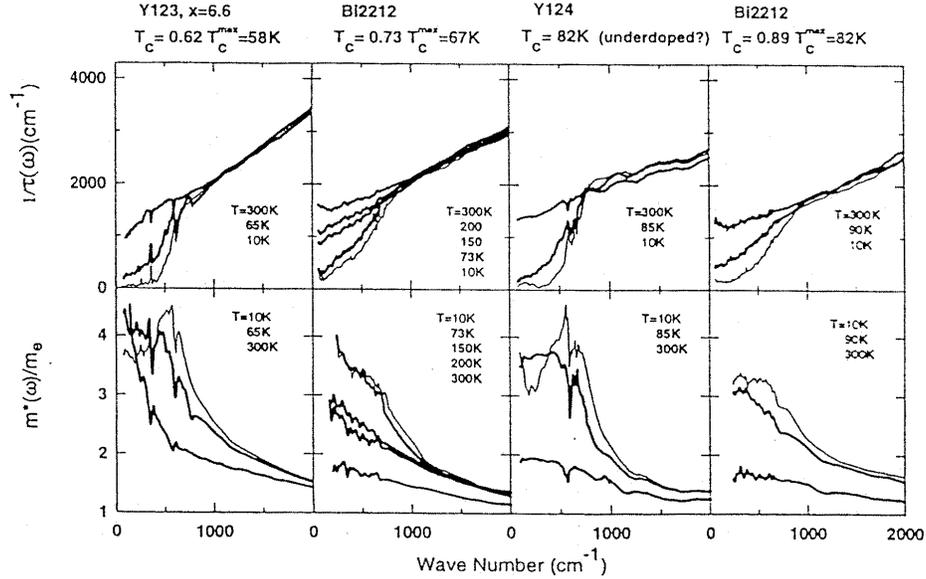}}
\caption{The transport scattering rate $1/\protect\tau (\protect\omega )$
(in the text $\Gamma _{tr}(\protect\omega )$) and the transport effective
mass $m^{\ast }(\protect\omega )/m_{e}$ (in the text $m_{tr}(\protect\omega
)/m_{\infty }$) for series of underdoped HTSC oxides. $\Gamma _{tr}(\protect%
\omega )$ is temperature independent above $1000$ $cm^{-1}$ but it is
depressed at low $T$ and low $\protect\omega $- from \protect\cite{Puchkov}.}
\label{GammaExpFig}
\end{figure}

Here we discuss briefly the normal state $\sigma (\omega )$ in the low
frequency region $\omega <1$ $eV$ where the \textit{intraband} effects
dominate the quasiparticle scattering. In the low $\omega $ regime the
processing of the data in the metallic state of HTSC oxides is usually done
by using the generalized Drude formula for the inplane conductivity $\sigma
(\omega )=\sigma _{1}+i\sigma _{2}$ \cite{Schlesinger}, \cite{Dolgov}, \cite
{Shulga}
\begin{equation}
\sigma _{ii}(\omega )=\frac{\omega _{p,ii}^{2}}{4\pi }\frac{1}{\Gamma
_{tr}(\omega ,T)-i\omega m_{tr}(\omega )/m_{\infty }}.  \label{Eq4}
\end{equation}
$i=a,b$ enumerates the plane axis, $\Gamma _{tr}(\omega ,T)$ and $%
m_{tr}(\omega )$ are the transport scattering rate and optic mass,
respectively. Sometimes in the analysis of experimental data the effective
transport scattering rate $\Gamma _{tr}^{\ast }(\omega ,T)$ and the
effective plasma frequency $\omega _{p}^{\ast }(\omega )$ are used, which
are defined by
\begin{equation}
\Gamma _{tr}^{\ast }(\omega ,T)=\frac{m}{m_{tr}(\omega )}\Gamma _{tr}(\omega
,T)=\frac{\omega \sigma _{1}(\omega )}{\sigma _{2}(\omega )},
\label{Gam-trans}
\end{equation}
and
\begin{equation}
\omega _{p}^{\ast 2}(\omega )=\frac{m}{m_{tr}(\omega )}\omega _{p}^{2}.
\label{pl-trans}
\end{equation}
For best optimally doped HTSC systems the best fit for $\Gamma
_{tr}^{\ast }(\omega ,T)$ is given by $\Gamma _{tr}^{\ast }(\omega
,T)\approx \max \{\alpha T,\beta \omega \}$ in the temperature and
frequency range from very low ($\sim 100$ $K$) up to $2000$ $K,$
where $\alpha ,\beta $ are of the order one - see
Fig.~\ref{GammaExpFig}. These results tell us that the
quasiparticle liquid, which is responsible for transport
properties in HTSC, is not a simple (weakly interacting) Fermi
liquid. We remind the reader that in the usual (canonical) normal
Fermi liquid\textbf{\ }with the Coulomb interaction on has $\Gamma
_{tr}(\omega ,T)\sim \Gamma _{tr}^{\ast }(\omega ,T)\sim \Gamma
(\omega ,T)\sim $ $\max \{T^{2},\omega ^{2}\}$ at low $T$ and
$\omega $, which means that quasiparticles are well defined
objects near (and at) the Fermi surface since $\omega \gg \Gamma
(\omega ,T)$. In case of HTSC oxides with $\Gamma (\omega ,T)\sim
\max (\omega ,T)$ in the broad regions and the quasiparticles
decay rapidly and therefore are not well defined objects. At these
temperatures and frequencies the simple canonical Landau
quasiparticle concept fails. The latter behavior can be due to the
strong electron-electron inelastic scattering, or due to the
quasiparticle scattering on phonons (or on other bosonic
excitations). It is important to stress, that quasiparticles

interacting with phonons at finite $T$ are not described with the standard
Fermi liquid, in particular at $T>\Theta _{D}/5$, since the scattering rate
is larger than the quasiparticle energy, i.e. one has $\Gamma \sim \max
(\omega ,T)$. Such a system is well described by the \textit{%
Migdal-Eliashberg theory}\textbf{\ }whenever $\omega _{D}\ll
E_{F}$ is fulfilled, which in fact treats quasiparticles beyond
the original Landau quasiparticle concept. Note, that even when
the original Landau quasiparticle concept fails the transport
properties may be described by the Boltzmann equation, which is a
wider definition of the Landau-Fermi liquid.

We point out, that in a number of articles it was
\textit{incorrectly} assumed that $\Gamma (\omega ,T)\approx
\Gamma _{tr}(\omega ,T)\approx \Gamma _{tr}^{\ast }(\omega ,T)$
holds in HTSC oxides . The above discussed experiments (see
Fig.~\ref{GammaExpFig}) give that $\Gamma _{tr}^{\ast }(\omega
,T)$ is linear in the broad region of $\omega $ and $T$ up to
$2500$ $K$ - see in Fig.~\ref{GammaExpFig}. However, if $\Gamma
(\omega ,T)$ is due to the EPI it saturates at the maximum phonon
frequencies $\omega _{\max }^{ph}(\leq 1000$ $K)$. By assuming
also that $\Gamma ^{EPI}(\omega ,T)\approx \Gamma
_{tr}^{EPI}(\omega ,T)$ holds for all $\omega $, in a number of
papers it was concluded, that the EPI does not contribute to the
inelastic scattering of quasiparticles and to the Cooper pairing
in HTSC oxides. Does it hold $\Gamma ^{EPI}(\omega ,T)\approx
\Gamma _{tr}^{EPI}(\omega ,T)$ in HTSC oxides? The answer is NO.

\begin{figure}[tbp]
\resizebox{1.0\textwidth}{!} {
\includegraphics*[width=8cm]{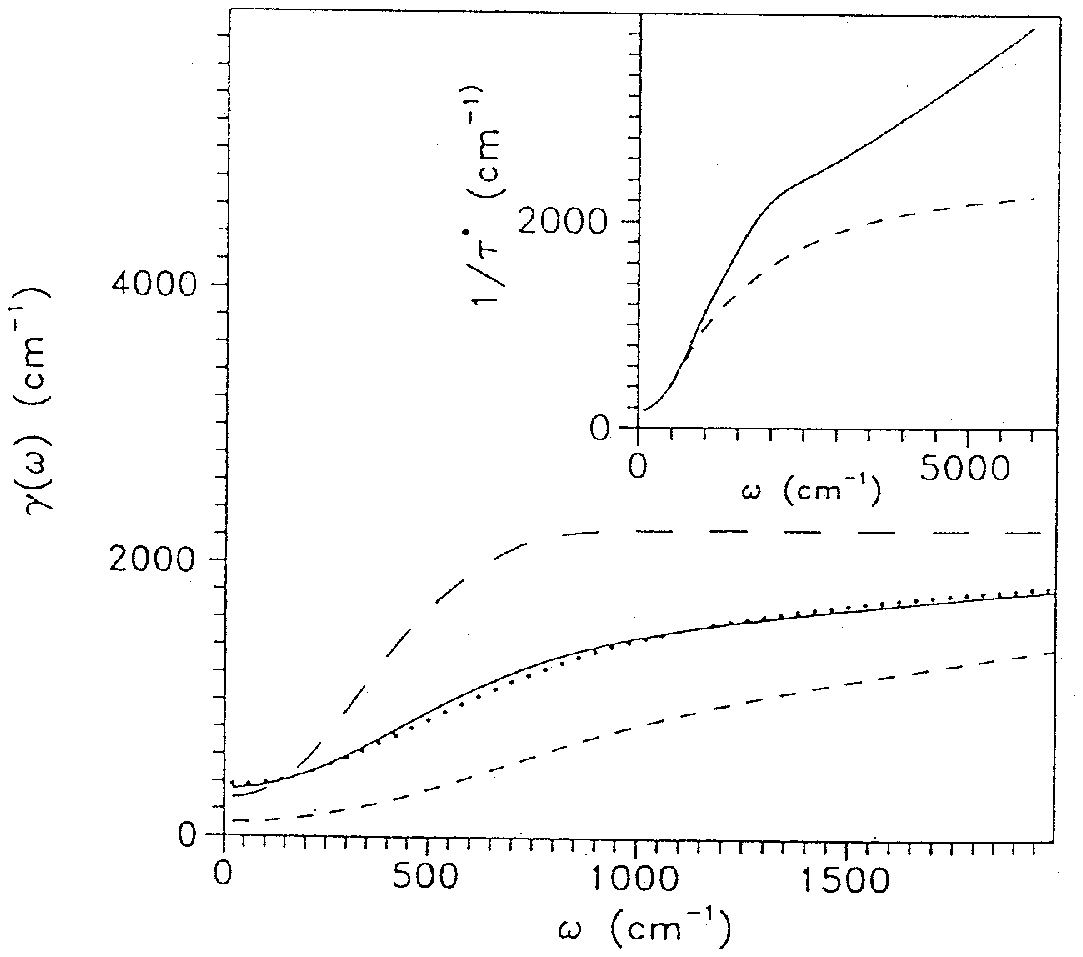}
\includegraphics*[width=6cm]{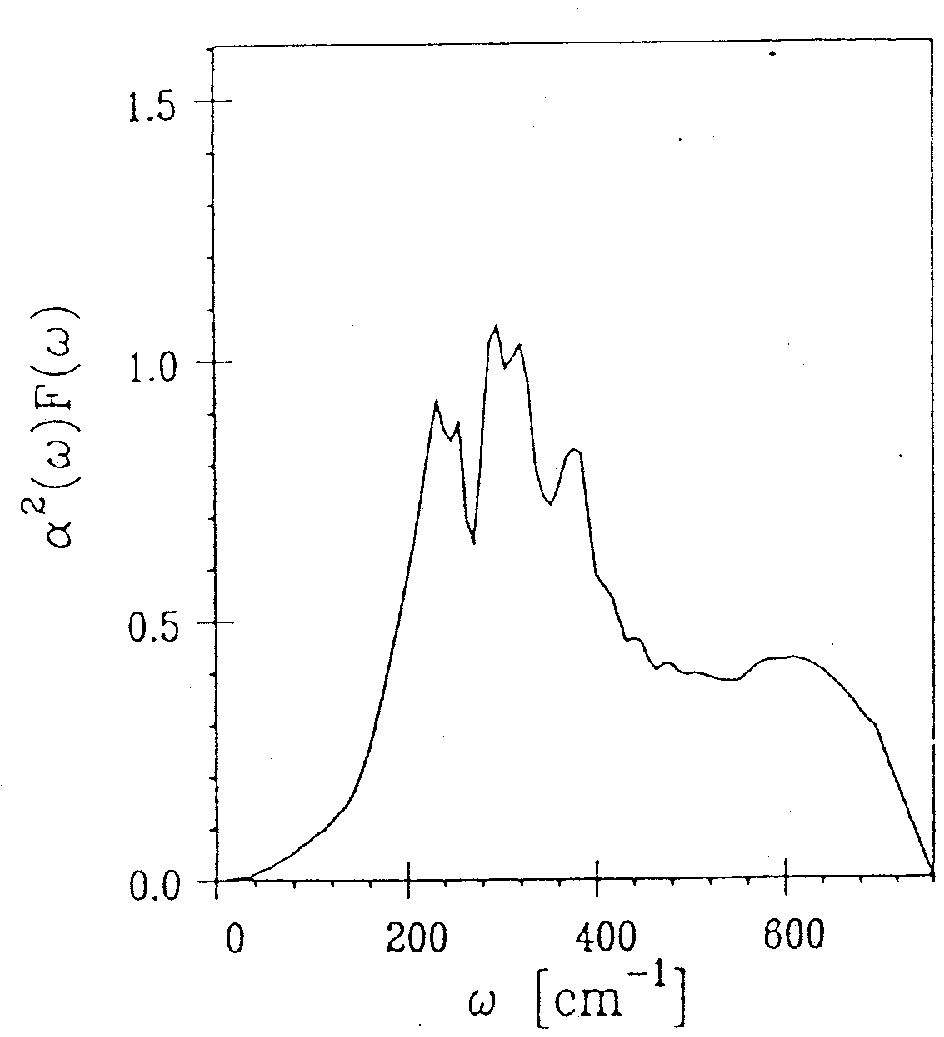}}
\caption{The theoretical predictions for the frequency dependence of the
various relaxation rates $\protect\gamma (=\Gamma )$ with $\protect\alpha %
^{2}F(\protect\omega )$ - right: the generalized Drude fit for $\Gamma _{tr}(%
\protect\omega )$ - solid line; $\Gamma _{tr}^{\ast }(\protect\omega )$ -
short-dashed line; $\Gamma (\protect\omega )$ - long dashed line; $\Gamma
_{tr}(\protect\omega )$ calculated - dotted line. In the inset the
calculated $\Gamma _{tr}^{\ast }(\protect\omega )(=1/\protect\tau ^{\ast }(%
\protect\omega ))$ with (solid line) and without (dashed line) the interband
contributions with $\protect\alpha ^{2}F(\protect\omega )$ from right and at
$T=100$ $K$ - from \protect\cite{Shulga}. }
\label{TheoGammaFig}
\end{figure}

$\sigma (\omega )$ of HTSC oxides was theoretically analyzed
\cite{Dolgov}, \cite{Shulga} in terms of the EPI, where it was
found that $\Gamma _{tr}(\omega ,T)$ and $m_{tr}(\omega ,T)$
depend on the transport spectral function $\alpha
_{tr,EP}^{2}F(\omega )$ - see more in \cite{KulicReview}. Their
analysis is based on: (i) the assumption that $\alpha
_{tr,EP}^{2}F(\omega )\approx \alpha _{EP}^{2}F(\omega )$ - the
Eliashberg spectral function: (ii) the shape of $\alpha
_{EP}^{2}F(\omega )$ is extracted from various tunnelling
conductivity measurements, \cite{Tun2}, \cite{Tun3}, \cite{Tun4},
\cite{Gonnelli}, which makes a rather large EPI coupling constant
and the critical temperature $\lambda =2\int_{0}^{\infty }d\omega
\alpha ^{2}(\omega )F(\omega )/\omega \approx 2$ and $T_{c}\approx
90$ $K$, respectively; (iii) the plasma frequency is taken to be
$\omega _{pl}=3$ $eV$. It was obtained that $\Gamma
_{tr}^{EP}(\omega ,T)\sim \omega $ in a very broad $\omega
$-interval (up to $250$ $meV$), which is much
larger than the maximum phonon frequency $\omega _{\max }^{ph}\approx 80$ $%
mev$. This is illustrated in Fig.~\ref{TheoGammaFig}. Moreover,
$\Gamma _{tr}^{EP}(\omega ,T)$ \textit{differs significantly} from
the quasiparticle
scattering rate $\Gamma ^{EP}(\omega ,T)=-2Im\Sigma (\omega )$ \cite{Dolgov}%
, \cite{Shulga}. We see from Fig.~\ref{TheoGammaFig} that $\Gamma
^{EP}(\omega ,T)$ is much steeper function than $\Gamma _{tr}^{EP}(\omega
,T) $ and the former saturates at much lower frequency - of the order of the
maximum phonon frequency $\omega _{\max }^{ph}$.

Note, that $\Gamma _{tr}^{\ast ,EP}(\omega ,T)=(m/m_{tr}(\omega ))\Gamma
_{tr}^{EP}(\omega ,T)$ is also quasi-linear function in a very broad region $%
150$ $K<\omega <3000$ $K$ - see Fig.~\ref{TheoGammaFig}. The slope of $%
\Gamma _{tr}^{\ast ,EP}(\omega ,T)$ is of the order of one, in accordance
with experiments results \cite{Dolgov}, \cite{Shulga}, and it (and $\Gamma
_{tr}^{EP}(\omega ,T)$) saturates at $\omega _{sat}\simeq -Im\Sigma
_{tr}(\omega _{sat})\gg \omega _{\max }^{ph}$ only. The transport spectral
function $\alpha _{tr}^{2}(\omega )F(\omega )$ can be also extracted from
the transport scattering rate $\Gamma _{tr}(\omega ,T=0)$ - see \cite
{KulicReview}, \cite{MaksimovReview}, since the theory gives that
\begin{equation}
\Gamma _{tr}(\omega ,T=0)=\frac{2\pi }{\omega }\int_{0}^{\omega }d\Omega
(\omega -\Omega )\alpha _{tr}^{2}(\Omega )F(\Omega ).  \label{Gam-trEq}
\end{equation}
However, real measurements are performed at finite$T(>T_{c}$) where $\alpha
_{tr}^{2}(\omega )F(\omega )$ is the solution of the Fredholm integral
equation (of the first kind) . Such an inverse problem at finite
temperatures in HTSC oxides is studied first in \cite{Dolgov} (see also \cite
{Shulga}), where the smeared structure of $\alpha _{tr}^{2}(\omega )F(\omega
)$ in $YBa_{2}Cu_{3}O_{7-x}$\ was obtained, which is in qualitative
agreement with the shape of the phonon density of states $F(\omega )$. At
finite $T$ the problem is more complex because the fine structure of $\alpha
_{tr}^{2}(\omega )F(\omega )$ gets blurred as the calculations in \cite
{Kaufmann} show. The latter gave that $\alpha _{tr}^{2}(\omega )F(\omega )$
ends up at $\omega _{\max }\approx 70-80$ $meV$, which is the maximal phonon
frequency in HTSC oxides. This result indicates strongly that the EPI in
HTSC oxides is dominant in the IR optics. We point out, that if $R(\omega )$
(and $\sigma (\omega )$) are due to some bosonic process with large
frequency cutoff $\omega _{c}$ in the spectrum, as it is the case with the
spin-fluctuation (SFI) scattering where $\omega _{c}\approx 400$ $meV$, the
extracted $\alpha _{tr}^{2}(\omega )F(\omega )$ should end up at this high $%
\omega _{c}$. The latter is \textit{not seen} in optic measurements at $%
T>T_{c}$, which tells us that the SFI scattering, with $\alpha
_{tr}^{2}(\omega )F(\omega )\sim g_{sf}^{2}Im\chi _{s}(\omega )$
and with the cutoff $\omega _{c}\geq 400$ $meV$, is rather weak
and ineffective in optics of HTSC oxides..

We stress that the extraction of $\Gamma _{tr}$ from $R(\omega )$ is subtle
procedure, because it depends also on the assumed value of $\varepsilon
_{\infty }$. For instance, if one takes $\varepsilon _{\infty }=1$ then $%
\Gamma _{tr}^{EP}$ is linear up to very high $\omega $, while for $%
\varepsilon _{\infty }>1$ the linearity of $\Gamma _{tr}^{EP}$
saturates at lower $\omega $. Since $\Gamma _{tr}^{EP}(\omega
,T)$, extracted in \cite {Puchkov}, and recently also in
\cite{Timusk-Nature}, is linear up to very high $\omega $ it may
be that the ion background and interband transitions (contained in
$\varepsilon _{\infty }$) are not properly taken into account in
these papers. As a curiosity in a number of papers, even in the
very cited ones such as \cite{Puchkov}, \cite{Timusk-Nature},
there is no information which value for $\varepsilon _{\infty }$
they take. We stress again, that the behavior of $\Gamma
_{tr}(\omega )$ is linear up to much higher frequencies for
$\varepsilon _{\infty }=1$ than for $\varepsilon _{\infty }\approx
4-5$ - the characteristic value for HTSC, giving a lot of room for
inadequate interpretations of results. In that respect, the recent
elipsometric optic measurements on YBCO \cite{BorisMPI} confirm
the results of the previous ones \cite{Bozovic4} that $\varepsilon
_{\infty }\geq 4$ and that $\Gamma _{tr}^{EP}$ saturates at lower
frequency than it was the case in Ref. \cite
{Puchkov}. We stress again that the reliable estimation of the value and $%
\omega ,T$ dependence of $\Gamma _{tr}(\omega )$ and $m(\omega )$ can be
done, not from the reflectivity measurements \cite{Puchkov}, \cite
{Timusk-Nature}, but from elipsometric ones only \cite{Bozovic4}, \cite
{BorisMPI}.

\begin{figure}[tbp]
\resizebox{.8\textwidth}{!} {
\includegraphics*[width=8cm]{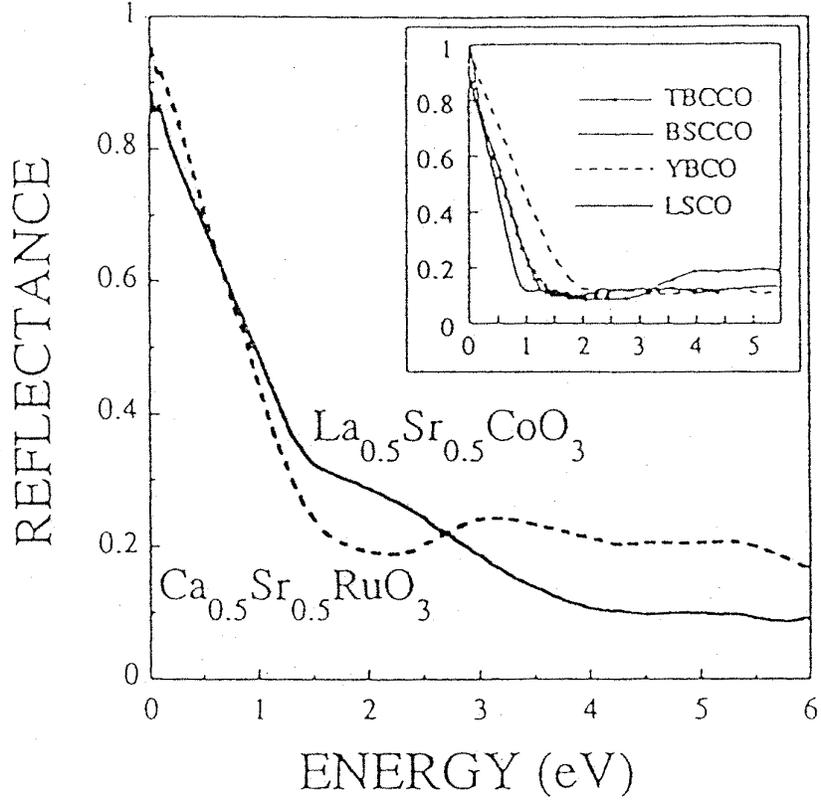}}
\caption{Broad range specular reflectance spectra of $%
Ca_{0.5}Sr_{0.5}RuO_{3} $ (broken line) and $La_{0.5}Sr_{0.5}CoO_{3}$ (solid
line). Inset spectra of $Tl_{2}Ba_{2}Ca_{2}Cu_{3}O_{10}$, $%
Bi_{2}Sr_{2}CaCu_{2}O_{8}$, $YBa_{2}Cu_{3}O_{7}$ and $%
La_{1.85}Sr_{0.15}CuO_{4}$. From \protect\cite{Bozovic}. }
\label{BozovFiga}
\end{figure}

In concluding this part we stress two facts: (\textbf{1}) The large
difference in the $\omega ,T$ behavior of $\Gamma _{tr}^{EP}(\omega ,T)$ and
$\Gamma ^{EP}(\omega ,T)$ is not a specificity of HTSC oxides but it is
realized also in a number of $LTSC$ materials. In fact this is a common
behavior even in simple metals, such as Al, Pb, as shown in \cite{MaksSav},
where $\Gamma ^{EP}(\omega ,T)$ saturates at much lower (Debay) frequency
than $\Gamma _{tr}^{EP}(\omega ,T)$ and $\Gamma _{tr}^{\ast ,EP}(\omega ,T)$
do. In that respect the difference between simple metals and HTSC oxides is
in the scale of phonon frequencies, i.e. $\omega _{\max }^{ph}\sim 100$ $K $
in simple metals, while $\omega _{\max }^{ph}\sim 1000$ $K$ in HTSC oxides.
Having in mind these well established and well understood facts, it is very
surprising that even nowadays, 18 years after the discovery of HTSC oxides,
the principal and quantitative difference between $\Gamma $ and $\Gamma _{tr}
$ is neglected in the analysis of experimental data. For instance, by
neglecting the pronounced (qualitative and quantitative) difference between $%
\Gamma _{tr}(\omega ,T)$ and $\Gamma (\omega ,T)$, in the recent papers \cite
{Timusk-Nature}, \cite{Norman-Nature} were made far reaching, but
unjustified, conclusions that the magnetic pairing mechanism prevails; (%
\textbf{2}) It is worth of mentioning, that quite similar (to HTSC oxides)
properties, of $\sigma (\omega )$, $R(\omega )$ and $\rho (T)$ were observed
in experiments \cite{Bozovic} on isotropic metallic oxides $%
La_{0.5}Sr_{0.5}CoO_{3}$ and $Ca_{0.5}Sr_{0.5}RuO_{3}$ - see Fig.~\ref
{BozovFiga}. We stress that in these compounds there are no signs of
antiferromagnetic fluctuations (which are present in HTSC oxides) and the
peculiar behavior is probably due to the EPI.

It is worth of mentioning that after the discovery of HTSC in 1986 a number
of controversial results related to $\sigma (\omega )$ were published,
followed by a broad spectrum of results and interpretations, from standard
approaches up to highly exotic ones. For example, the reported experimental
values for $\omega _{pl}$ were in the surprisingly large range $(0.06-25)$ $%
eV$, causing a number of exotic (and confusing) theoretical models for
electronic dynamics - see more in \cite{Bozovic4}. (The similar situation
was with ARPES measurements - see below.) So, one should be very cautious in
interpreting experimental and theoretical results. In that respect, recent
experiments related to the \textit{optical sum-rule} is an additional
example for controversies in this field coming from inadequate
interpretations of results. This is the reason why we devote more space to
the problem of ''violation" of partial sum-rule.

There are two kinds of sum rules which are used in interpreting results on $%
\sigma (\omega )$. The first one is the \textit{total sum rule} and in the
normal state it reads

\begin{equation}
\int_{0}^{\infty }\sigma _{1}^{N}(\omega )d\omega =\frac{\omega _{pl}^{2}}{8}%
=\frac{\pi ne^{2}}{2m},  \label{Eq5}
\end{equation}
while in the \textit{superconducting state} \cite{Tinkham} it is given by
the Tinkham-Ferrell-Glover (TFG) sum-rule

\begin{equation}
\frac{c^{2}}{8\lambda _{L}^{2}}+\int_{+0}^{\infty }\sigma _{1}^{S}(\omega
)d\omega =\frac{\omega _{pl}^{2}}{8}.  \label{Eq6}
\end{equation}
Here,$\ n$ - is the total electron density, $e$ - the electron charge, $m$ -
the bare electron mass, $\lambda _{L}$ - the London penetration depth. The
first term $c^{2}/8\lambda _{L}^{2}$ is due to the appearance of the
superconducting condensate (ideal conductivity) which contributes $\sigma
_{1,cond}^{S}(\omega )=(c^{2}/4\lambda _{L}^{2})\delta (\omega )$. The total
sum rule represents the fundamental property of matter - the conservation of
the electron number. To calculate it one should use the total Hamiltonian $%
\hat{H}_{tot}=\hat{T}_{e}+\hat{H}_{int}$ by taking into account all
electrons, bands and their interactions $\hat{H}_{int}$ (Coulomb, EPI, with
impurities,etc.). Here $T_{e}$ is the kinetic energy of bare electrons
\begin{equation}
\hat{T}_{e}=\sum_{\sigma }\int d^{3}x\psi _{\sigma }^{\dagger }(x)\frac{%
\mathbf{\hat{p}}^{2}}{2m}\psi _{\sigma }(x)=\sum_{\mathbf{p},\sigma }\frac{%
\mathbf{p}^{2}}{2m_{e}}c_{\mathbf{p}\sigma }^{\dagger }c_{\mathbf{p}\sigma }.
\label{Eq7}
\end{equation}

The \textit{partial sum rule} is related to the energetics in the conduction
(valence) band\textit{. }Usually it is derived by using the Hamiltonian of
the valence electrons
\begin{equation}
\hat{H}_{v}=\hat{T}_{v}+\hat{V}_{v,Coul}=\sum_{\mathbf{p},\sigma }\epsilon _{%
\mathbf{p}}c_{v,\mathbf{p}\sigma }^{\dagger }c_{v,\mathbf{p}\sigma }+\hat{V}%
_{v,Coul},  \label{Hval}
\end{equation}
which contains the band-energy (with dispersion $\epsilon _{\mathbf{p}}$)
and the Coulomb interaction of valence electrons $\hat{V}_{v,Coul}$. In the
normal state the \textit{partial sum-rule} reads \cite{Kubo} (for general
form of $\epsilon _{\mathbf{p}}$)
\begin{equation}
\int_{0}^{\infty }\sigma _{1,v}^{N}(\omega )d\omega =\frac{\pi e^{2}}{2V}%
\sum_{\mathbf{p}}\frac{\langle n_{v,\mathbf{p}}\rangle _{H_{v}}}{m_{\mathbf{p%
}}}\equiv \frac{\omega _{pl,v}^{2}(T)}{8}  \label{Eq8}
\end{equation}
where $n_{v,\mathbf{p}}=c_{\mathbf{p}\sigma }^{\dagger }c_{\mathbf{p}\sigma }
$ and the reciprocal mass is given by $1/m_{\mathbf{p}}=\partial
^{2}\epsilon _{\mathbf{p}}/\partial p_{x}^{2}$. To simplify further
discussion we assume for $\epsilon _{\mathbf{p}}=-2t(\cos p_{x}a+\cos p_{y}a)
$ the \textit{tight-binding model with nearest neighbors} (n.n.) where $1/m_{%
\mathbf{p}}=-2ta^{2}\cos p_{x}a$. In practice measurements are performed up
to finite $\omega $ and the integration over $\omega $ goes up to some
cutoff frequency $\omega _{c}$ (of the order of band plasma frequency). It
is straightforward to show that one has (for the n.n. tight-binding model)
\begin{equation}
\int_{0}^{\omega _{c}}\sigma _{1,v}^{N}(\omega )d\omega =\frac{\pi e^{2}a^{2}%
}{2}\langle -T_{v}\rangle \equiv \frac{\omega _{pl,v}^{2}(T)}{8}  \label{Eq9}
\end{equation}
where $\langle -T_{v}\rangle _{H_{v}}=-\sum_{\mathbf{p}}\epsilon _{\mathbf{p}%
}\langle n_{v}\rangle _{H_{v}}$ and by $\omega _{pl,v}^{2}$ is defined the
band plasma frequency.

In that case the \textit{partial sum-rule in the superconducting state}
reads
\begin{equation}
\frac{c^{2}}{8\lambda _{L}^{2}}+\int_{+0}^{\omega _{c}}\sigma
_{1,v}^{S}(\omega )d\omega =\frac{\pi e^{2}a^{2}}{2}\langle -T_{v}\rangle .
\label{Eq10}
\end{equation}

\begin{figure}[tbp]
\resizebox{0.8\textwidth}{!} {
\includegraphics*[width=6cm] {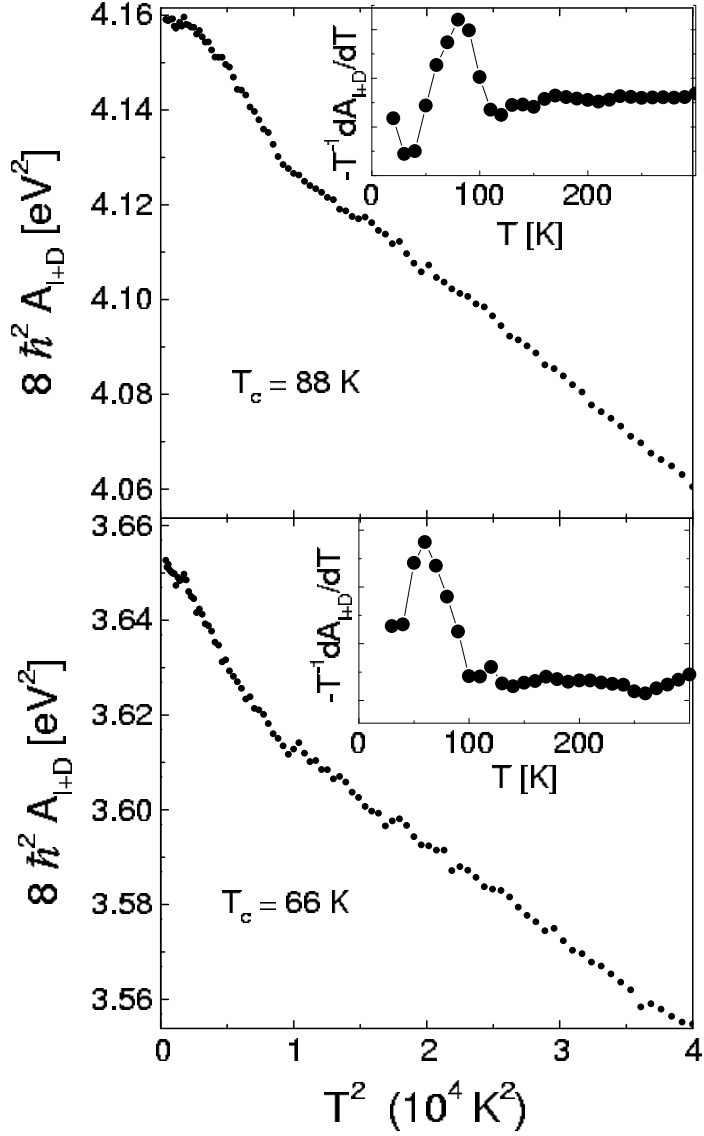}}
\caption{Measured T-dependence of $\bar{A}_{L}(0,\protect\omega _{c})$ and $%
\bar{A}_{H}(\protect\omega _{c},2\protect\omega _{c})$ for $\protect\omega %
_{c} \approx 1.25 eV$ of $Bi2212$ ($T_{c}=88$ $K$). The data from
\protect\cite{Molegraaf}}
\label{MolegFig}
\end{figure}

The sum-rule was studied intensively in optimally and underdoped $%
Bi_{2}Sr_{2}CaCu_{2}O_{8-x}$ and in $YBa_{2}Cu_{3}O_{7-x}$ for the \textit{%
intraplane conductivity}, where the whole frequency region is
separated into the
low (''intraband'')- and high (''interband'')-frequency parts $A_{L}$ and $%
(A_{H}+A_{VH})$, respectively
\begin{equation}
\bar{A}_{L}(0,\omega _{c})+A_{H}(\omega _{c},\alpha \omega
_{c})+A_{VH}(\alpha \omega _{c},\infty )=\omega _{pl}^{2}/8  \label{Eq11}
\end{equation}
with $\bar{A}_{L}(0,\omega _{c})=A_{L}(0,\omega _{c})+\delta _{SN}\omega
_{pl,S}^{2}$ where $\delta _{SN}=1$ in superconducting state and $A(\omega
_{1},\omega _{2})=\int_{\omega _{1}}^{\omega _{2}}\sigma _{1}(\omega
)d\omega $. The temperature dependence of $A_{L}$ and $A_{H}$ in the above
HTSC oxides was studied in \cite{Molegraaf} and \cite{BorisMPI}, by assuming
that ''intraband'' effects are exhausted for $\omega _{c}\approx 1.25$ $eV$
and the main temperature dependence of the high-frequency region comes for $%
\alpha =2$ in Eq.(\ref{Eq13}), i.e. for $1.25$ $eV<\omega <2.5$
$eV$, while the temperature dependence of the very high energy
part $A_{VH}$ is negligible. It was found that
$\bar{A}_{L}(0,\omega _{c})$ grows (quadratically with T) about
6\% between 300 and 4 K, while $A_{H}(\omega _{c},\alpha \omega
_{c})$ decreases with decreasing $T^{2}$. In the superconducting
state there is a small extra increase of $\bar{A}_{L}(0,\omega
_{c})$. The results are shown in Fig.~\ref{MolegFig}.

In connection with this experiment let us stress that in the BCS
superconductor the TFG sum-rule is practically satisfied if the integration
goes up to $\omega _{s}\approx (4-6)\Delta $, where $\Delta $ is the
superconducting gap. This means that in the BCS\ superconductors the
spectral weight appearing in the condensate (at $\omega =0$) is transferred
from the region $0^{+}-\omega _{s}$. However, the experiment in \cite
{Molegraaf} shows a transfer of the spectral-weight from the high ($\omega >1
$ $eV$) to low energies - see below. This fact was interpreted by some
researchers \cite{Hirsch}, \cite{Pepin} as a ''violation'' of the TFG
sum-rule, i.e. that there is more spectral weight in the condensate (at $%
\omega =0$) than\ it is expected from the TFG sum-rule and effectively means
the decrease of the kinetic energy in the superconducting state. This is in
contrast to the increase of the kinetic energy in the BCS superconducting
state. We are going to discuss this problem in details and to demonstrate
that the analysis in terms of the kinetic energy only, is untenable.

What is the origin of the spectral-weight transfer, especially in the
superconducting state of HTSC oxides? Here, we shall study the \textit{%
inplane} $\sigma _{a-b}(\omega )$ only, since the origin of the
quasiparticle dynamics along the c-axis is still unclear.

The \textit{first theoretical interpretation} of the spectral-weight
transfer was based on the partial sum-rule in which the temperature
dependence is related to the temperature change of the kinetic energy $%
\langle -T_{v}\rangle $ (or for more realistic spectrum of $\omega
_{pl,v}^{2}(T)$) - see Eq.(\ref{Eq8}-\ref{Eq10}). In this framework the
extra increase of $\bar{A}_{L}(0,\omega _{c})$ is related to the lowering of
the band kinetic energy in the superconducting state \cite{Hirsch}, \cite
{Pepin}. If this would be true, then the lowering of the band kinetic energy
(per particle) is approximately $(\langle T_{v}\rangle _{N,T>T_{c}}-\langle
T_{v}\rangle _{S,T<T_{c}})/N\sim 1$ $meV$, what is approximately by factor
ten larger than the superconducting condensation energy. Note, that in the
weak coupling BCS theory of superconductivity the kinetic energy is
increased in the superconducting state. So, the alleged large lowering of
the kinetic energy in the superconducting state is interpreted as a result
of some exotic pairing mechanism in which the kinetic energy (or $\omega
_{pl,v}^{2}(T)$ for more general spectrum) is significantly lowered but the
potential energy is increased in the superconducting state, contrary to the
case of BCS approach. However, this interpretation misses a very important
contribution to the partial sum rule, which is due to the large and strongly
$T,\omega $ dependent transport scattering rate $\Gamma _{tr}(T,\omega )$.

Before discussing the partial sum rule more adequately, let us
mention that the separation of the valence-band kinetic energy
from the potential one in strongly correlated systems is
\textit{not well defined procedure}. For instance, in the Hubbard
model with $U>>t$ and nearest neighbor hoping $t$ one has (see
below and also in \cite{KulicReview}) the low-energy (valence)
Hamiltonian $H_{v}$ is given
\begin{equation}
H_{v}=-t\sum_{i,\delta ,\sigma }X_{i}^{\sigma 0}X_{i+\delta }^{0\sigma },
\label{Eq12}
\end{equation}
where the Hubbard operators $X_{i}$ describe the motion of composite
quasiparticles with excluded doubly occupancy - see more in Section 4. They
have complicated non-canonical (anti)commutation rules, which means that Eq.(%
\ref{Eq12}) mixes the kinetic energy with the (kinematical) potential energy
of band (valence) quasiparticles.

\begin{figure}[tbp]
\resizebox{.8\textwidth}{!} {
\includegraphics*[width=6cm]{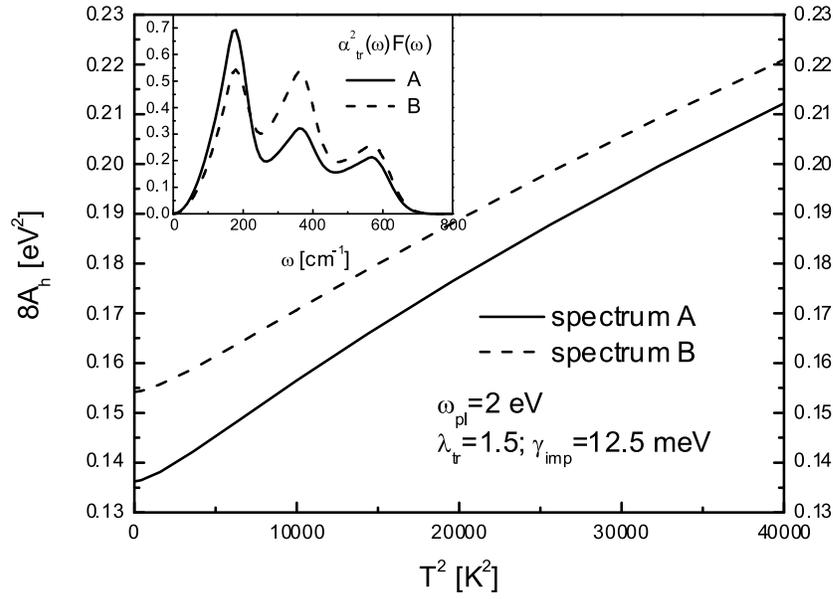}}
\caption{ Calculated theoretical T-dependence of high-energy part of the sum
rule $\bar{A}_{H}(\protect\omega _{c}= 1.25 eV,2\protect\omega _{c}= 2.5 eV)$
by taking into account the electron-phonon interaction. The data from
\protect\cite{Maks-Karakaz}}
\label{KarakozFig}
\end{figure}

The \textit{second theoretical approach is }proposed recently \cite
{Maks-Karakaz}, which is in principle exact, is based on the fact that in
HTSC oxides there is strong electron scattering - direct or via phonons, on
impurities, etc. So, the presence of an inelastic (and elastic) scattering
prevents the interpretation of the partial sum rule in terms of the band
kinetic energy only. As an illustrative example for this assertion may serve
the scattering of electrons on impurities, where the intraband contribution
to $\sigma _{1,v}(\omega )$ is given by
\begin{equation}
\sigma _{1,v}(\omega )=\frac{\omega _{pl,v}^{2}}{4\pi }\frac{\Gamma _{i,tr}}{%
\omega ^{2}+\Gamma _{i,tr}^{2}}.  \label{Eq13}
\end{equation}
Here, $\Gamma _{i,tr}/2=1/\tau _{i,tr}$ is the quasiparticle transport
relaxation rate due to impurities. In this case the partial sum-rule reads

\begin{equation}
\int_{0}^{\omega _{c}}\sigma _{1,v}^{N}(\omega )d\omega =\frac{\omega
_{pl,v}^{2}}{8}(1-\frac{2\Gamma _{i,tr}}{\omega _{c}}).  \label{Eq14}
\end{equation}
This result means that the intraband sum-rule can be satisfied in the
presence of impurities only for $\omega _{c}\rightarrow \infty $. The
similar conclusion holds in the case of inelastic scattering via phonons
although in that case $\Gamma _{tr}$ is $\omega $- and T-dependent. The
similar reasoning holds for interband transitions which in the presence of
scattering have also the low-frequency tail. Since in HTSC oxides $\Gamma
_{tr}(\omega )$ is dominantly due to the EPI and reaches values up to $100$ $%
meV$, there is no other way to study the partial sum-rule (the value of $%
\int_{0}^{\omega _{c}}\sigma _{1,v}(\omega )d\omega $) than to calculate $%
\sigma _{1,v}(\omega )$ directly from a microscopic model. The
latter must incorporate relevant scattering mechanisms and bands.
Such calculations were done in \cite{Maks-Karakaz} by taking into
account the large EPI interaction. By using the EPI spectral
function $\alpha ^{2}(\omega )F(\omega )$ from tunnelling
measurements and by assuming $\alpha _{tr}^{2}(\omega )F(\omega
)\approx \alpha ^{2}(\omega )F(\omega )$ the authors have
calculated $A_{H}(\omega_{c},2\omega _{c})$ (and $A_{L}(0,\omega
_{c})$) in the normal state and found a good agreement with
experiments \cite {Molegraaf} - see Fig.~\ref{KarakozFig}. We
stress that the recent elipsometric measurements of the dielectric
function $\varepsilon (\omega )$ \cite {BorisMPI} confirms this
theoretical prediction \cite{Maks-Karakaz}.

From the above analysis we conclude that: (i) the interpretation of the
partial sum-rule in HTSC oxides only in terms of the kinetic energy (or $%
\omega _{pl,v}^{2}(T)$) is physically unjustified; (ii) \textit{the EPI
interaction is strong and dominating scattering mechanism in the optical
properties of the normal state}. Reliable calculations of the partial
sum-rule in the superconducting state are still missing, since in that case
one should know much more details on the superconducting order $\Delta (%
\mathbf{k},\omega )$ and $\Gamma _{tr}(\mathbf{k},\omega )$, which are at
present too ambitious task.

\subsubsection{Resistivity $\rho (T)$}

A lot of experimental and theoretical works were devoted to the temperature
dependence of resistivity $\rho (T)$ in HTSC oxides. General properties of
the resistivity in HTSC oxides are the following: $\mathbf{(1)}$ The
resistivity is very anisotropic in single crystals where one has $%
r_{c}\equiv (\rho _{c}(T)/\rho _{a-b}(T))\gg 1$ at $T$ above T$_{c}$ - see
\cite{Gray}, i.e. $r_{c}\approx 300$ in $La_{1.85}Sr_{0.15}CuO_{4}$ and $%
Nd_{1.85}Ce_{0.15}CuO_{4}$, $r_{c}\approx 20-150$ in $YBa_{2}Cu_{3}O_{7-x}$,
$r_{c}\approx 10^{5}$ in $Bi_{2}Sr_{2}CaCu_{2}O_{8}$ depending also on the
sample preparation, temperature etc. The anisotropy of the in-plane
resistivity is much less, i.e. $r_{a}\equiv (\rho _{aa}(T)/\rho
_{bb}(T))\sim 1-2$, depending also on the sample preparation, temperature
etc.; $\mathbf{(2)}$ The in-plane resistivity $\rho _{a-b}(T)$ at room
temperature is more than two orders of magnitude higher than that of the
metallic $Cu$ (where $\rho _{Cu}(T_{room})\approx 1.5$ $\mu \Omega cm$),
i.e. $\rho _{a-b}(T)$ of HTSC oxides lies more in the semiconductor range
and $\rho _{a-b}(T)\gg \rho _{Cu}(T)$; $\mathbf{(3)}$ $\rho _{a-b}(T)\sim T$
for $T>T_{c}$, which deviates at $T>(800-1000)$ $K$ and saturates at even
higher temperatures, depending on samples etc.; $\mathbf{(4)}$ $\rho _{a-b}$
varies from $\rho _{a-b}(T)\sim T$ \ (with small residual resistivity) in
optimally doped systems being $\rho _{a-b}(T)\sim T^{3/2}$ in overdoped
systems, as experiments on $La_{2-x}Sr_{x}CuO_{4}$ show \cite{Resistivity}; $%
\mathbf{(5)}$ In most samples of HTSC oxides the $c$-axis resistivity $\rho
_{c}(T)$ shows a non-metallic behavior especially in samples with huge
anisotropy along the $c$-axis, growing by decreasing temperature, i.e. $%
(d\rho _{c}(T)/dT)<0$, being superconducting below T$_{c}$.

We discuss briefly the \textit{in-plane resistivity} $\rho _{a-b}(T)$ only,
because its temperature behavior is a direct consequence of the quasi-$2D$
motion of quasiparticles and of the inelastic scattering which they suffer.
At present there is no consensus on the origin of the linear temperature
dependence of the inplane resistivity $\rho _{a-b}(T)$ in the normal state.
As it is stressed several times many researchers are (erroneously) believing
that such a behavior can not be due to the EPI? The inadequacy of this claim
was already demonstrated by analyzing the dynamical conductivity $\sigma
(\omega )$. The inplane resistivity in HTSC oxides is usually analyzed by
the Kubo approach, or by the Boltzmann equation. In the latter case $\rho (T)
$ is given by
\begin{equation}
\rho (T)=\frac{4\pi }{\omega _{p}^{2}}\Gamma _{tr}(T)  \label{Eq15}
\end{equation}
\begin{equation}
\Gamma _{tr}(T)=\frac{\pi }{T}\int_{0}^{\infty }d\omega \frac{\omega }{\sin
^{2}(\omega /2T)}\alpha _{tr}^{2}(\omega )F(\omega ),  \label{Eq16}
\end{equation}
where $\alpha _{tr}^{2}(\omega )F(\omega )$ is the EPI transport
spectral function. It is well-known that at $T>\Theta _{D}/5$ and
for the Debay spectrum one has
\begin{equation}
\rho (T)\simeq 8\pi ^{2}\lambda _{tr}^{EP}\frac{k_{B}T}{\hbar \omega _{p}^{2}%
}=\rho ^{\prime }T.  \label{Eq17}
\end{equation}
In HTSC oxides the reach and broad spectrum of $\alpha _{tr}^{2}(\omega
)F(\omega )$ is favorable for such a linear behavior. The measured transport
coupling constant $\lambda _{tr}$ contains in principle all scattering
mechanisms, although usually some of them dominate. For instance, the
proponents of the spin-fluctuations mechanism assume that $\lambda _{tr}$ is
entirely due to the scattering on spin fluctuations. However, by taking into
account specificities of HTSC oxides the experimental results for the
inplane resistivity $\rho _{a-b}(T)$ can be satisfactory explained by the $%
EPI$ mechanism. From tunnelling experiments \cite{Tun1}, \cite{Tun2}, \cite
{Tun3}, \cite{Tun4}, \cite{Gonnelli} one obtains that $\lambda \approx 2-3$
and if one assumes that $\lambda _{tr}\approx \lambda $ and $\omega
_{pl}\geq (3-4)$ $eV$ (the value obtained from the band-structure
calculations) then Eq.(\ref{Eq17}) describes the experimental situation
rather well. The plasma frequency $\omega _{pl}$ which enters Eq.(\ref{Eq17}%
) can be extracted from optic measurements ($\omega _{pl,ex}$), i.e. from
the width of the Drude peak at small frequencies. However, since $\lambda
_{tr}\approx 0.25\omega _{pl}^{2}(eV)\rho ^{\prime }(\mu \Omega cm/K)$\
there is an experimental constraint on $\lambda _{tr}$. The experiments \cite
{Bozovic4} give that $\omega _{pl}\approx (2-2.5)$ $eV$ \ and $\rho ^{\prime
}\approx 0.6$ in oriented YBCO films, and $\rho ^{\prime }\approx 0.3$ in
single crystals of BISCO. These results makes a limit on $\lambda
_{tr}\approx 0.9-0.4$.

So, in order to explain $\rho (T)$ with small $\lambda _{tr}$ and high $%
T_{c} $ (which needs large $\lambda $) by the EPI it is necessary to have $%
\lambda _{tr}\leq (\lambda /3)$. This means that in HTSC oxides
the $EPI$ is reduced in transport properties where $\lambda
_{tr}\ll \lambda $. This reduction of $\omega _{p}^{2}$ and
$\lambda _{tr}$ means that the$y$ contain renormalization (with
respect to the $LDA$ results) due to various quasiparticle
scattering processes and interactions, which do not enter in the
$LDA$ theory. In subsequent chapters we shall argue that the
strong suppression of $\lambda _{tr}$ may have its origin in
strong electronic correlations \cite{Kulic1}, \cite{Kulic2},
\cite{Kulic3}.

\textit{In conclusion}, optic and resistivity measurements in normal state
of HTSC oxides are much more in favor of the EPI than against it. However,
some intriguing questions still remains to be answered: (\textbf{i}) which
are the values of $\lambda _{tr}$ and $\omega _{pl}$: (ii) why one has $%
\lambda _{tr}\ll \lambda $: (iii) what is the role of the Coulomb
scattering in $\sigma (\omega )$ and $\rho (T)$. The ARPES
measurements (see discussion below) give evidence for the
appreciable Coulomb scattering at higher frequencies, where
$\Gamma (\omega )\approx \Gamma _{0}+\lambda _{c}\omega $ for
$\omega
>\omega _{\max }^{ph}$ with $\lambda _{c}\approx 0.4$. So, in
spite of the fact that the EPI is suppressed in transport
properties it is sufficiently strong in order to dominate in some
temperature regime. It may happen that at higher temperatures the
Coulomb scattering dominates in $\rho (T)$, which certainly does
not disqualify the EPI as the pairing mechanism in HTSC oxides.
For better understanding of $\rho (T)$ we need a controllable
theory for the Coulomb scattering in strongly correlated systems,
which is at present lacking.

\subsection{Raman scattering in HTSC oxides}

If the elementary excitation involved in the Raman scattering are electronic
we deal with the \textit{electronic Raman effect}, while if an optical
phonon is involved we deal with the \textit{phonon Raman effect}. The Raman
scattering in the normal and superconducting state of HTSC oxides is an
important spectroscopic tool which gives additional information on
quasiparticle properties - the electronic Raman scattering, as well as on
phonons and their renormalization by electrons - the phonon Raman scattering.

\subsubsection{Electronic Raman scattering}

The Raman measurements on various HTSC oxides show a remarkable correlation
between the Raman cross-section $\tilde{S}_{\exp }(\omega )$ and the optical
conductivity $\sigma _{a-b}(\omega )$, i.e.
\begin{equation}
\tilde{S}_{\exp }(\omega )\sim \lbrack 1+n_{B}(\omega )]\langle \mid \gamma
_{sc}(\mathbf{q})\mid ^{2}\rangle _{F}\omega \sigma _{a-b}(\omega ),
\label{Eq18}
\end{equation}
\newline
where $n_{B}(\omega )$ is the Bose function and $\gamma
_{sc}(\mathbf{q})$ screened Raman vertex - see more in
\cite{KulicReview}. Previously it was demonstrated that $\sigma
_{a-b}(\omega )$ depends on the transport scattering rate $\Gamma
_{tr}(\omega ,T)$ where $\Gamma _{tr}(\omega ,T)\sim
T$ and $n_{B}(\omega )\sim T/\omega $ for $\omega <T$, thus giving $\tilde{S}%
(\mathbf{q},\omega )\approx Const_{1}$ in that range. For $\omega >T$ one
has $\omega \sigma _{a-b}(\omega )\approx Const$ giving also $\tilde{S}(%
\mathbf{q},\omega )\approx Const_{2}$. We have also demonstrated that the $%
EPI$ with the very broad spectral function $\alpha ^{2}F(\omega )$
(see Fig.~\ref
{TunnelFig} below) explains in a natural way $\omega ,T$ dependence of $%
\sigma _{a-b}(\omega )$ and $\Gamma _{tr}(\omega ,T)$. So, the Raman spectra
in HTSC oxides can be explained by the EPI in conjunction with strong
correlations. This conclusion is supported by calculations of the Raman
cross-section \cite{Rashkeev} which take into account the EPI with $\alpha
^{2}F(\omega )$ extracted from the tunnelling measurements on $%
YBa_{2}Cu_{3}O_{6+x}$ and $Bi_{2}Sr_{2}CaCu_{2}O_{8+x}$ \cite{Tun1}. They
are in a good qualitative agreement with experimental results - see more in
\cite{KulicReview}.

\begin{figure}[tbp]
\resizebox{.8\textwidth}{!} {
\includegraphics*[width=8cm]{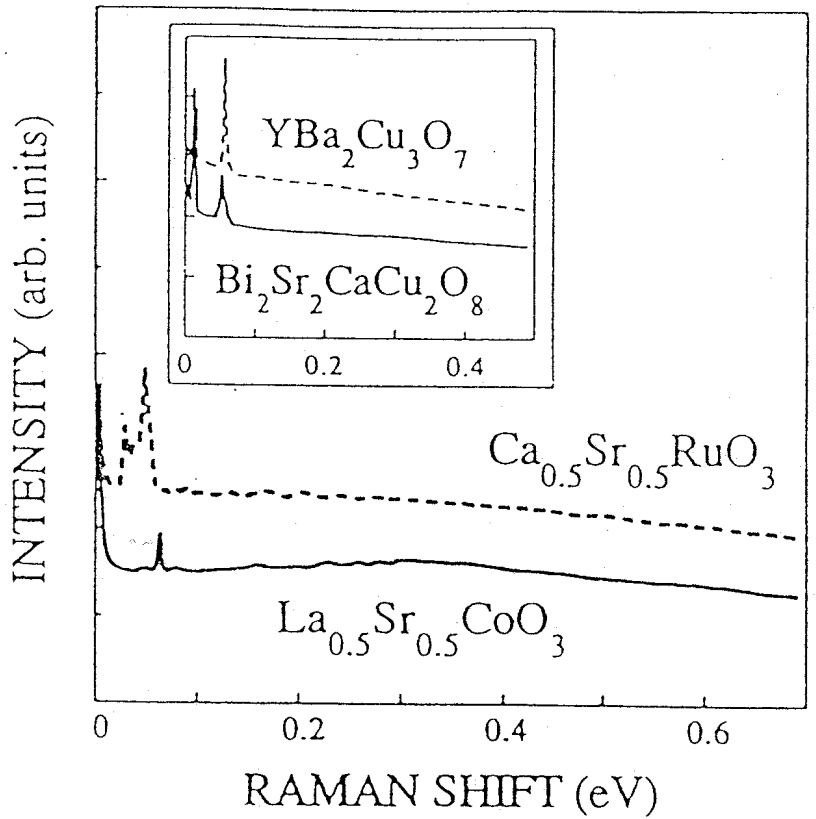}}
\caption{Broad range Raman scattering spectra of $Ca_{0.5}Sr_{0.5}RuO_{3}$
(broken line) and $La_{0.5}Sr_{0.5}CoO_{3}$ (solid line). Inset spectra of $%
Tl_{2}Ba_{2}Ca_{2}Cu_{3}O_{10}$, $Bi_{2}Sr_{2}CaCu_{2}O_{8}$, $%
YBa_{2}Cu_{3}O_{7}$ and $La_{1.85}Sr_{0.15}CuO_{4}$. From \protect\cite
{Bozovic}. }
\label{BozovFigb}
\end{figure}

We stress again, that quite similar (to HTSC oxides) properties of the
electronic Raman scattering (besides $\sigma (\omega )$, $R(\omega )$ and $%
\rho (T)$) were observed in experiments \cite{Bozovic} on isotropic metallic
oxides $La_{0.5}Sr_{0.5}CoO_{3}$ and $Ca_{0.5}Sr_{0.5}RuO_{3}$ - see Fig.~%
\ref{BozovFigb}. To repeat again, in these compounds there are no signs of
antiferromagnetic fluctuations (which are present in HTSC oxides) and the
peculiar behavior is probably due to the EPI.

\subsubsection{Phonon Raman scattering}

\paragraph{Normal state}

\begin{figure}[tbp]
\resizebox{.9\textwidth}{!} {
\includegraphics*[width=10cm]{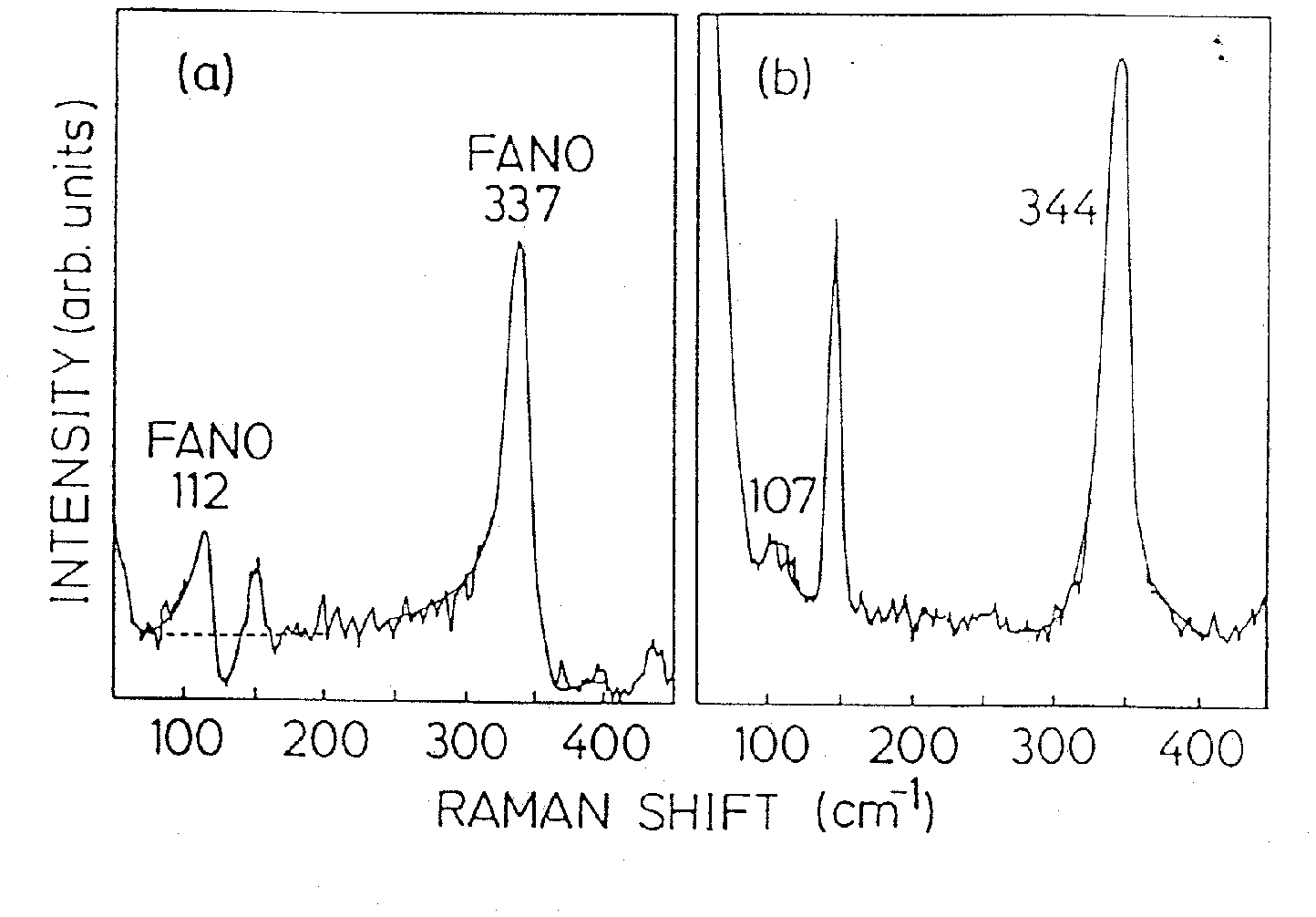}}
\caption{Fano resonance in $YBa_{2}Cu_{3}O_{7-x}$. The asymmetry is seen for
112 and 337 $cm^{-1}$ phonons in the superconductor ($x=0$). The
semiconductor ($x=1$) has Lorenzian line shapes. From \protect\cite{Thomsen2}%
.}
\label{FanoFig}
\end{figure}

- The effect of the EPI on the Raman scattering is characterized
by the \textit{Fano asymmetry parameter} $q(\omega )$ - see more
in \cite {KulicReview}. If it is finite the line shape is
\textit{asymmetric} - the \textit{Fano effect (resonance)}. By
decreasing $q(\omega )$ the phonon line shape becomes more
asymmetric, which means stronger EPI (in case when continuum
states are due to conduction carriers). The Fano resonance is
experimentally found in HTSC oxide $YBa_{2}Cu_{3}O_{7-\delta }$
\cite
{Thomsen2}, where the line asymmetry is clearly seen in optimally doped ($%
\delta \ll 1$) systems, while it is absent in the insulating state ($\delta
=1$). The existence of the Fano (asymmetric) line shape in HTSC oxides is a
direct proof that the discrete phonon level interacts with continuum of
states, which are conduction electrons in the metallic state - see Fig.~\ref
{FanoFig}.

\paragraph{Superconducting state}

\begin{figure}[tbp]
\resizebox{.9\textwidth}{!} {
\includegraphics*[width=10cm]{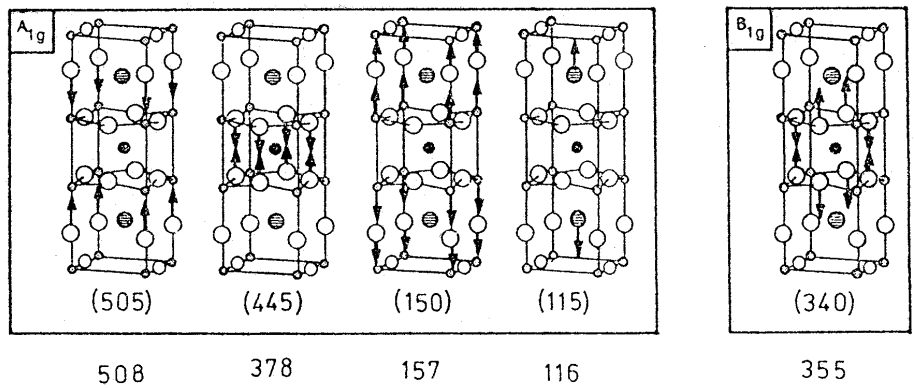}}
\caption{Assignation of $A_{g}$ modes according to calculations in
\protect\cite{Kress}. In brackets are experimental phonon frequencies in $%
cm^{-1}$. $115$ $cm^{-1}$ is the $Ba$ mode, $150$ $cm^{-1}$ is the $Cu2$
mode,$340$ $cm^{-1}$ ($B_{1g}$ mode) and $445$ $cm^{-1}$ modes are due to
vibration of $O(2,3)$ ions in the $CuO_{2}$ , while $505$ $cm^{-1}$ mode is
due to $O4$ ions. From \protect\cite{Gajic}.}
\label{A1gB1gFig}
\end{figure}

- It is well known that the renormalization of phonon frequencies
and their life-times by superconductivity in $LTSC$ materials is
rather small - around
one percent. The smallness of the effect is characterized by the parameter $%
\Delta /E_{F}$ which is very small in low temperature superconductors.
However, $\Delta /E_{F}$ is much larger in HTSC oxides and already from that
point of view one expects much stronger renormalization effects. At the very
beginning several Raman active phonon modes, with frequencies $%
128,153,333,437$ and $501$ $cm^{-1}$, were detected in
$YBa_{2}Cu_{3}O_{7}$ and these modes are totally symmetric modes
(with respect to the orthorhombic point group $D_{2h} $).
($1cm^{-1}=29.98GHz=0.123985meV=1.44K$) However, by using the
approximate tetragonal symmetry (with the point group $D_{4h}$)
the mode at $\omega _{B_{1g}}=333$ $cm^{-1}$ transforms according
to the $B_{1g}$ representation, while the other modes according to
the $A_{1g}$ one - see Fig.~\ref{A1gB1gFig}. The \textit{Fano
resonance} (asymmetric line shape) of the $B_{1g}$ mode indicates
an appreciable coupling of the lattice to the continuum, which in
fact corresponds to the charge carries. It is interesting to note
that the $A_{1g}$ modes in $YBCO$ are weakly affected in the
presence of superconductivity, while the $B_{1g}$ mode
\textit{softens} by $9$ $cm^{-1}$ (by approximately $3$ $\%$)
\cite{Kranz2}. It is well established also that this softening is
due to superconductivity and not due to, for instance, structural
changes, because it disappears in magnetic fields higher than
$H_{c2}$ .

The frequency shift $\delta \omega _{\lambda }$ and the phonon line width $%
\Gamma _{\lambda }$ in the superconducting state have been studied
numerically in \cite{Zwicknagl} for the case of the isotropic $s-wave$
superconducting gap ($\Delta (\mathbf{k})=\Delta =const$) and for strong
coupling superconductivity. They have predicted the \textit{phonon-softening}
and line-width \textit{narrowing}\textbf{\ }for $\omega _{0}<2\Delta $,
while for $\omega _{0}>2\Delta $ there is a \textit{phonon-hardening} and
line-width broadening. These predictions are surprisingly in agreement with
experiments \cite{Kranz2}, in spite of the assumed isotropic $s-wave$
pairing what is contrary to the experimentally well established $d-wave$
pairing in $YBCO$. Later calculations of the renormalization of the $B_{1g}$
Raman phonon mode in the presence of the weak coupling $d-wave$
superconductivity \cite{Deveraux3} show that if one assumes that $\omega
_{B_{1g}}<2\Delta _{\max }$ there is phonon softening accompanied with the
line broadening below T$_{c}$. The latter is possible because of the gapless
character (on a part of the Fermi surface) of $d-wave$ pairing. In that
respect the calculations of the phonon renormalization based on the strong
coupling $d-wave$ superconductivity are of significant interest and still
awaiting.

\begin{figure}[tbp]
\resizebox{.8\textwidth}{!} {
\includegraphics*[width=8 cm]{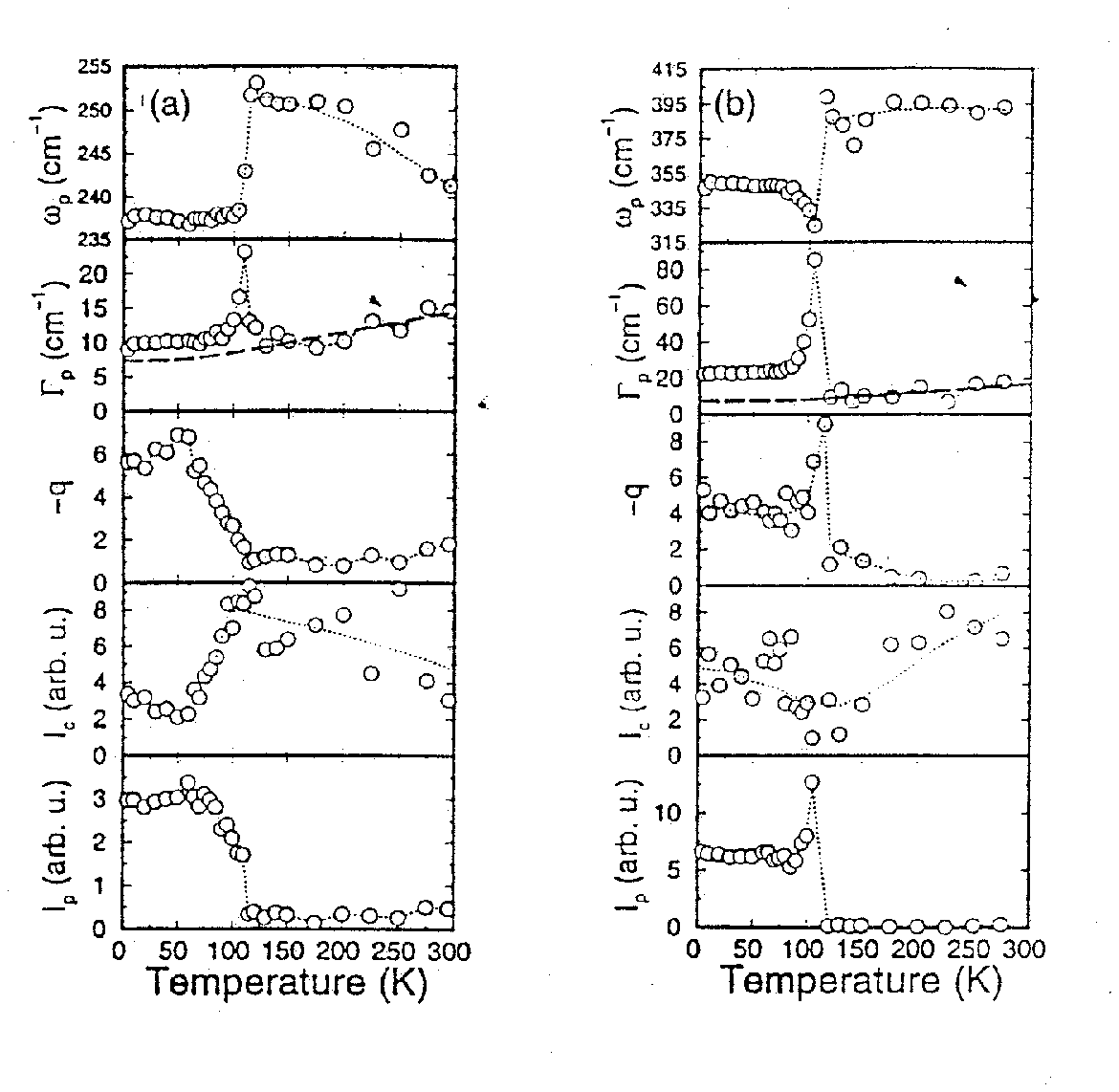}}
\caption{The fitted frequency $\protect\omega _{p}$, line-width $\Gamma _{p}$%
, asymmetry parameter $q$, and the phonon intensity $I_{p}$ of the $Hg-1234$
Raman spectra in the $A_{1g}$ mode measured in $x\prime x\prime $
polarization with $647.1$ $nm$ laser line: $(a)$ at $240$ $cm^{-1}$; $(b)$
at $390$ $cm^{-1}$ - from \protect\cite{PhoRaman}. }
\label{Hg-1234Fig}
\end{figure}

Recent report \cite{PhoRaman} on the superconductivity-induced strong phonon
renormalization of the $A_{1g}$ phonons at $240$ and $390$ $cm^{-1}$ (by $6$
and $18$ $\%$ respectively) in $HgBa_{2}Ca_{3}Cu_{4}O_{10+x}$ ($T_{c}=123$ $K
$) - the so called $Hg-1234$) compound, renders an additional evidence for
the strong EPI in HTSC oxides. In \cite{PhoRaman} the EPI coupling constant
is estimated to be rather large for the $A_{1g}$ phonons ($\lambda
_{A_{1g}}\approx 0.08$). Since there are $60$ phonon modes in $%
HgBa_{2}Ca_{3}Cu_{4}O_{10+x}$ they are capable to produce large
EPI coupling constant $\lambda =\sum_{\nu =1}^{60}\lambda _{\nu
}>1$ - see Fig.~\ref {Hg-1234Fig}(a-b).
A conservative estimation of the upper limit of $\lambda _{\max }$ gives $%
\lambda _{\max }\approx 60\times 0.08=4.8$ which is, of course, far from the
realistic value of $\lambda \leq 2$. In any case this analysis confirm that
the EPI of some Raman modes in HTSC oxides is strong. To this point, very
recent Raman scattering measurements on the $(Cu,C)-1234$ compound with $%
T_{c}=117$ $K$ reveal strong superconductivity induced phonon self-energy
effects \cite{Hadjiev2}. The $A_{1g}$ phonons at $235$ cm$^{-1}$ and $360$ cm%
$^{-1}$ (note $\omega _{pl}<2\Delta _{0}$), which involve vibrations of the
plane oxygen with some admixture of $Ca$ displacements, exhibit pronounced
Fano line shape (in the normal and superconducting state) with the following
interesting properties in the superconducting state: \textit{(i)} the phonon
intensity is increased substantially; \textit{(ii)} both phonons soften;
\textit{(iii)} the phonon line width (of both phonons) increases
dramatically below T$_{c}$ passing through a maximum slightly below T$_{c}$,
and decreases again at low $T$ but remaining broader than immediately below T%
$_{c}$. This line broadening is difficult to explain by $s-wave$ pairing,
where the line narrowing is expected, but it can be explained by
superconducting pairing with nodes in the quasiparticle spectrum, for
instance by $d-wave$ pairing \cite{Jiang}. The large EPI coupling constants
for these two modes are estimated from the asymmetric Fano line shape, which
gives $\lambda _{235}=0.05$ and $\lambda _{360}=0.07$ (note in $YBCO$ $%
\lambda _{A_{1g}}=0.01$ for $\omega _{A_{1g}}=440$ $cm^{-1}$ and $\lambda
_{B_{1g}}=0.02$ for $\omega _{B_{1g}}=340$ $cm^{-1}$, rather small values)
giving the upper value for the total coupling constant $\lambda _{\max }=4$.
This result gives additional important evidence for the strong $EPI$ in HTSC
oxides.

\subsubsection{Electron-phonon coupling in Raman scattering}

We would like to stress the importance of the (phonon) Raman scattering
measurements for the theory of the EPI in HTSC oxides. The \textit{covalent
part} of the EPI is due to the strong \textit{covalency} of the $Cu$ and $O$
orbitals in the $CuO_{2}$ planes. In that case the EPI coupling constant is
characterized by the parameter (''field'') $E^{cov}\sim \partial
t_{p-d}/\partial R\sim q_{0}t_{p-d}$, where $t_{p-d}$ is the hopping
integral between $Cu(d_{x^{2}-y^{2}})$ and $O(p_{x,y})$ orbitals and the
length $q_{0}^{-1}$ characterizes the spacial exponential fall-off of the
hopping integral $t_{p-d}$. The covalent EPI is unable to explain the strong
phonon renormalization (the self-energy features) in the $B_{1g}$ mode in $%
YBa_{2}Cu_{3}O_{7}$ by superconductivity, since in this mode the O-ions move
along the $c-axis$ in opposite directions, while for this mode $\partial
t_{p-d}/\partial R$ is zero in the first order in the phonon displacement.
Therefore the EPI in this mode must be due to the\textit{\ ionic contribution%
} to the $EP$ interaction which comes from the change in the
Madelung energy as it was first proposed in \cite{Jarlborg},
\cite{Barisic1}. Namely, the Madelung interaction creates an
electric field perpendicular to the $CuO_{2}$ planes, which is due
to the surrounding ions which form an asymmetric environment. In
that case the site energies $\epsilon _{i}^{0}$ contain the matrix
element $\epsilon ^{ion}=\langle \psi _{i}\mid \phi
(\mathbf{r})\mid \psi _{i}\rangle $, where $\mid \psi _{i}\rangle
$ is the atomic wave function at the $i$-th site, while the
potential $\phi (\mathbf{r})$ steams from surrounding ions. In
simple and transition metals the surrounding ions are well
screened and therefore the change of $\epsilon ^{ion}$ in the
presence of phonons is negligible, contrary to HTSC oxides which
are almost \textit{ionic compounds} (along the $c$-axis) where the
change of $\epsilon
^{ion}$ is appreciable and characterized by the field strength $%
E^{ion}=V/a_{n}$. Here, $V$ is the characteristic potential due to
surrounding ions and $a_{n}$ is the distance of the neighboring ions.
Immediately after the discovery of HTSC oxides in many papers \cite{Weber},
\cite{Mattheiss}, \cite{Alligia} it was (incorrectly) assumed that the
covalent part dominates the EPI in these materials. The calculation of $T_{c}
$ by considering only covalent effects \cite{Weber}, \cite{Mattheiss} gave
rather small T$_{c}$ ($\sim 10-20$ $K$ in $YBCO$, and $20-30$ $K$ in $%
La_{1.85}Sr_{0.15}CuO_{4}$). It turns out that in HTSC oxides the opposite
inequality $E^{ion}\gg E^{cov}$ is realized for most c-axis phonon modes, on
which basis the renormalization of the Raman $B_{1g}$ mode can be explained
- see more in \cite{KulicReview}. This is supported by detailed theoretical
studies in for the $YBCO$ compound \cite{Barisic1}, \cite{Barisic2}, where\
it is calculated the change in the \textit{ionic Madelung energy} due to the
out of plane oxygen vibration in the $B_{1g}$ mode. Similarly as in $YBCO$,
the large superconductivity-induced phonon self-energy effects in $%
HgBa_{2}Ca_{3}Cu_{4}O_{10+x}$ and in $(Cu,C)Ba_{2}Ca_{3}Cu_{4}O_{10+x}$ for
the $A_{1g}$ modes are also due to the ionic (Madelung) coupling. In these
modes oxygen ions move also along the $c-axis$ and the ionicity of the
structure is involved in the EPI. This type of the (\textit{long-range})%
\textit{\ }EPI is absent in usual isotropic metals ($LTSC$
superconductors), where the large Coulomb screening makes it to be
local. Similar ideas are recently incorporated into the Eliashberg
equations in \cite{Weger}, \cite {Abrikosov4}. The weak screening
along the $c$-axis, which is due to the very small hopping
integral for carrier motion, is reflected in the very small plasma
frequency $\omega _{p}^{(c)}$along this axis. Since for some
optical phonon modes one has $\omega _{ph}>\omega _{p}^{(c)}$ then
nonadiabatic effects in the screening are important. The latter
can give rise to much larger EPI coupling constant for this modes
\cite{Falter97}, \cite{Falter98}.

In conclusion, the electron and phonon Raman scattering measurements in the
normal and superconducting state of HTSC oxides give the following important
results: $(a)$ phonons interact strongly with the electronic continuum, i.e.
the EPI is substantial; $(b)$ the ionic contribution (the Madelung energy)
to the $EPI$ interaction for c-axis phonon modes gives substantional
contribution to the (large) EPI coupling constant ($\lambda >1$).

\subsection{Tunnelling spectroscopy in HTSC oxides}

Tunnelling methods are important tools in studying the electronic
density of states $N(\omega )$ in superconductors and in the past
they have played very important role in investigating of low
T$_{c}$-superconductors. By measuring the current-voltage ($I-V$)
characteristic in typical tunnelling junctions (with large
tunnelling barrier) it was possible from the tunnelling
conductance $G(V)(=dI/dV)$ to determine $N(\omega )$ and the
superconducting gap as a function of temperature, magnetic field
etc. Moreover, by measuring of $G(V)$ at voltages $eV>\Delta $ in
the $NIS$ (normal metal - isolator - superconductor) junctions it
was possible to determine the Eliashberg spectral function $\alpha
^{2}F(\omega )$ (which is due to some bosonic mechanism of
quasiparticle scattering) and finally to confirm (definitely) the
phonon mechanism of pairing in $LTSC$ materials, except maybe
heavy fermions \cite{Wolf}. We shall discuss here only the results
for $\alpha ^{2}F(\omega )$ obtained from $I-V$ measurements,
while a more extensive discussion of other aspects is given in
\cite{KulicReview}.

\subsubsection{$I-V$\ characteristic and $\protect\alpha ^{2}F(\protect\omega
)$}

\begin{figure}[tbp]
\includegraphics*[width=8cm]{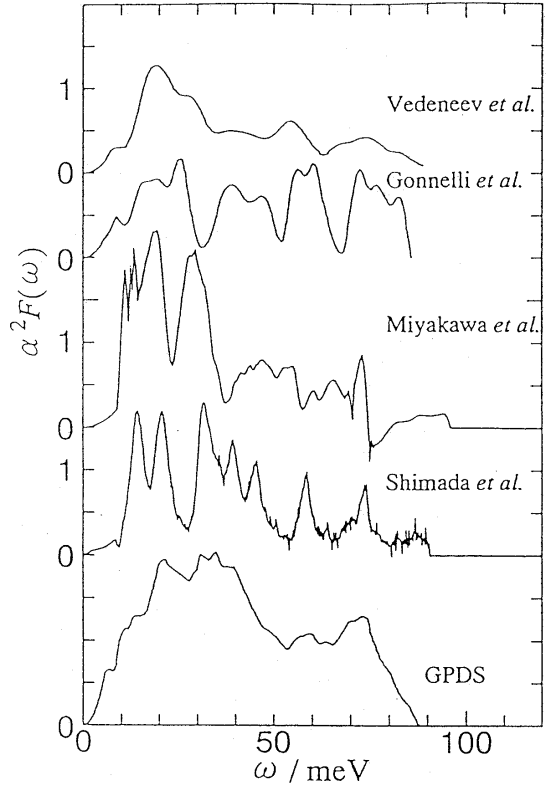}
\caption{The spectral function $\protect\alpha ^{2}F(\protect\omega )$
obtained from measurements of $G(V)$ by various groups on various junctions:
Vedeneev et al. \protect\cite{Tun1}, Gonnelli et al. \protect\cite{Gonnelli}%
, Miyakawa et al. \protect\cite{Tun3}, Shimada et al.\protect\cite{Tun2}.
The generalized density of states GPDS\ for $Bi2212$ is plotted at the
bottom - from \protect\cite{Tun2}.}
\label{TunnelFig}
\end{figure}

If one considers a $NIS$ contact where the left ($L$) and right ($R$) banks
of the contact can be normal ($N$) metal or superconductor ($S$),
respectively, with very small transparency then tunnelling effects are
studied in the framework of the tunnelling Hamiltonian $\hat{H}_{T}=\sum_{%
\mathbf{k},\mathbf{p}}(T_{\mathbf{k},\mathbf{p}}c_{\mathbf{k}L}^{\dagger }c_{%
\mathbf{q}R}+h.c)$. In that case the single-particle tunnelling current is
given by the formula \cite{Mahan}
\[
I_{qp}(V)=2e\sum_{\mathbf{k},\mathbf{p}}\mid T_{\mathbf{k},\mathbf{p}}\mid
^{2}\times
\]
\begin{equation}
\times \int_{-\infty }^{\infty }d\omega A_{N}(\mathbf{k},\omega )A_{S}(%
\mathbf{p},\omega +eV)[n_{F}(\omega )-n_{F}(\omega +eV)].  \label{Eq19}
\end{equation}
The single-particle spectral function $A_{N(S)}(\mathbf{k},\omega )$ is
related to the imaginary part of the retarded single particle Green's
function, i.e. $A(\mathbf{k},\omega )=-ImG^{ret}(\mathbf{k},\omega )/\pi $,
while the tunnelling matrix element $\mid T_{\mathbf{k},\mathbf{p}}\mid ^{2}$
is derived in the quantum-mechanical theory of tunnelling through the
barrier - see \cite{KulicReview}. Note, in the superconducting state $A(%
\mathbf{k},\omega )$ depends on the superconducting gap function $\Delta (%
\mathbf{k},\omega )$, which is on the other hand a functional of the
spectral function $\alpha ^{2}F(\omega )$. The fine structure in the second
derivative $d^{2}I/dV^{2}$ at voltages above the superconducting gap is
related to the spectral function $\alpha ^{2}F(\omega )$. For instance,
plenty of break-junctions made from $Bi-2212$ single crystals \cite{Tun1}
show that negative peaks in $d^{2}I/dV^{2}$, although broadened, coincide
with the peaks in the generalized phonon density of states $G_{ph}(\omega )$
measured by neutron scattering - see more in \cite{KulicReview}. Note, the
reported broadening of these peaks might be partly due to $d-wave$ pairing
in HTSC oxides. The tunnelling density of states $N_{T}(V)\sim dI/dV$ shows
a gap structure and it was found that $2\bar{\Delta}/T_{c}=6.2-6.5$, where $%
T_{c}=74-85$ $K$ $\ $and $\bar{\Delta}$ is some average value of the gap. By
assuming $s-wave$ superconductivity \cite{Tun1} and by solving the $MR$
problem (inversion of Eliashberg equations), the spectral function $\alpha
^{2}F(\omega )$ is obtained which gives $\lambda \approx 2.3$. Note, in
extracting $\lambda $ \cite{Tun1} the standard value of the effective
Coulomb parameter $\mu ^{\ast }\approx 0.1$ is assumed. Although this
analysis \cite{Tun1} was done by assuming $s-wave$ pairing it is
qualitatively valuable procedure also in the case of $d-wave$ pairing,
because one expects that $d-wave$ pairing does not spoil significantly the
global structure of $d^{2}I/dV^{2}$ at $eV>\Delta $, but introducing mainly
a broadening of peaks. The latter effect can be partly due to an
inhomogeneity of the gap. The results obtained in \cite{Tun1} were
\textit{reproducible} on more than 30 junctions, while in $Bi(2212)-GaAs$ and $%
Bi(2212)-Au$ planar tunnelling junctions similar results were. Several
groups \cite{Tun3}, \cite{Tun4}, \cite{Gonnelli} have obtained similar
results for the shape of the spectral function $\alpha ^{2}F(\omega )$ from
the $I-V$ measurements on various HTSC oxides as shown in Fig.~\ref
{TunnelFig}. The results shown in Fig.~\ref{TunnelFig} leave no much doubts
on the effectiveness of the EPI in pairing mechanism of HTSC oxides. In that
respect recent tunnelling measurements on $Bi_{2}Sr_{2}CaCu_{2}O_{8}$ \cite
{Tsuda} are impressive, since the Eliashberg spectral function $\alpha
^{2}F(\omega )$ was extracted from the measurements of $d^{2}I/dV^{2}$. The
obtained $\alpha ^{2}F(\omega )$ has several peaks in the broad frequency
region up to $80$ $meV$ - see Fig.~\ref{TunnelFig} (curve Shimada et al.),
which coincide rather well with the peaks in the phonon density of states $%
F(\omega )$. Moreover, the authors of \cite{Tsuda} were able to
extract the coupling constant for modes laying in (and around)
these peaks and their contribution to T$_{c}$. \ They managed to
extract the EPI coupling constant, which is unexpectedly very
large, i.e. $\lambda (=2\int d\omega \alpha ^{2}F(\omega )/\omega
)=\sum \lambda _{i}\approx 3.5$. Since almost all phonon mode
contributes to $\lambda $, this means that on the average each
particular phonon mode is moderately coupled to electrons thus
keeping the lattice stable. Additionally, they have found that
some low-frequency phonon modes corresponding to $Cu,Sr$ and $Ca$
vibrations are rather strongly coupled to electrons, similarly as
the high frequency oxygen vibrations along the $c$-axis do. These
results confirm the importance of the axial modes in which the
change of the Madelung energy is involved, thus supporting the
idea conveyed through this article of the importance of the ionic
Madelung energy in the EPI interaction of HTSC oxides.

In conclusion, the common results for all reliable tunnelling
measurements in HTSC oxides, including $Ba_{1-x}K_{x}BiO_{3}$ too
\cite{Huang}, \cite {Jensen}, is that no particular mode can be
singled out in the spectral function $\alpha ^{2}F(\omega )$ as
being the only one which dominates in pairing mechanism. This
important result means that the high T$_{c}$ is not attributable
to a particular phonon mode in the EPI mechanism, since all phonon
modes contribute to $\lambda $. Having in mind that the phonon
spectrum in HTSC oxides is very broad (up to $80$ $meV$ ), then the large $%
EPI$ coupling constant ($\lambda \approx 2$) in HTSC oxides is not
surprising at all. We stress, that compared to neutron scattering
experiments the tunnelling experiments are superior in determining
the EPI spectral function $\alpha ^{2}F(\omega )$.

\subsection{Isotope effect in HTSC oxides}

The isotope effect has played an important role in elucidating the pairing
mechanism in $LTSC$ materials. Note, the standard BCS theory predicts that
for the pure phonon-mediated mechanism of pairing the isotope coefficient $%
\alpha =-d\ln T_{c}/d\ln M$, where $M$ is the ionic mass, takes
its canonical value $\alpha =1/2$. However, later on it was clear
that $\alpha $ can take values less (even negative) then its
canonical value in the phonon-mediated mechanism of pairing if
there is pronounced Coulomb pseudopotential $\mu \ast $ - see more
in \cite{KulicReview}.

\subsubsection{Experiments on the isotope coefficient $\protect\alpha $}

\begin{figure}[tbp]
\resizebox{1.0 \textwidth}{!} {
\includegraphics*[width=8cm]{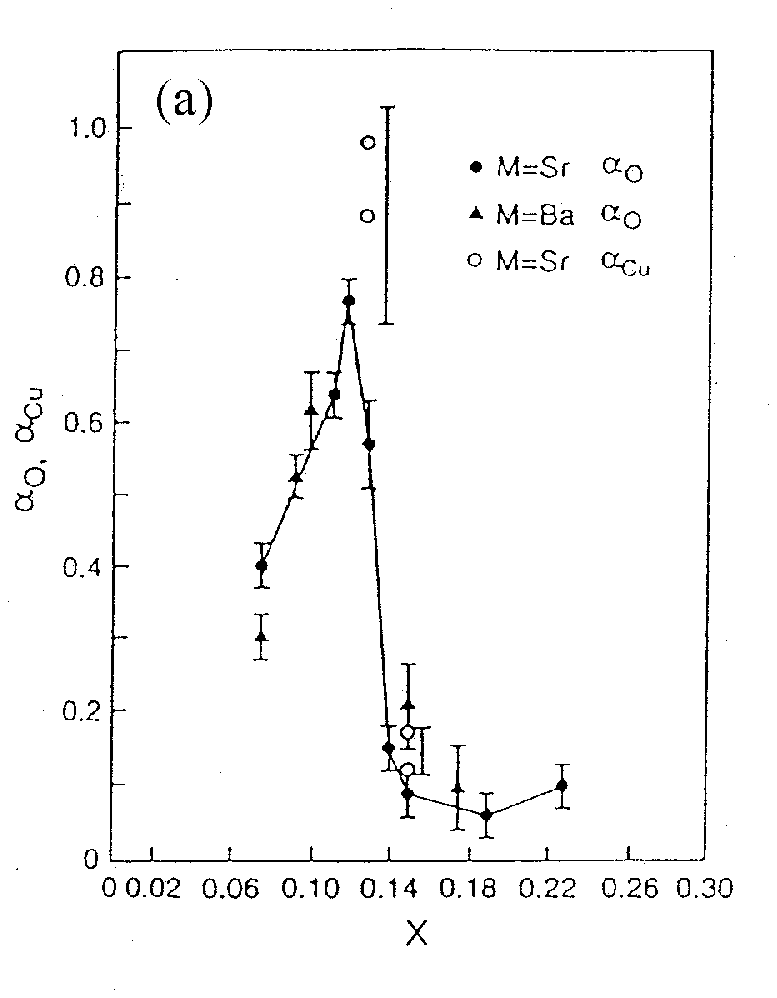}
\includegraphics*[width=14cm]{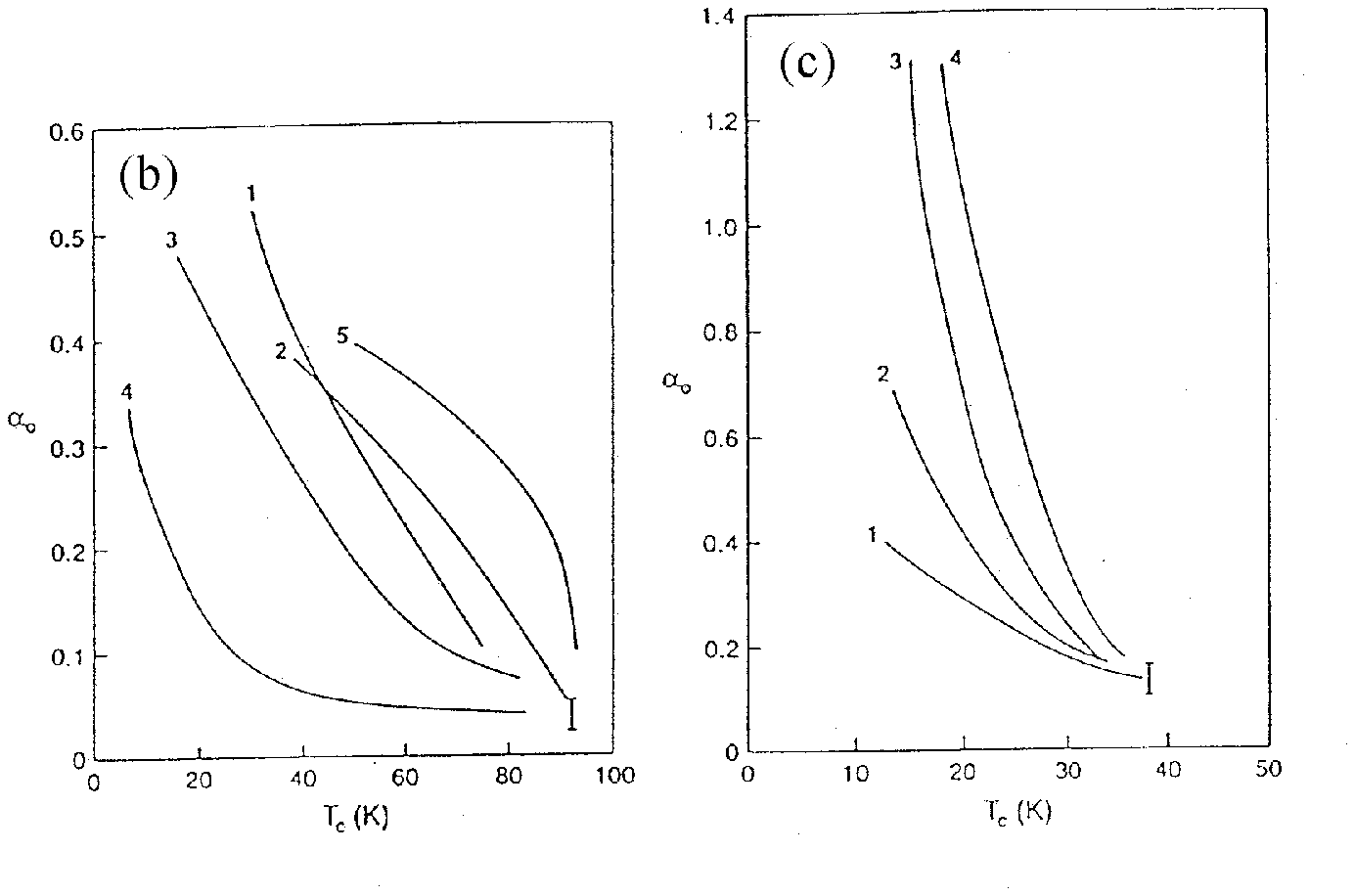}}
\caption{The oxygen isotope exponent $\protect\alpha _{O}$ for: (a) $%
La_{2-x}Sr_{x}CuO_{4} $ as a function of $Sr$ concentration - from
\protect\cite{Franck}. The oxygen isotope exponent $\protect\alpha _{O}$ as
a function of T$_{c}$ for: (b) $YBa_{2}Cu_{3}O_{7}$. 1: $(Y_{1-x}%
\Pr_{x})Ba_{2}Cu_{3}O_{7}$; 2: $YBa_{2-x}La_{x}Cu_{3}O_{7}$; 3: $%
YBa_{2}(Cu_{1-x}Co_{x})_{3}O_{7}$; 4: $YBa_{2}(Cu_{1-x}Zn_{x})_{3}O_{7}$; 5:
$YBa_{2}(Cu_{1-x}Fe_{x})_{3}O_{7}$. (c) $La_{1.85}Sr_{0.15}CuO_{4}$. 1: $%
La_{1.85}Sr_{0.15}(Cu_{1-x}Ni_{x})O_{4}$; 2: $%
La_{1.85}Sr_{0.15}(Cu_{1-x}Zn_{x})O_{4}$; 3: $%
La_{1.85}Sr_{0.15}(Cu_{1-x}Co_{x})O_{4}$; 4: $%
La_{1.85}Sr_{0.15}(Cu_{1-x}Fe_{x})O_{4}$ - from \protect\cite{Franck}. }
\label{AlfaFig}
\end{figure}

A lot of measurements of $\alpha _{O}$ and $\alpha _{Cu}$ were performed on
various hole-doped and electron-doped HTSC oxides and we give a brief
summary of the main results \cite{Franck}: (\textbf{1})\textbf{\ }The $O$
isotope coefficient $\alpha _{O}$ strongly depends on the hole concentration
in the hole-doped materials where in each group of HTSC oxides ($%
YBa_{2}Cu_{3}O_{7-x}$, or $La_{2-x}Sr_{x}CuO_{4}$ etc.) a small\textbf{\ }%
oxygen isotope effect is observed in the optimally doped\textbf{\ }(maximal $%
T_{c}$)\textbf{\ }samples. For instance $\alpha _{O}\approx 0.02-0.05$ in $%
YBa_{2}Cu_{3}O_{7}$ with $T_{c,\max }\approx 91$ $K$, $\alpha _{O}\approx
0.1-0.2$ in $La_{1.85}Sr_{0.15}CuO_{4}$ with $T_{c,\max }\approx 35$ $K$; $%
\alpha _{O}\approx 0.03-0.05$ in $Bi_{2}Sr_{2}CaCu_{2}O_{8}$ with $T_{c,\max
}\approx 76$ $K$; $\alpha _{O}\approx 0.03$ and even negative ($-0.013$) in $%
Bi_{2}Sr_{2}Ca_{2}Cu_{2}O_{10}$ with $T_{c,\max }\approx 110$ $K$; the
experiments on $Tl_{2}Ca_{n-1}BaCu_{n}O_{2n+4}$ ($n=2,3$) with $T_{c,\max
}\approx 121$ $K$ are still unreliable and $\alpha _{O}$ is unknown; $\alpha
_{O}<0.05$ in the electron-doped $(Nd_{1-x}Ce_{x})_{2}CuO_{4}$ with $%
T_{c,\max }\approx 24$ $K$. (\textbf{2}) For hole concentrations
away from the optimal one, T$_{c}$ decreases while $\alpha _{O}$
increases and in some cases reaches large value $\alpha
_{O}\approx 0.5$ - see Fig.~\ref{AlfaFig} for $La$ compounds. This
holds not only for parent compounds but also for
systems with substitutions, like $(Y_{1-x-y}\Pr_{x}Ca_{y})Ba_{2}Cu_{3}O_{7}$%
, $Y_{1-y}Ca_{y}Ba_{2}Cu_{4}O_{4}$ and $Bi_{2}Sr_{2}Ca_{1-x}Y_{x}Cu_{2}O_{8}$%
. Note, the decrease of T$_{c}$ is not a prerequisite for the increase of $%
\alpha _{O}$. This became clear from the $Cu$ substituted experiments $%
YBa_{2}(Cu_{1-x}Zn_{x})_{3}O_{7}$ where the decrease of T$_{c}$
(by increasing of the $Zn$ concentration) is
followed by only small increase of $\alpha _{O}$ \cite{Franck2}. Only in the
case of very low $T_{c}<20$ $K$ then$\ \alpha _{O}$ becomes large, i.e. $%
\alpha _{O}>0.1$. (\textbf{3)} The largest\textbf{\ }$\alpha _{O}$ is
obtained even in the optimally doped compounds like in systems with
substitution, such as $La_{1.85}Sr_{0.15}Cu_{1-x}M_{x}O_{4}$, $M=Fe,Co$,
where $\alpha _{O}\approx 1.3$ for $x\approx 0.4$ $\%$. (\textbf{4}) In $%
La_{2-x}M_{x}CuO_{4}$ there is a $Cu$\textbf{\ }isotope effect\textbf{\ }%
which is of the order of the oxygen one, i.e. $\alpha _{Cu}\approx \alpha
_{O}$ giving $\alpha _{Cu}+\alpha _{O}\approx 0.25-0.35$ for optimally doped
systems ($x=0.15$). In the case when $x=0.125$ with $T_{c}\ll T_{c,\max }$
one has$\ \alpha _{Cu}\approx 0.8-1$ with $\alpha _{Cu}+\alpha _{O}\approx
1.8$. The appreciate copper isotope effect in $La_{2-x}M_{x}CuO_{4}$ tells
us that vibrations of other than oxygen ions could be important in giving
high T$_{c}$. The latter property is more obvious from tunnelling
measurements, which are discussed above. (\textbf{5}) There is \textit{%
negative} Cu isotope effect in the oxygen-deficient system
$YBa_{2}Cu_{3}O_{7-x}$ where $\alpha _{Cu}$ is
between $-0.14$ and $-0.34$ if T$_{c}$ lies in the $60$ $K$ plateau. (%
\textbf{6}) There are reports on \textit{small negative}\textbf{\
}$\alpha _{O}$ in some systems like $YSr_{2}Cu_{3}O_{7}$ with
$\alpha _{O}\approx -0.02$ and in $BISCO-2223$ ($T_{c}=110$ $K$)
where $\alpha _{O}\approx -0.013 $ etc. However, the systems with
negative $\alpha _{O}$ present considerable experimental
difficulties, as it is pointed out in \cite{Franck}.

The above enumerated results, despite experimental difficulties, are more in
favor than against of the hypothesis that the EPI interaction is strongly
involved in the pairing mechanism of HTSC oxides. By assuming that the
experimental results on the isotope effect reflect an intrinsic property of
HTSC oxides one can rise a question: which theory can explain these results?
Since at present there is no consensus \ on the pairing mechanism in HTSC
materials there is also no definite theory for the isotope effect. Besides
the calculation of the coupling constant $\lambda $ any microscopic theory
of pairing is confronted also with the following questions: $\mathbf{(a)}$
why is the isotope effect small in optimally doped systems and $\mathbf{(b)}$
why $\alpha $ increases rapidly by further under(over)doping of the system?

It should be stressed, that at present all theoretical approaches
are semi-microscopic, but what is interesting most of them
indicate that in order to explain the rather unusual isotope
effect in HTSC materials one should invoke the \textit{forward
scattering peak} in the EPI \cite {KulicReview}.

In conclusion, experimental investigations of the isotope effect
in HTSC oxides have shown the importance of the EPI interaction in
the pairing mechanism.

\subsection{ARPES experiments in HTSC oxides}

\subsubsection{Spectral function $A(\vec{k},\protect\omega)$ from ARPES}

The \textit{angle-resolved photoemission spectroscopy} (ARPES) is nowadays a
leading spectroscopy method in the solid state physics. The method consists
in shining light (photons) with energies between $20-1000$ $eV$ on the
sample and by detecting momentum ($\mathbf{k}$)- and energy($\omega $%
)-distribution of the outgoing electrons. The resolution of ARPES is
drastically increased in the last decade with the energy resolution of $%
\Delta E\approx 2$ $meV$ (for photon energies $\sim 20$ $eV$) and
angular resolution of $\Delta \theta \approx 0.2{{}^{\circ }}$.
The ARPES method is surface sensitive technique, since the average
escape depth ($l_{esc}$)\ of the outgoing electrons is of the
order of $l_{esc}\sim 10$ \AA . Therefore, one needs very good
surfaces in order that the results be representative for the bulk
sample. In that respect the most reliable studies were done on the
bilayer $Bi_{2}Sr_{2}CaCu_{2}O_{8}$ ($Bi2212$) and its single
layer counterpart $Bi_{2}Sr_{2}CuO_{6}$ ($Bi2201$), since these
materials contains weakly coupled $BiO$ planes with the longest
interplane separation in the HTSC oxides. This results in a
\textit{natural cleavage} plane making these materials superior to
others in ARPES experiments. After a drastic improvement of sample
quality in others families of HTSC materials, became the ARPES
technique a central method in theoretical considerations.
Potentially, it gives information on the quasiparticle Green's
function, i.e. on the quasiparticle spectrum and life-time
effects. The ARPES can indirectly give information on the momentum
and energy dependence of the pairing potential. Furthermore, the
electronic spectrum of the HTSC oxides is highly \textit{quasi-2D}
which allows an unambiguous determination of the momentum of the
initial state from the measured final state momentum, since the
component parallel to the surface is conserved in photoemission.
In this case the ARPES probes (under some favorable conditions)
directly the single particle spectral function
$A(\mathbf{k},\omega )$.

In the following we discuss only those ARPES experiments which
give us evidence for the importance of the EPI in HTSC oxides -
see detailed reviews in \cite{Shen}, \cite{Campuzano}.

The \textit{photoemission} measures a nonlinear response function
of the electron system, since the photo-electron current $\langle
\mathbf{j}(1)\rangle $ at the detector is proportional to the
incident photon flux (square of the vector potential
$\mathbf{A}$), i. e. schematically one has
\begin{equation}
\langle \mathbf{j}(1)\rangle \sim \langle \mathbf{j}(\bar{2})\mathbf{j}(1)%
\mathbf{j}(\bar{3})\rangle \mathbf{A(}\bar{2}\mathbf{)A}(\bar{3}),
\label{one-step}
\end{equation}
and\ integration over bar ($1=(\mathbf{x},t$) indices is understood. The
correlation function $\langle \mathbf{j}(\bar{2})\mathbf{j}(1)\mathbf{j}(%
\bar{3})\rangle $ describes all processes related to electrons, such as
photon absorption, electron removal and electron detection, are treated as a
single coherent process. In this case the bulk, surface and evanescent
states, as well as surface resonances should be taken into account - the so
called \textit{one-step model}.

Under some conditions the one-step model can be simplified by an
approximative, but physically plausible, \textit{three-step model}. In this
model the photoemission intensity
\begin{equation}
I_{tot}(\mathbf{k},\omega )=I\cdot I_{2}\cdot I_{3}  \label{Itot}
\end{equation}
is the product of three independent terms: (\textbf{1}) $I$ - describes
optical excitation of the electron in the bulk; (\textbf{2}) $I_{2}$ - the
scattering probability of the travelling electrons; (\textbf{2}) $I_{3}$ -
the transmission probability through the surface potential barrier. The
central quantity in the three-step model is $I(\mathbf{k},\omega )$. To
calculate it one assumes the \textit{sudden approximation, }i.e. that the
outgoing electron is moving so fast that it has no time to interact with the
photo-hole - see more in \cite{Shen},\cite{Campuzano}. It turns out that $I(%
\mathbf{k},\omega )$ can be written in the form \cite{Shen}, \cite{Campuzano}
(for $\mathbf{k=k}_{\parallel }$)
\begin{equation}
I(\mathbf{k},\omega )\simeq I_{0}(\mathbf{k},\upsilon )f(\omega )A(\mathbf{k}%
,\omega ).  \label{Eq20}
\end{equation}
$I_{0}(\mathbf{k},\upsilon )\sim \mid \langle \psi _{f}\mid \mathbf{pA\mid }%
\psi _{i}\rangle \mid ^{2}$ where $\langle \psi _{f}\mid \mathbf{pA\mid }%
\psi _{i}\rangle $ is the dipole matrix element and depends on
$\mathbf{k}$, polarization and energy $\upsilon $ of the incoming
photons. $f(\omega )=1/(1+\exp \{\omega /T\})$ is the Fermi
function and $A(\mathbf{k},\omega
)=-\mathrm{Im}G(\mathbf{k},\omega )/\pi $ is the quasiparticle
spectral
function. In reality because of finite resolution of experiments, in $%
\mathbf{k}$ and $\omega $, $I(\mathbf{k},\omega )$ should be convoluted by
the $\omega $-convolution function $R(\omega )$ and $\mathbf{k}$-convolution
function $Q(\mathbf{k})$. It must be also added the extrinsic background $B$%
, which is due to secondary electrons (those which escape from the sample
after having suffered inelastic scattering events coming out with reduced
kinetic energy).

By measuring $A(\mathbf{k},\omega )$ one can determine $\Sigma (\mathbf{k}%
,\omega )=\Sigma _{1}(\mathbf{k},\omega )+i\Sigma _{2}(\mathbf{k},\omega )$%
\begin{equation}
A(\mathbf{k},\omega )=-\frac{1}{\pi }\frac{\Sigma _{2}(\mathbf{k},\omega )}{%
[\omega -\xi _{0}(\mathbf{k})-\Sigma _{1}(\mathbf{k},\omega )]^{2}+[\Sigma
_{2}(\mathbf{k},\omega )]^{2}}.  \label{Eq21}
\end{equation}
$\xi _{0}(\mathbf{k})=\epsilon _{\mathbf{k}}-\mu $ is the bare quasiparticle
energy. For instance in the case of the Landau-Fermi liquid $A(\mathbf{k}%
,\omega )$ can be separated into the coherent and incoherent part
\begin{equation}
A(\mathbf{k},\omega )=Z_{\mathbf{k}}\frac{\Gamma _{\mathbf{k}}}{(\omega -\xi
(\mathbf{k}))^{2}+\Gamma _{\mathbf{k}}^{2}}+A_{inch}(\mathbf{k},\omega ),
\label{Eq22}
\end{equation}
where $Z_{\mathbf{k}}=1/(1-\partial \Sigma _{1}/\partial \omega )$, $\xi (%
\mathbf{k})=Z_{\mathbf{k}}(\xi _{0}(\mathbf{k})+\Sigma _{1})$ and $\Gamma _{%
\mathbf{k}}=Z_{\mathbf{k}}\mid \Sigma _{2}\mid $ calculated at $\omega =\xi (%
\mathbf{k})$. For small $\omega $ one has $\xi (\mathbf{k})>>\mid \Sigma
_{2}\mid $ and $\Gamma _{\mathbf{k}}\sim \lbrack (\pi T)^{2}+\xi ^{2}(%
\mathbf{k})]$.

In some period of the HTSC era there were a number of
controversial ARPES results and interpretations, due to bad
samples and to the euphoria with exotic theories. For instance, a
number of (now well) established results were \textit{questioned}
in the first ARPES measurements, such as: the shape of the Fermi
surface, which is correctly predicted by the LDA band-structure
calculations; bilayer splitting in $Bi2212$, etc.

\begin{figure}[tbp]
\resizebox{.9 \textwidth}{!}
{\includegraphics*[width=10cm]{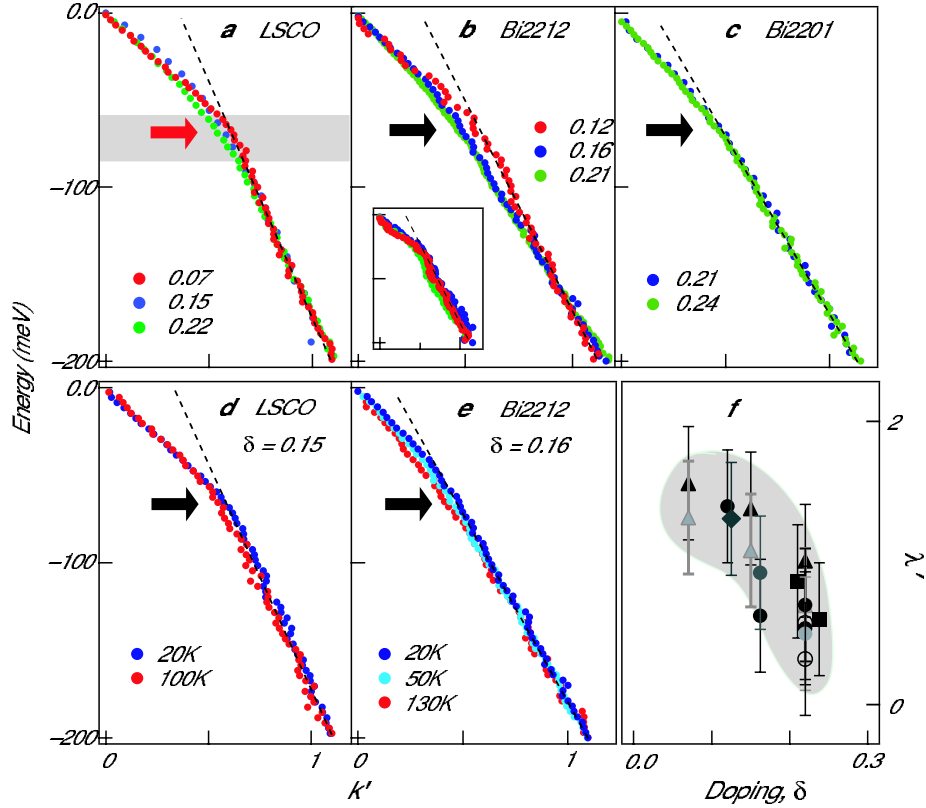} } \caption{Quasiparticle
dispersion of $Bi2212$, $Bi2201$ and $LSCO$ along the nodal
direction, plotted vs the momentum $k$ for $(a)-(c)$ different
doings, and $(d)-(e)$ different $T$; black arrows indicate the
kink energy; the red arrow indicates the energy of the
$q=(\protect\pi,0)$ oxygen stretching phonon mode; inset of $(e)$-
T-dependent $\Sigma ^{\prime }$ for
optimally doped $Bi2212$; $(f)$ - doping dependence of $\protect\lambda%
^{\prime}$ along $(0,0)-(\protect\pi,\protect\pi)$ for the
different HTSC oxides. From \protect\cite{Lanzara}}
\label{ARPESLanzFig}
\end{figure}

\begin{figure}[tbp]
\resizebox{.8 \textwidth}{!} {
\includegraphics*[width=8 cm]{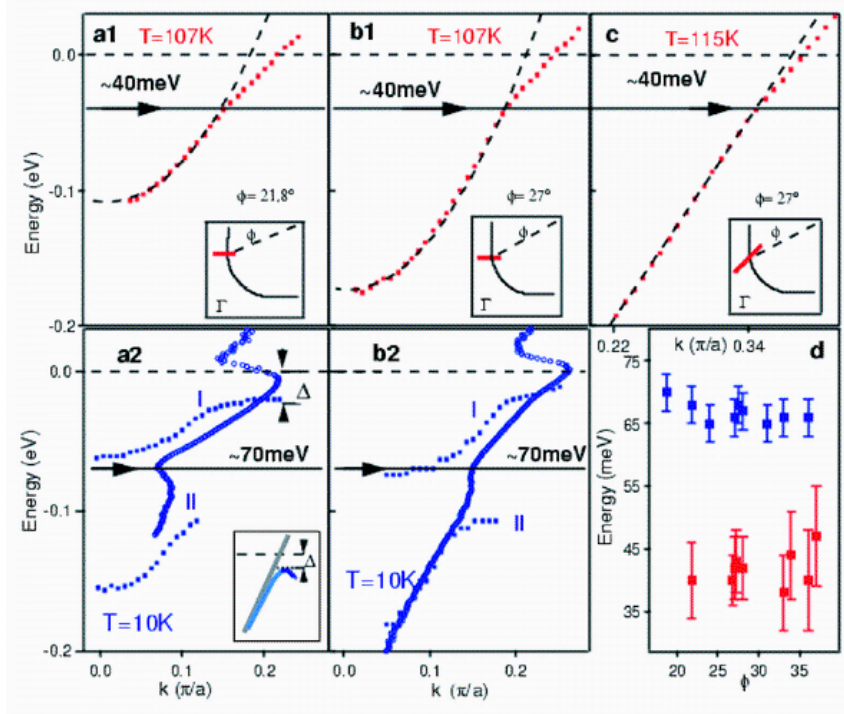}}
\caption{Quasiparticle dispersion $E(k)$ in the normal state (a1,
b1, c), at 107 K and 115 K, along various directions $\phi$ around
the anti-nodal point. The kink at $E=40meV$ is shown by the
horizontal arrow. (a2 and b2) is $E(k)$ in the superconducting
state at 10 K with the shifted kink to $70meV$. (d) kink positions
as a function of $\phi$ in the anti-nodal region. From
\protect\cite{Cuk}} \label{40meVFig}
\end{figure}

We summarize here important ARPES results which were obtained
recently, first in the \textit{normal state} \cite{Shen}, \cite
{Campuzano}: (\textbf{N1}) There is well defined Fermi surface in
the metallic state - with the topology predicted by the LDA;
(\textbf{N2}) the spectral line are broad with $\mid \Sigma
_{2}(\mathbf{k},\omega )\mid \sim \omega $ (or $\sim T$ for
$T>\omega $); (\textbf{N3}) there is a bilayer band splitting in
$Bi2212$ (at least in the overdoped state); (\textbf{N4}) at
temperatures $T_{c}<T<T^{\ast }$ and in the underdoped HTSC oxides
there is a d-wave like pseudogap $\Delta _{pg}(\mathbf{k})\sim
\Delta _{pg,0}(\cos k_{x}-\cos k_{y})$ in the
quasiparticle spectrum; (\textbf{N5}) the pseudogap $%
\Delta _{pg,0}$ increases by lowering doping; (\textbf{N6}) there
is evidence for the strong EPI interaction and
\textit{characteristic phonon} energy $\omega _{ph}$
 - at $T>T_{c}$. Namely, in all HTSC oxides which
are superconducting there are \textit{kinks} in the quasiparticle
dispersion in the nodal direction (along the $(0,0)-(\pi ,\pi )$
line) at around $\omega _{ph}^{(70)}\sim (60-70)$ $meV $
\cite{Lanzara} - see Fig.~\ref {ARPESLanzFig}, and around the
anti-nodal point $(\pi,0)$ at 40 meV \cite{Cuk} - see Fig.~\ref
{40meVFig}.

In the \textit{superconducting state} ARPES results are the following \cite
{Shen}, \cite{Campuzano}: (\textbf{S1}) there is an anisotropic
superconducting gap in most HTSC compounds, predominately of d-wave like, $%
\Delta _{sc}(\mathbf{k})\sim \Delta _{0}(\cos k_{x}-\cos k_{y})$ with $%
2\Delta _{0}/T_{c}\approx 5-6$; (\textbf{S2}) the dramatic changes
in the spectral shapes near the point $(\pi ,0)$, i.e. a
\textit{sharp quasiparticle peak} develops at the lowest binding
energy followed by a dip and a broader hump, giving rise to the so
called \textit{peak-dip-hump structure}; (\textbf{S3}) the kink at
$(60-70)$ $meV$ is surprisingly \textit{unshifted} in the
superconducting state -\cite{Lanzara}. To remind
the reader the standard Eliashberg theory the kink should be shifted to $%
\omega _{ph}+\Delta _{0}$. (\textbf{S4}) the anti-nodal kink at
$\omega _{ph}^{(40)}\sim 40$ $meV$ is shifted in the
superconducting state by $\Delta _{0}$, i.e. $\omega
_{ph}^{(40)}\rightarrow \omega _{ph}^{(40)}+\Delta
_{0}=(65-70)meV$ since $\delta = (25-30)meV$ - see \cite{Cuk}.

\subsubsection{Theory of the ARPES kink}

We would like to point out that the breakthrough-experiments done
by the Shen
group \cite{Lanzara}, \cite{Cuk} shown in Fig.~\ref{ARPESLanzFig},
Fig.~\ref{40meVFig}, with the properties (%
\textbf{N6}), (\textbf{S3}) and (\textbf{S4}) - which we call the
\textit{ARPES\ shift-puzzle}, are the \textit{smoking-gun}
experiments for the microscopic theory of HTSC oxides. Namely, any
theory which reflects to explain the pairing in HTSC oxides must
solve the \textit{shift-puzzle}.

In that respect the recent theory \cite{Kulic-Dolgov}, which is based on the
existence of the forward scattering peak (which is due to strong correlation
in the EPI) in the EPI - the \textit{FSP model}, was able to explain this
puzzle in a consequent way. The FSP model (see more in \cite{KulicReview},
\cite{Kulic-Dolgov}) contains the following basic ingredients: \textit{(i)}
the electron-phonon interaction is dominant in HTSC and its spectral
function $\alpha ^{2}F(\mathbf{k},\mathbf{k}^{\prime },\Omega )\approx
\alpha ^{2}F(\varphi ,\varphi ^{\prime },\Omega )$ ($\varphi $ is the angle
on the Fermi surface) has a pronounced forward scattering peak due to strong
correlations. Its width is very narrow $\mid \mathbf{k}-\mathbf{k}^{\prime
}\mid _{c}\ll k_{F}$ even for overdoped systems \cite{Kulic1}, \cite{Kulic2}%
, \cite{Kulic3}. In the leading order $\alpha ^{2}F(\varphi ,\varphi
^{\prime },\Omega )\sim \delta (\varphi -\varphi ^{\prime })$; \textit{(ii)}
the dynamical part (beyond the Hartree-Fock) of the Coulomb interaction is
characterized by the spectral function $S_{C}(\mathbf{k},\mathbf{k}^{\prime
},\Omega )$. The ARPES shift puzzle implies that $S_{C}$ is \textit{either
peaked} at small transfer momenta $\mid \mathbf{k}-\mathbf{k}^{\prime }\mid $%
, \textit{or it is so small} that the shift is weakly affected and
is beyond the experimental resolution of ARPES. We assume that the
former case is realized; \textit{(iii)} The scattering potential
on non-magnetic impurities has pronounced forward scattering peak,
which is also due to strong correlations \cite{Kulic1},
\cite{Kulic2}, \cite{Kulic3}. The latter is characterized by two
rates $\gamma_{1(2)}$. The case $\gamma_{1}=\gamma_{2}$ mimics the
extreme forward scattering, which does not affect pairing. On the
other hand , $\gamma_{2}=0$ describes the isotropic exchange
scattering - see discussion in

The Green's function is given by $G_{k}=1/(i\omega _{k}-\xi _{\mathbf{k}%
}-\Sigma _{k}(\omega ))$ $=-(i\tilde{\omega}_{k}+\xi _{\mathbf{k}})/(\tilde{%
\omega}_{k}^{2}+\xi _{\mathbf{k}}^{2}+\tilde{\Delta}_{k}^{2})$ where in the $%
k=(\mathbf{k},\omega )$. In the FSP model the equations for $\omega _{k}$
and $\tilde{\Delta}_{k}$ are \cite{Kulic-Dolgov}

\begin{equation}
\tilde{\omega}_{n,\varphi }=\omega _{n}+\pi T\sum_{m}\frac{\lambda
_{1,\varphi }(n-m)\tilde{\omega}_{m,\varphi }}{\sqrt{\tilde{\omega}%
_{m,\varphi }^{2}+\tilde{\Delta}_{m,\varphi }^{2}}}+\Sigma _{n,\varphi }^{C},
\label{Eq23}
\end{equation}

\begin{equation}
\tilde{\Delta}_{n,\varphi }=\pi T\sum_{m}\frac{\lambda _{2,\varphi }(n-m)%
\tilde{\Delta}_{m,\varphi }}{\sqrt{\tilde{\omega}_{m,\varphi }^{2}+\tilde{%
\Delta}_{m,\varphi }^{2}}}+\tilde{\Delta}_{n,\varphi }^{C},  \label{Eq24}
\end{equation}
where
\[
\lambda _{1(2),\varphi }(n-m)=\lambda _{ph,\varphi }(n-m)+\delta _{mn}\gamma
_{1(2),\varphi }
\]
with the electron-phonon coupling function
\begin{equation}
\lambda _{ph,\varphi }(n)=2\int_{0}^{\infty }d\Omega \alpha _{ph,\varphi
}^{2}F_{\varphi }(\Omega )\frac{\Omega }{\Omega ^{2}+\omega _{n}^{2}}.
\label{Spec-fun}
\end{equation}

Since the EPI and $\Sigma _{n,\varphi }^{C}$ in Eq.(\ref{Eq23}-\ref{Eq24})
has a \textit{local form} as a function of the angle $\varphi $, then the
equation for $\tilde{\omega}_{n,\varphi }$ has also local form, which means
that the different points on the Fermi surface are decoupled. In that case $%
\tilde{\omega}_{n,\varphi }$ depends on the local value of the gap $\tilde{%
\Delta}_{n,\varphi }\approx \Delta _{0}\cos 2\varphi $. \textit{Just this
property is important in solving the ARPES shift puzzle}. So, in the \textit{%
nodal point (}$\varphi =\pi /4$\textit{)} one has $\tilde{\Delta}_{n,\varphi
}=0$ and the quasiparticle spectrum given by $E-\xi _{\mathbf{k}}-\Sigma
_{k}(E,\tilde{\Delta}_{n,\varphi }=0)=0$ is \textit{unaffected} by
superconductivity, i.e. the kink is unshifted. This is exactly what is seen
in the experiment of the Shen group \cite{Lanzara} - see Fig.~\ref
{ARPESLanzFig}. In the case of the \textit{antinodal point} ($\varphi
\approx \pi /2$) there is a singularity at $40meV$in the quasiparticle spectrum ($%
E_{\sin g}$) in the normal state - see Fig.~\ref{40meVFig}. The
analytic and numeric calculations of Eq.(\ref{Eq23}) show that
this singularity is \textit{shifted} by $\Delta _{0}$ in
superconducting state, i.e. $E_{\sin g}\rightarrow E_{\sin
g}+\Delta _{0}$. This is exactly what is seen in the recent
experiment on BISCO \cite{Cuk} - see Fig.~\ref{40meVFig}, where
the singularity of the normal state spectrum
at $40$ $meV$ is shifted to $(65-70)$ $meV$ in the superconducting state, since $%
\Delta _{0}\approx (25-30)$ $meV$. The FSP model explains in the
natural way also the \textit{peak-dip-hump structure }in
$A(\mathbf{k},\omega )$ - for more details see
\cite{Kulic-Dolgov}.

\subsubsection{ARPES and the EPI coupling constant $\lambda$}

One can rise the question - is it possible to extract the coupling
constant $\lambda $ from ARPES measurements. As we have seen
above, by
assuming that the three-step model holds, where $I(\mathbf{k},\omega )\sim A(%
\mathbf{k},\omega )$, then one possibility is by measuring the kink in the
quasiparticle renormalization, i.e. by measuring the real part of the self-energy
$\Sigma _{1}(\mathbf{k},\omega )$.
These measurements \cite{Lanzara}, \cite{Cuk} give $%
\lambda _{ARPES}^{(1)}\sim 1$ in both nodal and antinodal direction. Another
possibility is by measuring the width ($\Delta k_{FW}(\omega )$) of the
momentum distribution curves (MDCs) which give the imaginary part $\Sigma
_{2}(\omega ,T)$ via
\begin{equation}
\Sigma _{2}(\omega ,T)\approx \frac{1}{2}v_{F}\Delta k_{FW}(\omega ).
\label{Sig2}
\end{equation}

\begin{figure}[tbp]
\resizebox{.8 \textwidth}{!} {
\includegraphics*[width=8cm]{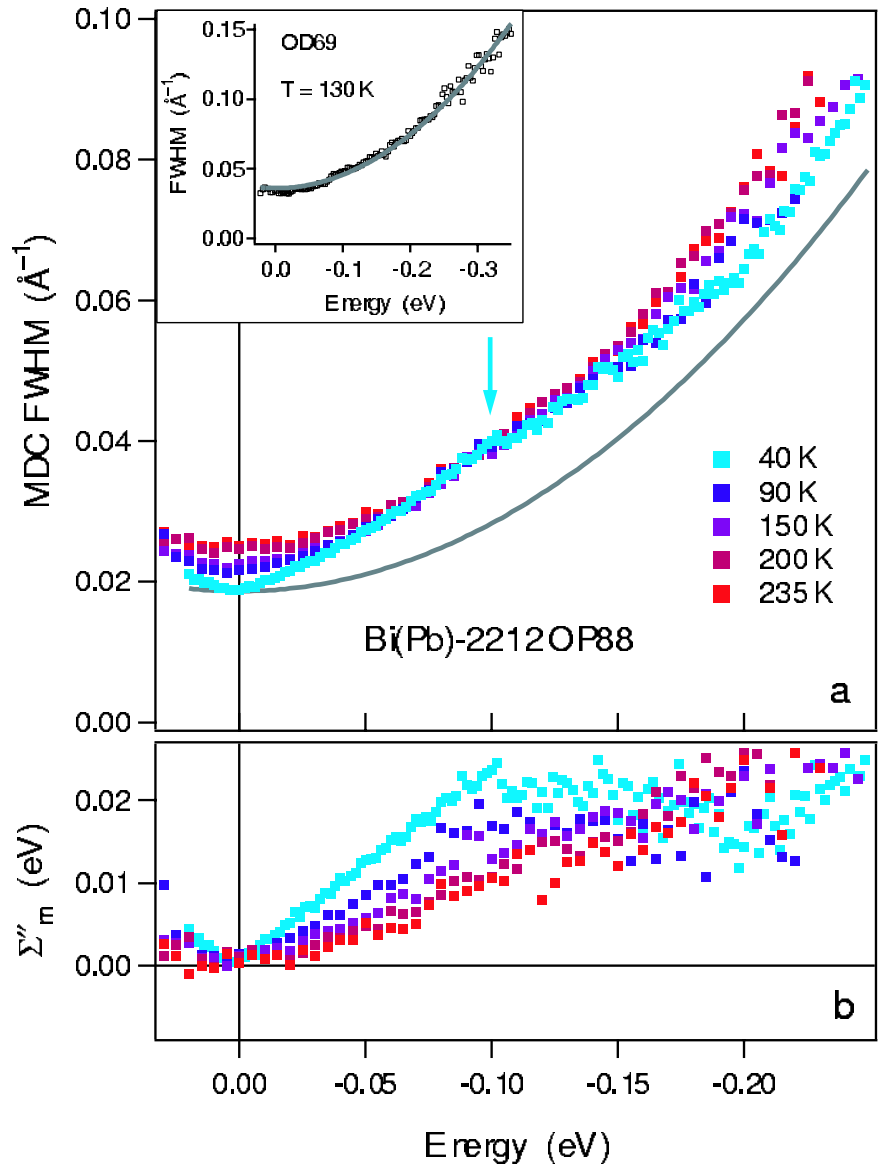}}
\caption{T- and $\protect\omega$-dependence of $\Sigma _{2}$ for the nodal
quasiparticles in optimally doped Bi(Pb)-2212. (\textbf{a}) - the full width
at half maximum of the ARPES intensity. The gray solid line is the Fermi
liquid parabola obtained by fitting the data for highly overdoped sample
(OD69) at 130 K (see inset). (\textbf{b}) - the bosonic part $\Sigma _{2,B}$
for various T.From \protect\cite{Kordyuk}}
\label{ARPESKordyukFig}
\end{figure}

In that respect very indicative are recent measurements \cite{Kordyuk} of $%
\Sigma _{2}(\omega ,T)$ around the \textit{nodal point} in a
number of HTSC compounds, such as the superstructure free
$Bi_{2-x}Pb_{x}Sr_{2}CaCu_{2}O_{8+\delta }$ ($Bi(Pb)-2212
$), $Bi_{2}Sr_{2}CaCu_{2}O_{8+\delta }$ ($Bi-2212$) and $%
Bi_{2}Sr_{2-x}La_{x}Cu_{2}O_{8+\delta }$ ($Bi-2201$). In the
analysis of ARPES spectra the authors in \cite{Kordyuk} have
assumed that there are two $\omega$-dependent contributions to
$\Sigma $ - the Fermi liquid contribution $\Sigma_{FL}$ and the
part $\Sigma_{B}$ due to the interaction via bosonic excitations
(let say phonons and spin-fluctuations), i.e. $\Sigma _{2}=\Sigma
_{2,FL}+\Sigma _{2,B}+\Sigma _{2,imp}$ - see
Fig.~\ref{ARPESKordyukFig}. We stress, that in \cite{Kordyuk} it
is assumed that $v_{F}=4$ $eV$ \AA.

In the framework of this procedure the theoretical analysis \cite
{Kulic-Dolgov-lambda} gives small coupling $\lambda
_{ARPES}^{(2)}<0.2$, which is extracted from the slope of $\Sigma
_{2}$ in the interval $0.05$ $eV<\omega <0.1$ $eV$ - see the gray
solid line in Fig.~\ref{ARPESKordyukFig}b). Such a small coupling
gives very small T$_{c}$ and none of pairing mechanisms is
effective. Furthermore, the small value of $\lambda
_{ARPES}^{(2)}$ (most measurements give $\lambda _{\max
,ARPES}^{(2)}<0.4$) - which is extracted from $\Sigma _{2}$, is a
generic
property of most ARPES measurements. This means, that we are \textit{%
confronted with a trilemma}: (\textbf{1}) to abandon the
boson-fermion separation procedure of $\Sigma (\omega)$ done in
\cite{Kordyuk}, (\textbf{2}) to abandon the Eliashberg theory,
(\textbf{3}) to abandon the interpretation of ARPES data within
the three-step model.

In some sense the situation in ARPES measurements with $\lambda
_{ARPES}^{(2)}\ll \lambda $ resembles the one in transport measurements
where $\lambda _{tr}\ll \lambda $, i.e. . This problem deserves further
investigation.

\section{EPI in HTSC oxides}

In the following we present briefly some elements of the general
theory of the strong EPI and its low-energy version. In the
latter, high-energy processes are integrated out and the
low-energy phenomena are governed by
the high-energy vertex functions $\Gamma _{c}$, the excitation potential $%
\Sigma _{0}$ (part of the self-energy due to the Coulomb
interaction) and the EPI coupling constants $g_{EP,ren}\sim \Gamma
_{c}$ - see more in \cite {KulicReview}. However, this procedure
was never performed in its full extent, because of difficulties to
calculate $\Sigma _{0}$ and $\Gamma _{c}$. Therefore, the EPI
coupling constant was as a rule calculated by some other methods.
Usually in $LTSC$ materials the EPI is calculated by using the
local-density functional ($LDA$) method, which is suitable for
ground state properties of crystals (matter) and which is based on
an effective electronic crystal potential $V_{g}$. Since in
principle $V_{g}$ may significantly differ from $\Sigma _{0}$ then
the $LDA$ calculated coupling constant $g_{EP}^{(LDA)}$ can be
also very different from the real coupling constant $g_{EP}$. The calculation of $%
g_{EP}^{(LDA)}$ is complicated, even in the $LDA$ method, and further
approximations are necessary, like for instance the rigid-ion ($RI$) and
rigid muffin-tin ($RMTA$) approximations. These approximation were justified
in simple metals. However, these approximations are inadequate for HTSC
oxides, because they fail to take into account correctly the \textit{%
long-range forces} (Madelung energy - see below) and \textit{strong
electronic correlations}. Strictly speaking the EPI does not have meaning in
the $LDA$ method - see more in \cite{KulicReview}, because the latter treats
the ground state properties of materials, while the EPI is due to excited
states and inelastic processes in the system.

\subsection{General strong coupling theory of the EPI}

It is based on the fully microscopic electron-ion Hamiltonian for the
interacting electrons and ions in a crystal - see for instance \cite
{ScalapRev}, \cite{DoMa}, \cite{DierkLowTemp}, and comprises electrons
interacting between themselves as well as with ions and ionic vibrations. In
order to describe superconductivity the Nambu-spinor $\hat{\psi}^{\dagger }(%
\mathbf{r})$ is introduced which operates in the electron-hole space\ $\
\hat{\psi}^{\dagger }(\mathbf{r})$\ $=(\hat{\psi}_{\uparrow }^{\dagger }(%
\mathbf{r})$ $\hat{\psi}_{\downarrow }(\mathbf{r}))$ (analogously for the
column $\hat{\psi}(\mathbf{r})$) with $\hat{\psi}_{\uparrow }(\mathbf{r})$, $%
\hat{\psi}_{\uparrow }^{\dagger }(\mathbf{r})$ as annihilation and creation
operators for spin up, respectively etc. The microscopic Hamiltonian which
in principle should describe the normal and superconducting state of the
system contains three parts $\hat{H}=\hat{H}_{e}+\hat{H}_{i}+\hat{H}_{e-i}$.
The \textit{electronic Hamiltonian }$\hat{H}_{e}$, which describes the
kinetic energy and the Coulomb interactions of electrons, is given in the
second-quantization by
\[
\hat{H}_{e}=\int d^{3}r\hat{\psi}^{\dagger }(\mathbf{r})\hat{\tau}%
_{3}\epsilon _{0}(\hat{p})\hat{\psi}(\mathbf{r})+
\]
\begin{equation}
+\frac{1}{2}\int d^{3}rd^{3}r^{\prime }\hat{\psi}^{\dagger }(\mathbf{r})\hat{%
\tau}_{3}\hat{\psi}(\mathbf{r})V_{c}(\mathbf{r-r}^{\prime })\hat{\psi}%
^{\dagger }(\mathbf{r}^{\prime })\hat{\tau}_{3}\hat{\psi}(\mathbf{r}^{\prime
}),  \label{Eq25}
\end{equation}
where $\epsilon _{0}(\hat{p})=\hat{p}^{2}/2m$ is the kinetic energy of
electron and $V_{c}(\mathbf{r-r}^{\prime })=e^{2}/\mid \mathbf{r-r}^{\prime
}\mid $ is the electron-electron Coulomb interaction. Note, that in the
electron-hole space the pseudo-spin (Nambu) matrices $\hat{\tau}_{i}$, $%
i=0,1,2,3$ are Pauli matrices.

The \textit{lattice Hamiltonian}\textbf{\ } (describes lattice vibrations $%
\hat{u}_{\alpha n}$ of ions enumerated by $n$) is given by
\[
\hat{H}_{i}=\frac{1}{2}\sum_{n}M(\frac{d\mathbf{\hat{u}}_{n}}{dt})^{2}+\frac{%
1}{2}\sum_{n,m,\alpha }V_{ii}(\mathbf{R}_{n}^{0}-\mathbf{R}_{m}^{0})+\frac{1%
}{2}\sum_{n,m}(\hat{u}_{\alpha n}-\hat{u}_{\alpha m})\nabla _{\alpha }V_{ii}(%
\mathbf{R}_{n}^{0}-\mathbf{R}_{m}^{0})+
\]

\begin{equation}
+\frac{1}{2}\sum_{n,m,\alpha ,\beta }(\hat{u}_{\alpha n}-\hat{u}_{\alpha m})(%
\hat{u}_{\beta n}-\hat{u}_{\beta m})\nabla _{\alpha }\nabla _{\beta }V_{ii}(%
\mathbf{R}_{n}^{0}-\mathbf{R}_{m}^{0})+\hat{H}_{i}^{anh}.  \label{Eq26}
\end{equation}
The first term in Eq.(\ref{Eq26}) is the kinetic energy of vibrating ions
(with charge $Ze$), $V_{ii}(\mathbf{R}_{n}^{0}-\mathbf{R}%
_{m}^{0})=Z^{2}e^{2}/\mid \mathbf{R}_{n}^{0}\mathbf{-R}_{m}^{0}\mid $ is the
bare ion-ion interaction in equilibrium, while the third and fourth terms
describe the change of $V_{ii}$ by lattice vibrations with the
ion-displacement is \textbf{$\hat{u}$}$_{n}=\mathbf{R}_{n}-\mathbf{R}%
_{n}^{0} $. The term $\hat{H}_{i}^{anh}$ describes higher anharmonic terms
with respect to $\hat{u}_{n}^{\beta }$. The theory which we describe below
holds for any kind of anharmonicity.

The \textit{electron-ion Hamiltonian} describes the interaction of electrons
with the equilibrium lattice and with its vibrations, respectively
\begin{equation}
\hat{H}_{e-i}=\sum_{n}\int d^{3}rV_{e-i}(\mathbf{r}-\mathbf{R}_{n}^{0})\hat{%
\psi}^{\dagger }(\mathbf{r})\hat{\tau}_{3}\hat{\psi}(\mathbf{r})+\int d^{3}r%
\hat{\Phi}(\mathbf{r})\hat{\psi}^{\dagger }(\mathbf{r})\hat{\tau}_{3}\hat{%
\psi}(\mathbf{r}),  \label{Eq27}
\end{equation}
\[
\hat{\Phi}(\mathbf{r})=\sum_{n}[V_{e-i}(\mathbf{r}-\mathbf{R}_{n}^{0}-%
\mathbf{\hat{u}}_{n})-V_{e-i}(\mathbf{r}-\mathbf{R}_{n}^{0}).
\]
Here, $V_{e-i}(\mathbf{r}-\mathbf{R}_{n}^{0})$ is the electron-ion potential
- see \cite{KulicReview}. The second term which depends on the lattice
distortion operator $\hat{\Phi}(\mathbf{r})$ describes the interaction of
electrons with harmonic ($\sim \hat{u}_{\alpha n}$) (or anharmonic $\sim
\hat{u}_{\alpha n}^{k}$, $k=2,3...$) lattice vibrations.

Based on the above Hamiltonian one can in principle calculate the electron
and phonon Green's functions
\begin{equation}
\hat{G}(1,2)=-\langle T\hat{\psi}(1)\hat{\psi}^{\dagger }(2)\rangle
\label{Gel}
\end{equation}
and
\begin{equation}
\tilde{D}(1-2)=-\langle T\hat{\Phi}(1)\hat{\Phi}(2)\rangle ,  \label{Dph}
\end{equation}
respectively. The solution of these equations is written in the form of
Dyson's equations
\begin{equation}
\hat{G}^{-1}(1,2)=\hat{G}_{0}^{-1}(1,2)-\hat{\Sigma}(1,2)  \label{G-Day}
\end{equation}
and
\begin{equation}
\tilde{D}^{-1}(1,2)=\tilde{D}_{0}^{-1}(1,2)-\tilde{\Pi}(1,2),  \label{D-Day}
\end{equation}
where $\hat{G}_{0}^{-1}(1,2)$ and $\tilde{D}_{0}^{-1}(1,2)$ are the bare
inverse electron and phonon Green's function, respectively. The nontrivial
effects of interactions are hidden in the self-energies $\hat{\Sigma}(1,2)$
and $\tilde{\Pi}(1,2)$. Here, $1=(\mathbf{r}_{1},\tau _{1})$, where $\tau
_{1}$ is the imaginary time. The calculation of $\hat{\Sigma}$ is simplified
by using the \textit{Migdal adiabatic approximation} \cite{Migdal}, which
incorporates the experimental fact that in most metals the characteristic
phonon (Debye) energy of lattice vibrations $\omega _{D}$ is much smaller
than the characteristic electronic Fermi energy $E_{F}$ ($\omega _{D}\ll
E_{F}$). Using this fact Migdal formulated a theorem which claims that in
the self-energy $\Sigma $ one should keep explicitly terms linear in the
phonon propagator $\tilde{D}$ only. As the result one obtains the \textit{%
Migdal-Eliashberg theory} for
\begin{equation}
\hat{\Sigma}=\hat{\Sigma}_{c}+\hat{\Sigma}_{EP},  \label{Sitot}
\end{equation}
where

\begin{equation}
\hat{\Sigma}_{c}(1,2)=-V_{c}^{sc}(1,\bar{1})\hat{\tau}_{3}\hat{G}(1,\bar{2})%
\hat{\Gamma}_{c}(\bar{2},2;\bar{1}).  \label{Eq28}
\end{equation}
$V_{c}^{sc}(1,2)=V_{c}(1,\bar{2})\varepsilon _{e}^{-1}(\bar{2},2)$ is the
screened Coulomb interaction. The part which is due to the EPI has the
following form
\begin{equation}
\hat{\Sigma}_{EP}(1,2)=-V_{EP}(\bar{1},\bar{2})\hat{\Gamma}_{c}(1,\bar{3};%
\bar{1})\hat{G}(\bar{3},\bar{4})\hat{\Gamma}_{c}(\bar{4},2;\bar{2}),
\label{Eq29}
\end{equation}
where
\[
V_{EP}(1,2)=\varepsilon _{e}^{-1}(1,\bar{1})\tilde{D}(\bar{1},\bar{2}%
)\varepsilon _{e}^{-1}(\bar{2},2)
\]
is the screened EPI and $\varepsilon _{e}$ is the electronic dielectric
function. Note, $\hat{\Sigma}_{EP}(1,2)$ depends now quadratically on the
vertex function $\hat{\Gamma}_{c}$, due to the adiabatic theorem. If $\hat{%
\Gamma}_{c}$ (which is a functional of $\hat{G}$) is known then the
quasiparticle dynamics can be in principle determined. In that respect the
central question is: (\textbf{1}) how to calculate $\hat{\Gamma}_{c}$ -
which contains all information on Coulomb interaction and electronic
correlations? This is a difficult task and practically never realized in its
full extent for real systems. However, this program is realized recently in
the t-J model with the EPI interaction in the framework of the X-method -
see below and \cite{KulicReview}; (\textbf{2}) how to calculate the
effective EPI potential $V_{EP}$ $\sim g_{EP}^{2}/\varepsilon _{e}^{2}$, or
more precisely the coupling constant $g_{EP}$ and the electronic dielectric
function $\varepsilon _{e}$? In absence of a better theory these quantities
are usually calculated in the framework of the LDA band-structure theory.

\subsection{LDA calculations of $\protect\lambda$ in HTSC oxides}

The LDA method considers electrons in the ground state (there is a
generalization to finite $T$), whose energy can be calculated by knowing the
spectrum \{$\epsilon _{k}$\} of the Kohn-Sham (Schr\"{o}dinger like)
equation
\begin{equation}
\lbrack \frac{\mathbf{\hat{p}}^{2}}{2m}+V_{g}(\mathbf{r})]\psi _{k}(\mathbf{r%
})=\epsilon _{k}\psi _{k}(\mathbf{r}),  \label{Eq30}
\end{equation}
which depends on the \textit{effective one-particle potential}
\begin{equation}
V_{g}(\mathbf{r})=V_{ei}(\mathbf{r})+V_{H}(\mathbf{r})+V_{XC}(\mathbf{r}).
\label{Eff-pot}
\end{equation}
Here. $V_{ei}$ is the electron-lattice potential, $V_{H}$ is the Hartree
term and $V_{XC}$ describes exchange-correlation effects - see \cite
{KulicReview}. Because the EPI depends on the excited states (above the
ground state) of the system this means, that in principle the LDA method can
not describe it - see \cite{KulicReview}. However, by using an analogy with
the microscopic Migdal-Eliashberg theory one can define the EPI coupling
constant $g^{(Mig)}=g\Gamma _{c}/\varepsilon $ also in the LDA theory - see
\cite{KulicReview}. It reads

\begin{equation}
g_{\alpha ,ll^{\prime }}^{(LDA)}(\mathbf{k},\mathbf{k}^{\prime
})=\sum_{n}g_{\alpha ,nll^{\prime }}^{(LDA)}(\mathbf{k},\mathbf{k}^{\prime
})=\langle \psi _{\mathbf{k}l}\mid \sum_{n}\frac{\delta V_{g}(\mathbf{r})}{%
\delta R_{n\alpha }}\mid \psi _{\mathbf{k}^{\prime }l^{\prime }}\rangle ,
\label{Eq31}
\end{equation}
where $n$ means summation over the lattice sites, $\alpha =x,y,z$ and the
wave function $\psi _{\mathbf{k}l}$ is the solution of the Kohn-Sham
equation. Formally one has $\delta V_{g}/\delta \mathbf{R}_{n}=\Gamma
_{LDA}\varepsilon _{e}^{-1}\nabla V_{ei}$. Even in such a simplified
approach it is difficult to calculate $g_{\alpha ,ll^{\prime
}}^{(LDA)}=g_{\alpha ,n}^{RMTA}+g_{\alpha ,n}^{nonloc}$ because it contains
the short-range (local) coupling
\begin{equation}
g_{\alpha ,n}^{RMTA}\sim g_{\alpha ,n}^{RMTA}(\mathbf{k},\mathbf{k}^{\prime
})\sim \langle Y_{lm}\mid \hat{r}_{\alpha }\mid Y_{l^{\prime }m^{\prime
}}\rangle   \label{RMT}
\end{equation}
with $\Delta l=1$, and the long-range coupling
\begin{equation}
g_{\alpha ,n}^{nonloc}(\mathbf{k},\mathbf{k}^{\prime })\sim \langle
Y_{lm}\mid (\mathbf{R}_{n}^{0}-\mathbf{R}_{m}^{0})_{\alpha }\mid
Y_{l^{\prime }m^{\prime }}  \label{Nonloc}
\end{equation}
with $\Delta l=0$. In most calculations the local term $g_{\alpha ,n}^{RMTA}$
is calculated only, which is justified in simple metals only but not in the
HTSC oxides. In HTSC oxides the latter gives very small EPI coupling $%
\lambda ^{RMTA}\sim 0.1$, which is apparently much smaller than the
experimental value $\lambda >1$ giving rise to the pessimistically small T$%
_{c}$ \cite{Mazin}. The small $\lambda ^{RMTA}$ was also one of the reasons
for abandoning the EPI as pairing mechanism in HTSC oxides. At the beginning
of the HTSC era the electron-phonon spectral function $\alpha ^{2}F(\omega )$%
\ for the case $La_{2-x}Sr_{x}CuO_{4}$ was calculated in \cite{Weber} by
using the first-principles band structure calculations and the nonorthogonal
tight-binding theory of lattice dynamics. It was obtained $\lambda =2.6$ and
for assumed $\mu ^{\ast }=0.13$ gave $T_{c}=36$ $K$. However, these
calculations predict a lattice instability for the oxygen breathing mode
near $La_{1.85}Sr_{0.15}CuO_{4}$ that is never observed. Moreover, the same
method was applied\ to $YBa_{2}Cu_{3}O_{7}$\ in \cite{Weber-Mattheiss} where
it was found $\lambda =0.5$ which leads at most to $T_{c}=(19-30)$ $K$.\ In
fact the calculations in \cite{Weber}, \cite{Weber-Mattheiss} do not take
into account the Madelung coupling (i.e. neglect the matrix elements with $%
\Delta l=0$). \ \ \ \ \ \ \ \ \ \ \ \ \ \ \ \

However, because of the weak screening of the ionic (long-range) Madelung
coupling in HTSC oxides - especially for vibrations along the c-axis, it is
necessary to include the nonlocal term $g_{\alpha ,n}^{nonloc}$. This goal
was achieved in the LDA approach by the Pickett's group \cite{Krakauer},
where the EPI coupling for $La_{2-x}M_{x}CuO_{4}\ $is calculated in the
\textit{frozen-in phonon} ($FIP$) method. They have obtained $\lambda =1.37$
and $\omega _{\log }\approx 400$ $K$ and for $\mu ^{\ast }=0.1$ one has $%
T_{c}=49$ $K$ ($T_{c}\approx \omega _{\log }\exp \{-1/[(\lambda /(1+\lambda
))-\mu ^{\ast }]\}$). For more details see Ref. \cite{KulicReview} and
references therein. We point out, that some calculations which are based on
the tight-binding parametrization of the band structure in $%
YBa_{2}Cu_{3}O_{7}$ gave rather large EPI coupling $\lambda \approx 2$ and $%
T_{c}=90$ $K$\

Recently, a new \textit{linear-response full-potential
linear-muffin-tin-orbital} ($LR$-$LMTO$) method for the calculation of $%
\lambda ^{LDA}$ was invented in \cite{Savrasovi}. It is very efficient in
explaining the physics of elemental metals, like $Al,Cu,Mo,Nb,Pb,Pd,Ta$ and $%
V$ with disagreements by only $10-30\%$ of theoretical and experimental
results (obtained from tunnelling and resistivity measurements) for the EPI
coupling constants $\lambda $ and $\lambda _{tr}$. However, the $LR$-$LMTO$
method applied to the doped HTSC oxide $(Ca_{1-x}Sr_{x})_{1-y}CuO_{2}$ for $%
x\sim 0.7$ and $y\sim 0.1$ with $T_{c}=110$ $K$ gives surprisingly small $EPI
$ coupling $\lambda _{s}\approx 0.4$ for $s-wave$ pairing and $\lambda
_{d}\leq 0.3$ for $d-wave$ pairing \cite{SavAnder}. Although this finding,
that $\lambda _{d}$ is of the similar magnitude as $\lambda _{s}$ ($\lambda
_{d}\approx \lambda _{s}$), is interesting and encouraging it seems that
this method misses some ingredients of the ionic structure of the layered
structure \cite{Falter97}, \cite{Falter98}.

We point out, that the model calculations which take into account the
long-range ionic Madelung potential appropriately \cite{Jarlborg}, \cite
{Barisic1}, \cite{Zeyher1} gave also rather large coupling constant $\lambda
\sim 2$, what additionally hints to the importance of the long-range forces
in the EPI.

Since in HTSC oxides the plasma frequency along the c-axis, $\omega _{pl}^{c}
$, is of the order (or even less) of some characteristic c-axis vibration
mode, it is necessary to include the \textit{nonadiabatic effects} in the
EPI coupling constant, i.e. its frequency dependence $g_{\alpha ,n}\sim
g^{0}/\varepsilon _{cc}(\omega )$. This non-adiabaticity is partly accounted
for in the Falter group \cite{Falter97}, \cite{Falter98} by calculating the
electronic dielectric function along the c-axis $\varepsilon _{cc}(\mathbf{k}%
,\omega )$ in the RPA approximation. The result is that $g_{\alpha ,n}$ is
increased appreciable beyond its (well screened) metallic part, what gives a
large increase of the EPI coupling not only in the phonon modes but also in
the plasmon one. This question deserves much more attention than it was in
the past.

\subsection{Lattice dynamics and EPI coupling}

The calculation of the phonon frequencies $\omega _{ph}$, which are obtained
from
\begin{equation}
D_{0}^{-1}(\mathbf{q},\omega _{ph})-\hat{\Pi}(\mathbf{q},\omega _{ph})=0,
\label{Ph-mode}
\end{equation}
is in principle even more complicated problem than the calculations of the
electronic properties. It lies on the difficulty to calculate the phonon
polarization operator $\hat{\Pi}$ - see more in \cite{KulicReview}.
Schematically one has
\begin{equation}
\hat{\Pi}\sim (\nabla _{\alpha }V_{e-i})^{2}\hat{\chi}_{c},  \label{Pol}
\end{equation}
where $\hat{\chi}_{c}$ is the \textit{electronic charge susceptibility. }$%
V_{e-i}$ is the bare electron-lattice interactions. $\hat{\chi}_{c}$ is
schematically given by $\hat{\chi}_{c}=\hat{P}\hat{\varepsilon}_{e}^{-1}$
and the electronic \textit{polarization operator} $\hat{P}=\hat{G}\hat{\Gamma%
}_{c}\hat{G}$. As we see the phonon frequencies depends crucially on the
screening properties of electrons. The screening effects in HTSC oxides are
determined by the specificity of the metallic-ionic structure and strong
electronic correlations. At present there is a controllable theory for the
electronic properties in the t-J model \cite{Kulic1}, \cite{Kulic2}, \cite
{Kulic3}, \cite{Kulic4}, \cite{KulicReview} only, where these two
ingredients are successfully incorporated in the theory. However, until now
there is no controllable theory for the lattice dynamics which incorporates
these two ingredients, in spite the fact that the X-method (see below)
offers well defined and procedure. There were a number of interesting
attempts to calculate renormalization of some specific phonons \cite{Horsch}%
, \cite{Rosch}, \cite{Ishihara}, such as for instance of the half-breathing
phonon mode along the $(1,0,0)$ direction - which is strongly softened. In
spite of some alleged theoretical confirmation of the experimental softening
in YBCO and LASCO, none of these calculations are reliable, because none of
them take into account the screening due to strong correlations (the charge
vertex $\hat{\Gamma}_{c}$ and the dielectric function $\hat{\varepsilon}_{e}$%
) in a controllable way. That is the reason that all attempts until were
unable to extract the reliable magnitude of the coupling constant with a
specific phonon. Even more, by playing only with a single phonon mode, and
with a particular wave-vector in the Brillouin zone, \textit{one can not get
large EPI} and large $\lambda $ - see \cite{KulicReview}. The latter claim
is confirmed by tunnelling experiments, which demonstrate that almost all
phonons (for instance $39$ modes in $YBCO$) contribute to $\lambda $. No
particular mode can be singled out in the spectral function $\alpha
^{2}F(\omega )$ as being the only one which dominates in pairing mechanism
in HTSC oxides.

\section{Theory of strong electronic correlations}

The well established fact is that \textit{strong electronic correlations}
are pronounced in HTSC oxides, at least in underdoped systems. However, the
LDA theory fails to capture effects of strong correlations by treating they
as a local perturbation. This is, as we shall see later, an unrealistic
approximation in HTSC oxides, where strong correlations introduce
non-locality. The shortcoming of the $LDA$ is that in the half-filling case
(with $n=1$ and one particle per lattice site) it predicts metallic state
missing the existence of the \textit{Mott insulating state.} In the latter,
particles are localized at lattice sites independent of the (non)existence
of the $AF$ order and the localization is due to the large Coulomb repulsion
$U$ at a given lattice site, i.e. $U\gg W$ where $W$ is the band width. Some
properties in the metallic state can not be described by the simple
canonical Landau-Fermi liquid concept. For instance, recent ARPES
photoemission measurements \cite{Olson} on the hole doped samples show a
well defined Fermi surface in the one-particle energy spectrum, which
contains $1-\delta $ electrons in the Fermi volume ($\delta $ is the hole
concentration), but the band width is $(2-3)$ times smaller than the $LDA$
band structure calculations predict. The latter is consistent with the
Luttinger theorem as well as with the $LDA$ band structure calculations.
However, experimental data on the dynamical conductivity (spectral weight of
the Drude peak), Hall measurements etc. indicate that in transport
properties a low density of hole-like charge carriers (which is proportional
to $\delta $) participates predominantly. These carriers suffer strong
scattering and their inverse lifetime is proportional to the temperature (at
$T>T_{c}$) as we discussed earlier. It is worth of mentioning here that the
local moments on the $Cu$ sites, which are localized in the parent $AF$
compound, are counted as part of the Fermi surface area when the system is
doped by small concentration of holes in the metallic state. The latter fact
gives rise to a \textit{large Fermi surface} which scales with the number
(per site) of electrons $1-\delta $. At the same time the conductivity
sum-rule is proportional to the number of doped holes $\delta $, instead of $%
1-\delta $ as in the canonical Landau-Fermi liquid. These two properties
tell us that we deal with a \textit{correlated state}, and the latter must
be due to the specific electronic structure of HTSC oxides (cuprates). The
common ingredient of all cuprates is the presence of the $Cu$ atoms. In
order to account for the absence of $Cu^{3+}$ ionic configuration (the
charge transfer $Cu^{2+}\rightarrow Cu^{3+}$ costs large energy $U\sim 10$ $%
eV$, i.e. the occupation of the $Cu$ site with two holes with opposite spins
is unfavorable) P. W. Anderson \cite{Anderson} proposed the Hubbard model as
the basic model for quasiparticle properties in these compounds. For some
parameter values it can be derived from the (minimal) microscopic \textit{%
three-band model}. Besides the hopping $t_{pd}$ between the $d$-orbitals of $%
Cu$ and $p$-orbitals of $O$ ions (as well as $t_{pp}$) - the Emery model
\cite{Emery}, it includes also the strong Coulomb interaction $U_{Cu}$ on
the Cu ions as well as interaction between p- and d-electrons. The main two
parameters are $U_{Cu}\sim (6-10)$ $eV$ and the charge transfer energy $%
\Delta _{pd}\equiv \epsilon _{d}^{0}-\epsilon _{p}^{0}\sim (2.5-4)$ $eV$,
where $\epsilon _{d}^{0},\epsilon _{p}^{0}$ are energies of the d- and
p-level, respectively. In HTSC oxides the case $U_{Cu}>>\Delta _{pd}$ is
realized, i.e. they belong to the class of \textit{charge transfer materials}%
. This allows us to project the complicated three-band Hamiltonian onto the
low-energy sector, and to obtain an effective single-band Hubbard
Hamiltonian with an effective hopping parameter $t$ and the effective
repulsion $U\approx \Delta _{pd}$. It turns out that the case $U>>t$ is
realized, since $\Delta _{pd}\gg t=t_{pd}^{2}/\Delta _{pd}$. The effective
and minimal Hamiltonian which describes the low-energy physics of HTSC
oxides comprises also the long-range Coulomb interaction $\hat{V}_{C}$ and
the EPI $\hat{V}_{EPI}$ - see \cite{Kampf}, \cite{KulicReview}
\begin{equation}
\hat{H}=-\sum_{i,j,\sigma }t_{ij}c_{i\sigma }^{\dagger }c_{j\sigma
}+U\sum_{i}n_{i\uparrow }n_{i\downarrow }+\hat{V}_{C}+\hat{V}_{EPI}.
\label{Eq32}
\end{equation}
The effective repulsion $U\approx 4$ $eV$ has its origin in the
charge-transfer gap of the three-band model, while the nearest neighbor and
next-nearest neighbor hopping $t$ and $t^{\prime }$ , respectively are
estimated to be $t=0.3-0.5$ $eV$ and $t^{\prime }/t$ equal $-0.15$ in $La$
compounds and $-0.45$ $YBCO$. Since $(U/t)\gg 1$ the above Hamiltonian is
again in the regime of strong electronic correlations, where the doubly
occupancy of a given lattice site is strongly suppressed, i.e. $\langle
n_{i\uparrow }n_{i\downarrow }\rangle \ll 1$. The latter restricts charge
fluctuations of electrons (holes) on a given lattice site are allowed, since
$n_{i}=0,1$ is allowed only, while processes with $n_{i}=2$ are
(practically) forbidden. Note, that in (standard) weakly correlated metals
all charge fluctuation processes ($n_{i}=0,1,2$) are allowed, since $U\ll W$
in these systems. From the Hamiltonian in Eq.(\ref{Eq32}), which is the 2D
model for the low-energy physics in the $CuO_{2}$ plane, comes out that in
the \textit{undoped} system there is one particle per lattice - the so
called half-filled case (in the band language) with $\langle n_{i}\rangle =1$%
. It is an insulator because of large $U$ and even antiferromagnetic
insulator at $T=0$ $K$. The effective exchange interaction (with $J=4t^{2}/U$%
) between spins is Heisenberg-like. By doping the system by holes (with the
hole concentration $\delta (<1)$), means that particles are taken out from
the system in which case there is on the average $\langle n_{i}\rangle
=1-\delta $ particles per lattice site. Above some (small) critical doping $%
\delta _{c}\sim 0.01$ the $AF$ order is destroyed and the system is strongly
correlated metal. For some \textit{optimal} doping $\delta _{op}(\sim 0.1)$
the system is metallic with the large Fermi surface and can exhibit even
high-T$_{c}$ superconductivity in the presence of the EPI, as it will
demonstrated below. The latter interaction and its interplay with strong
correlations is the central subject in the following sections.

\subsection{X-method for strongly correlated systems}

Since $U>>t$ one can put with good accuracy $U\rightarrow \infty $, i.e. the
system is in the \textit{strongly correlated regime }where the doubly
occupancy $n_{i}=2$ is excluded. One of the ways to cope with such strong
correlations is to introduce the (fermionic like) creation and annihilation
operators ($\hat{X}_{i}^{\sigma 0}$ and $\hat{X}_{i}^{0\sigma }=(\hat{X}%
_{i}^{\sigma 0})^{\dagger }$)
\begin{equation}
\hat{X}_{i}^{\sigma 0}=c_{i\sigma }^{\dagger }(1-n_{i,-\sigma }),
\label{Eq33}
\end{equation}
which respect the condition $n_{i,\sigma }+n_{i,-\sigma }\leq 1$ on each
lattice site. The latter means that there is no more than one electron
(hole) at a lattice site, i.e. the doubly occupancy is forbidden. The
bosonic like operators
\begin{equation}
\hat{X}_{i}^{\sigma _{1}\sigma _{2}}=\hat{X}_{i}^{\sigma _{1}0}\hat{X}%
_{i}^{0\sigma _{2}}  \label{Boson}
\end{equation}
(with \ $\sigma _{1}\neq \sigma _{2}$) create a spin fluctuation at the $i-th
$ site. Here, the spin projection parameter $\sigma =\uparrow ,\downarrow $
and $-\sigma =\downarrow ,\uparrow $ and the operator $\hat{X}_{i}^{\sigma
\sigma }$ has the meaning of the electron (hole) number on the i-th site. In
the following we shall use the convention that when $\hat{X}_{i}^{\sigma
\sigma }\mid 1\rangle =1\mid 1\rangle $ there is a fermionic particle
(''electron'') on the $i$-th site, while for $\hat{X}_{i}^{\sigma \sigma
}\mid 0\rangle =0\mid 0\rangle $ the site is empty, i.e. there is a hole on
it. It is useful to introduce the hole number operator
\begin{equation}
\hat{X}_{i}^{00}=\hat{X}_{i}^{0\sigma }\hat{X}_{i}^{\sigma 0}  \label{hole}
\end{equation}
at a given lattice site, i.e. if $\hat{X}_{i}^{00}\mid 0\rangle =1\mid
0\rangle $ the $i$-th site is empty - there is one hole on it, while for $%
\hat{X}_{i}^{00}\mid 1\rangle =0\mid 1\rangle $ it is occupied by an
''electron'' and there is no hole. They fulfill the non-canonical
commutation relations

\begin{equation}
\left[ \hat{X}_{i}^{\alpha \beta },\hat{X}_{j}^{\gamma \lambda }\right]
_{\pm }=\delta _{ij}\left[ \delta _{\gamma \beta }\hat{X}_{i}^{\alpha
\lambda }\pm \delta _{\alpha \lambda }\hat{X}_{i}^{\gamma \beta }\;\right] .
\label{Eq34}
\end{equation}
Here, $\alpha ,\beta ,\gamma ,\lambda =0,\sigma $ and $\delta _{ij}$ is the
Kronecker symbol. The (anti)commutation relations in Eq.(\ref{Eq34}) are
rather different from the canonical Fermi and Bose (anti)commutation
relations.

Since $U\rightarrow \infty $ the doubly occupancy is excluded, i.e. $\hat{n}%
_{i\uparrow }\hat{n}_{i\downarrow }\mid \psi \rangle (=\hat{X}_{i}^{22}\mid
\uparrow \downarrow \rangle )=0$, and by construction the $\hat{X}$
operators satisfy the \textit{local constraint} (the completeness relation)
\begin{equation}
\hat{C}_{X}(i)\equiv \hat{X}_{i}^{00}+\sum_{\sigma =1}^{N}\hat{X}%
_{i}^{\sigma \sigma }=1\;.  \label{Eq35}
\end{equation}
This condition tells us that at a given lattice site there is either one
hole ($\hat{X}_{i}^{00}\mid hole\rangle =1\mid hole\rangle $) ore one
electron ($\hat{X}_{i}^{\sigma \sigma }\mid elec\rangle =1\mid elec\rangle $%
). Note, if Eq.(\ref{Eq35}) is obeyed then both commutation and
anticommutation relations hold also in Eq.(\ref{Eq34}) at the same lattice
site, which is due to the projection properties of the Hubbard operators $%
\hat{X}^{\alpha \beta }\hat{X}^{\gamma \mu }=\delta _{\beta \gamma }\hat{X}%
^{\alpha \mu }$.

For further purposes, i.e. for the study of low-energy excitations in a
controllable way, the number of spin projections is \textit{generalized}%
\textbf{\ }to be $N$ , i.e. $\sigma =1,2,...N$. This way of generalization
was very useful in describing heavy fermion physics, where for some $Ce$
compounds $N$ means the number of projections of the total angular momentum,
for instance when $j=5/2$ then $N=2j+1=6$ ($N\gg 1$). For some $Yb$
compounds one has $j=7/2$, i.e. $N=2j+1=8$ ($N\gg 1$). By projecting out the
doubly occupied (high energy) states from the Hamiltonian in Eq.(\ref{Eq32})
one obtains the generalized $t-J$ model. The details of this derivation are
given in Appendix and here we give the final expression for the Hamiltonian
which excludes the doubly occupancy

\begin{equation}
\hat{H}=\hat{H}_{t}+\hat{H}_{J}=-\sum_{i,j,\sigma }t_{ij}\hat{X}_{i}^{\sigma
0}\hat{X}_{j}^{0\sigma }\;+\sum_{i,j,}J_{ij}(\mathbf{S}_{i}\cdot \mathbf{S}%
_{j}-\frac{1}{4}\hat{n}_{i}\hat{n}_{j}\;)+\hat{H}_{3}.  \label{Eq36}
\end{equation}
The first term describes the hopping of the ''electron'' by taking into
account that the doubly occupancy of sites are excluded. The second term
describes the Heisenberg-like exchange energy of almost localized
''electrons''. $\hat{H}_{3}$ contains three-sites hopping which is usually
omitted believing it is not important. For effects related to charge
fluctuation processes it is plausible to omit it, while for spin-fluctuation
processes it may be questionable approximation. The spin and number
operators can be expressed via the Hubbard operators \cite{Hubbard}
\begin{equation}
\mathbf{S}=\hat{X}_{i}^{\bar{\sigma}_{1}0}(\vec{\sigma})_{\bar{\sigma}_{1}%
\bar{\sigma}_{2}}\hat{X}_{i}^{0\bar{\sigma}_{2}};\ \hat{n}_{i}=\hat{X}_{i}^{%
\bar{\sigma}\bar{\sigma}}  \label{Spin}
\end{equation}
where summation over bar indices is understood.\

The basic idea behind the \textit{X-method}\textbf{\ } is that the Dyson's
equation for the electron Green's function can be effectively obtained by
introducing \textit{external potentials}\textbf{\ }(sources) $u^{\sigma
_{1}\sigma _{2}}(1)$. The source Hamiltonian $\hat{H}_{s}$ is used in the
form
\begin{equation}
\int \hat{H}_{s}\,d\tau =\int \sum_{\sigma _{1},\sigma _{2}}u^{\sigma
_{1}\sigma _{2}}(1)\hat{X}^{\sigma _{1}\sigma _{2}}(1)\,d1\equiv u^{\bar{%
\sigma}_{1}\bar{\sigma}_{2}}(\bar{1})\hat{X}^{\bar{\sigma}_{1}\bar{\sigma}%
_{2}}(\bar{1})\;,  \label{Eq37}
\end{equation}
where $1\equiv (\mathbf{l},\tau )$ and $\int (..)\,d1\equiv \int (..)\,d\tau
\sum_{\mathbf{l}}$ and $\tau $ is the Matsubara time. Here and in the
following, integration over bar variables($\bar{1},\bar{2}..$) and a
summation over bar spin variables($\bar{\sigma}..$) is understood. The
sources $u^{\sigma _{1}\sigma _{2}}(1)$ are useful in generating higher
correlation functions entering the self-energy. The electronic Green's
function is defined by \cite{KulicReview} ($\hat{T}$ is the time-ordering
operator)
\begin{equation}
G^{\sigma _{1}\sigma _{2}}(1,2)=\frac{-\langle \hat{T}\left( \hat{S}\hat{X}%
^{0\sigma _{1}}(1)\hat{X}^{\sigma _{2}0}(2)\right) \rangle }{\langle \hat{T}%
\hat{S}\rangle },  \label{Eq38}
\end{equation}
where $\hat{S}=\hat{T}\exp \{-\int \hat{H}_{s}(1)\,d1\}$ and the
corresponding Dyson's equation reads
\begin{equation}
\left[ G_{0,u}^{-1,\sigma _{1}\bar{\sigma}_{2}}(1,{\bar{2}})-\Sigma
_{G}^{\sigma _{1}\bar{\sigma}_{2}}(1,{\bar{2}})\right] G^{\bar{\sigma}%
_{2}\sigma _{2}}(\bar{2},2)=Q^{\sigma _{1}\sigma _{2}}(1)\delta (1-2),
\label{Eq39}
\end{equation}
\begin{equation}
G_{0,u}^{-1,\sigma _{1}\sigma _{2}}(1,2)=(-\frac{\partial }{\partial t_{1}}%
)\delta ^{\sigma _{1}\sigma _{2}}\delta (1-2)-u^{\sigma _{1}\sigma
_{2}}(1)\delta (1-2)  \label{Eq40}
\end{equation}
The so called Hubbard spectral weight is given by
\begin{equation}
Q^{\sigma _{1}\sigma _{2}}(1)=\delta ^{\sigma _{1}\sigma _{2}}\langle \hat{X}%
^{00}(1)\rangle +\langle \hat{X}^{\sigma _{1}\sigma _{2}}(1)\rangle .
\label{Hub-we}
\end{equation}
$\Sigma _{G}^{\sigma _{1}\sigma _{2}}(1,{2})$ is a functional of the Green's
function $G^{\sigma _{1}\sigma _{2}}(1,2)$. The latter describes the \textit{%
composite} (correlated) particle (in the language of the $SB$ theory it
describes the combined ''spinon + holon''). For further analysis it is
useful to introduce the quasiparticle Green's $g^{\sigma _{1}\sigma _{2}}$
(something analogous to the ''spinon'' Green's function in the $SB$
approach) and the vertex functions $\gamma _{\sigma _{3}\sigma _{4}}^{\sigma
_{1}\sigma _{2}}(1,2;3)$, respectively
\begin{equation}
g^{\sigma _{1}\sigma _{2}}(1,2)=G^{\sigma _{1}\overline{\sigma _{2}}%
}(1,2)Q^{-1,\overline{\sigma }_{2}\sigma _{2}}(2)  \label{Eq41}
\end{equation}
\begin{equation}
\gamma _{\sigma _{3}\sigma _{4}}^{\sigma _{1}\sigma _{2}}(1,2;3)=-\frac{%
\delta g^{-1,\sigma _{1}\sigma _{2}}(1,2)}{\delta u^{\sigma _{3}\sigma
_{4}}(3)},  \label{Eq42}
\end{equation}
respectively. Note, that $\gamma _{\sigma _{3}\sigma _{4}}^{\sigma
_{1}\sigma _{2}}(1,2;3)$ are the \textit{three-point vertex function}, which
also renormalizes the ionic EPI coupling constant - as we shall see below. $%
g^{\sigma _{1}\sigma _{2}}(1,2)$ is the solution of the equation
\begin{equation}
\left[ G_{0,u}^{-1,\sigma _{1}\bar{\sigma}_{2}}(1,{\bar{2}})-\Sigma
_{g}^{\sigma _{1}\bar{\sigma}_{2}}(1,{\bar{2}})\right] g^{\bar{\sigma}%
_{2}\sigma _{2}}(\bar{2},2)=\delta ^{\sigma _{1}\sigma _{2}}\delta (1-2),
\label{Eq43}
\end{equation}
where $\Sigma _{g}^{\sigma _{1}\sigma _{2}}(1,{2})$ depends on the
''quasiparticle'' Green's function $g^{\sigma _{1}\sigma _{2}}(1,2)$. Note,
that in Eq.(\ref{Eq43}) the Hubbard spectral weight $Q^{\sigma \sigma }$
disappears from the right hand side.

Since in the following we study \textit{nonmagnetic (paramagnetic)} normal
state one has $\Sigma _{g}^{\sigma \sigma }(1,2)\equiv \Sigma _{g}(1,2)$ for
$\sigma =1,..N$ , as well as $Q^{\sigma \sigma }($ $\equiv Q)$. In \cite
{Kulic1}, \cite{Kulic2}, \cite{Kulic3}, \cite{Kulic4} it is shown that in
that case $\Sigma _{g}$ can be expressed via two vertex functions - the
\textit{charge vertex}
\begin{equation}
\gamma _{c}(1,2;3)\equiv \gamma _{\bar{\sigma}\bar{\sigma}}^{\sigma \sigma
}(1,2;3)  \label{gamm-ch}
\end{equation}
and the \textit{spin vertex}
\begin{equation}
\gamma _{s}(1,2;3)\equiv \gamma _{\bar{\sigma}\sigma }^{\bar{\sigma}\sigma
}(1,2;3),  \label{gamm-sp}
\end{equation}
i.e. $\Sigma _{g}$ is given by
\[
\Sigma _{g}(1-2)=-\frac{t_{0}(1-2)}{N}Q(1)+\delta (1-2)\frac{J_{0}(1-%
\overline{1})}{N}<\hat{X}^{\sigma \sigma }(\overline{1})>-
\]
\[
-\frac{t_{0}(1-\overline{1})}{N}g(\overline{1}-\bar{2})\gamma _{c}(\overline{%
2},2;1)+
\]
\begin{equation}
\frac{+t_{2}(1,\overline{1},\overline{3})}{N}g(\bar{1}-\overline{2})\gamma
_{s}(\overline{2},2;\overline{3})+\Sigma _{Q}(1-2),  \label{Eq44}
\end{equation}
where $t_{2}(1,2,3)=\delta (1-2)t_{0}(1-3)-\delta (1-3)J_{0}(1-2)$. The
notation $t_{0}(1-2)$ (and $J_{0}(1-2)$) means $t_{0}(1-2)=t_{0,i_{1}j_{2}}%
\delta (\tau _{1}-\tau _{2})$. The first two terms in Eq.(\ref{Eq44})
represent an effective kinetic energy of quasiparticles in the lower Hubbard
band. As we shall see below they give rise to the band narrowing and to the
shift of the band center, respectively. The third and fourth terms describe
the kinematic and dynamic interaction of quasiparticles with charge and spin
fluctuations, respectively, while the very important term proportional to $%
\Sigma _{Q}(1,2)$ takes into account the counterflow of surrounding
quasiparticles which takes place in order to respect the local constraint
(absence of doubly occupancy). It reads
\[
\Sigma _{Q}(1,2)=\frac{t_{0}(1-\overline{1})}{N}\frac{g(\overline{1}-\bar{2})%
}{Q}[\frac{\delta Q(\bar{2})}{\delta u^{\bar{\sigma}\bar{\sigma}}(1)}-\frac{%
\delta Q^{\bar{\sigma}\sigma }(\bar{2})}{\delta u^{\bar{\sigma}\sigma }(1)}%
]g^{-1}(\bar{2}-2).
\]
$\Sigma $ depends on the vertex functions $\gamma _{c}(1,2;3)$ and $\gamma
_{s}(1,2;3)$ and it does not contain a small expansion parameter, like the
interaction energy in weakly interacting systems, because the hopping
parameter $t$ describes at the same time the kinetic energy and \textit{%
kinematic interaction} of quasiparticles. This means that there is no
controllable perturbation technique, due to the lack of small parameter.
There are various decoupling procedures and mean-field like techniques - the
path integral method, or $1/N$ expansion in various slave-boson approaches
\cite{KulicReview}.

What is the advantage of the \textit{X-method }expressed by Eq.(\ref{Eq41}-%
\ref{Eq44}). It turns out that it allows to formulate a controllable $1/N$
expansion for $\Sigma _{g}$ by including also \ the EPI \cite{Kulic1}, \cite
{Kulic2}, \cite{Kulic3}, \cite{Kulic4} - see below. For that purpose it is
necessary to generalize the \textit{local constraint}\textbf{\ }condition
\begin{equation}
\hat{C}_{X_{N}}(i)\equiv \hat{X}_{i}^{00}+\sum_{\sigma =1}^{N}\hat{X}%
_{i}^{\sigma \sigma }=\frac{N}{2},  \label{Eq45}
\end{equation}
where $N/2$ replace the unity in Eq.(\ref{Eq35}). It is apparent from Eq.(%
\ref{Eq35}) that for $N=2$ it coincides with Eq.(\ref{Eq35}) and has the
meaning that maximally half of all spin states at a given lattice site can
be occupied.

The spectral function $A(\mathbf{k},\omega )=-\mathrm{Im}G(\mathbf{k},\omega
)/\pi $ must obey the generalized Hubbard sum rule which respects the new
local constraint in Eq.(\ref{Eq45}).
\begin{equation}
\int d\omega A(\mathbf{k},\omega )=\frac{1+(N-1)\delta }{2}  \label{Eq46}
\end{equation}

The $N>2$ generalization of the local constraint allows us to make a \textit{%
controllable }$1/N$\textit{\ expansion} of the self-energy with respect to
the small quantity $1/N$ (when $N\gg 1$). Physically this procedure means
that we select a \textit{class of diagrams} in the self-energy and response
functions which might be important in some parameter regime. By careful
inspection of Eq.(\ref{Eq44}) one concludes that for large $N$ there is $1/N$
expansion for various quantities - for instance
\begin{equation}
g=g_{0}+\frac{g_{1}}{N}+...;Q=Nq_{0}+Q_{1}+...,  \label{Eq47}
\end{equation}
\begin{equation}
\Sigma _{g}=\Sigma _{0}+\frac{\Sigma _{1}}{N}+...;\gamma _{c}=\gamma _{c0}+%
\frac{\gamma _{c1}}{N}+...,  \label{Eq48}
\end{equation}
\[
\gamma _{s}(1,2;3)=N\delta (1-2)\delta (1-3)+\gamma _{s1}+...
\]
As a result of this expansion one obtains $\Sigma _{0}$ and $g_{0}$ in
\textit{leading }$O(1)$\textit{-order} - see details in \cite{KulicReview},
\cite{Kulic1}, \cite{Kulic2}, \cite{Kulic3}, \cite{Kulic4} ($\Sigma _{Q}$ in
Eq.(\ref{Eq44}) is of O(1/N) order)
\begin{equation}
g_{0}(\mathbf{k},\omega )\equiv \frac{G_{0}(\mathbf{k},\omega )}{Q_{0}}=%
\frac{1}{\omega -[\epsilon _{0}(\mathbf{k})-\mu ]},  \label{Eq49}
\end{equation}
where the \textit{quasiparticle energy} $\epsilon _{0}(\mathbf{k})=\epsilon
_{c}-q_{0}t_{0}(\mathbf{k})-\sum_{\mathbf{p}}J_{0}(\mathbf{k}+\mathbf{p}%
)n_{F}(\mathbf{p})$, and the \textit{level shift} $\epsilon _{c}=\sum_{%
\mathbf{p}}t_{0}(\mathbf{p})n_{F}(\mathbf{p})$. Here, $t_{0}(\mathbf{k})$
and $J_{0}(\mathbf{k})$ are Fourier transforms of $t_{0,ij}$ and $J_{0,ij}$,
respectively. For $t^{\prime }=0$ one has $t_{0}(\mathbf{k})=2t_{0}(\cos
k_{x}+\cos k_{y})$ and $J_{0}(\mathbf{k})=2J_{0}(\cos k_{x}+\cos k_{y})$. In
the equilibrium state ($u^{\sigma _{1}\sigma _{2}}\rightarrow 0$) and in
leading order one has
\begin{equation}
Q_{0}=<\hat{X}_{i}^{00}>=Nq_{0}=N\delta /2.  \label{Eq50}
\end{equation}
The chemical potential $\mu $ is obtained from the condition
\begin{equation}
1-\delta =2\sum_{\mathbf{p}}n_{F}(\mathbf{p}).  \label{chem-pot}
\end{equation}

Let us summarize the main results which were obtained by the \textit{X-method%
} in leading $O(1)$-order and compare these results with corresponding
results of the $SB$-method \cite{Kulic1}, \cite{Kulic2}, \cite{Kulic3}, \cite
{Kulic4}: ($\mathbf{1)}$ In the $O(1)$ order the Green's function $g_{0}(%
\mathbf{k},\omega )$ describes the \textit{coherent motion} of
quasiparticles whose contribution to the total spectral weight of the
Green's function $G_{0}(\mathbf{k},\omega )$ is $Q_{0}=N\delta /2$. Note, $%
G_{0}(\mathbf{k},\omega )=Q_{0}g_{0}(\mathbf{k},\omega )$ in leading order.
The dispersion of the quasiparticle energy is dominated by the exchange
parameter if $J_{0}>\delta t_{0}$. In the case when $J_{0}=0$ there is a
band narrowing by lowering the hole doping $\delta $, where the band width
is proportional to the hole concentration $\delta $, i.e. $W=z\cdot \delta
\cdot t_{0}$. ($\mathbf{2)}$ The X-method\textbf{\ }respects the local
constraint at each lattice site and in each step of calculations. ($\mathbf{%
3)}$ In the important paper \cite{Greco} - which is based on the theory
elaborated in \cite{Kulic1}, \cite{Kulic2}, \cite{Kulic3}, \cite{Kulic4}, it
is shown that in the superconducting state the anomalous self-energy (which
is of $O(1/N)$-order in the $1/N$ expansion) of the $X$- and $SB$-methods
\textit{differ} substantially. As a consequence, the $SB$-method \cite
{Kotliar2} predicts false superconductivity in the $t-J$ model (for $J=0$)
with large T$_{c}$ (due to the kinematical interaction), while the X-method
gives extremely small $T_{c}(\approx 0)$ \cite{Greco}. The reason\ for this
discrepancy between the two methods is that calculations done in the $SB$%
-method miss a class of compensating diagrams, which are on the other hand
taken automatically in the X-method. So, although the two approaches yield
some similar results in leading $O(1)$-order their implementation in the
next leading $O(1/N)$-order make that they are different. Note, that the $1/N
$ expansion in the X-method is well-defined and transparent. ($\mathbf{4)}$
By explicit calculation and comparison of the two methods in \cite{Kulic2}
it is shown that the renormalization of the EPI coupling constant is
different in the two approaches even in the large $N$-limit - see below; ($%
\mathbf{5}$) Very interesting behavior as a function of doping concentration
exhibits the optical conductivity $\sigma (\omega ,\mathbf{q}=0)\equiv
\sigma (\omega )$ which scales with the doping $\delta $. Note, the volume
below the Fermi surface in the case of strong correlations scales with $%
n=1-\delta $, like in the usual Fermi liquid. The above analysis clearly
demonstrate shows difference in response functions of strongly correlated
systems and the canonical Landau-Fermi liquid.

\subsection{Forward scattering peak in the charge vertex $\protect\gamma _{c}
$}

The three-point \textit{charge vertex}\textbf{\ }$\gamma _{c}(1,2;3)$ plays
important role in the renormalization all charge processes, such as the EPI,
Coulomb scattering and the scattering on non-magnetic impurities. It was
shown in \cite{Kulic1}, \cite{Kulic2}, \cite{Kulic3}, \cite{Kulic4}, \cite
{KulicReview} that $\gamma _{c}(1,2;3)$ can be calculated exactly in the
leading $O(1)$ order (and in all other orders) of the $t-J$ model. The
integral equation in the $O(1$) order reads

\[
\gamma (1,2;3)=\delta (1-2)\delta (1-3)+t(1-2)g_{0}(1,\bar{1})g_{0}(\bar{2}%
,1^{+})\gamma (\bar{1},\bar{2};3)
\]
\[
+\delta (1-2)t(1-\bar{1})g_{0}(\bar{1},\bar{2}))g_{0}(\bar{3},1)\gamma (\bar{%
2},\bar{3};3
\]
\begin{equation}
-J(1-2)g_{0}(1,\bar{1}))g_{0}(\bar{2},2)\gamma (\bar{1},\bar{2};3.
\label{Eq51}
\end{equation}
The analytical solution of Eq.(\ref{Eq51}) is given by \cite{KulicReview},
\cite{Kulic4}
\begin{equation}
\gamma _{c}(\mathbf{k},q)=1-\sum_{\alpha =1}^{6}\sum_{\beta =1}^{6}F_{\alpha
}(\mathbf{k})[\hat{1}+\hat{\chi}(q)]_{\alpha \beta }^{-1}\chi _{\beta 2}(q),
\label{Eq52}
\end{equation}
where
\begin{equation}
\chi _{\alpha \beta }(q)=\sum_{p}G_{\alpha }(p,q)F_{\beta }(\mathbf{p}),
\label{Eq53}
\end{equation}
\[
G_{\alpha }(p,q)=[1,t(\mathbf{p}+\mathbf{q}),\cos p_{x},\sin p_{x},\cos
p_{y},\sin p_{y}]\Pi (p,q),
\]
\[
F_{\alpha }(\mathbf{k})=[t(\mathbf{k}),1,2J_{0}\cos k_{x},2J_{0}\sin
k_{x},2J_{0}\cos k_{y},2J_{0}\sin k_{y}],
\]
where $\Pi (k,q)=-g(k)g(k+q)$ and $q=(\mathbf{q},iq_{n})$, $q_{n}=2\pi nT$, $%
p=(\mathbf{p},ip_{m})$, $p_{m}=\pi T(2m+1)$. Note, the frequency sum over $%
p_{m}$ in $\chi _{\alpha \beta }(q)$ in Eq.(\ref{Eq52}) involves only $\Pi $
and can easily be carried out $\sum_{p_{m}}\Pi (p,q)=[n_{F}(\xi _{\mathbf{q+p%
}})-n_{F}(\xi _{\mathbf{p}})]/[\xi _{\mathbf{p}}-\xi _{\mathbf{q+p}}-iq_{n}]$%
.

\begin{figure}[tbp]
\resizebox{.8 \textwidth}{!} {
\includegraphics*[width=10cm]{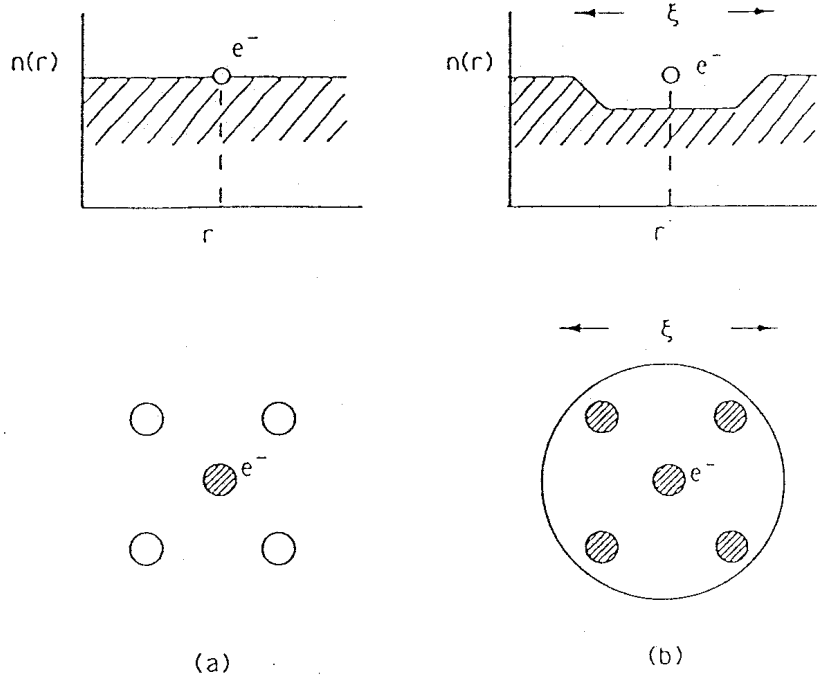}}
\caption{Schematic picture of electron correlation hole and the $E-P$
interaction for uncorrelated (weakly correlated) $(a)$ and strongly
correlated $(b)$ electron. In the case $(a)$ the electron does not perturb
the electronic density $n(r)$ and it interacts with the vibrations of a
single atom (shaded). In the case (b) an electron is accompanied by a large
correlation hole of size $\protect\xi \sim 1/\protect\delta $ ($\protect%
\delta $ is doping) and it will interact with atoms within this zone. From
\protect\cite{Grimaldi}.}
\label{CorrHoleFig}
\end{figure}

\begin{figure}[tbp]
\resizebox{0.8\textwidth}{!} {
\includegraphics*[width=8cm]{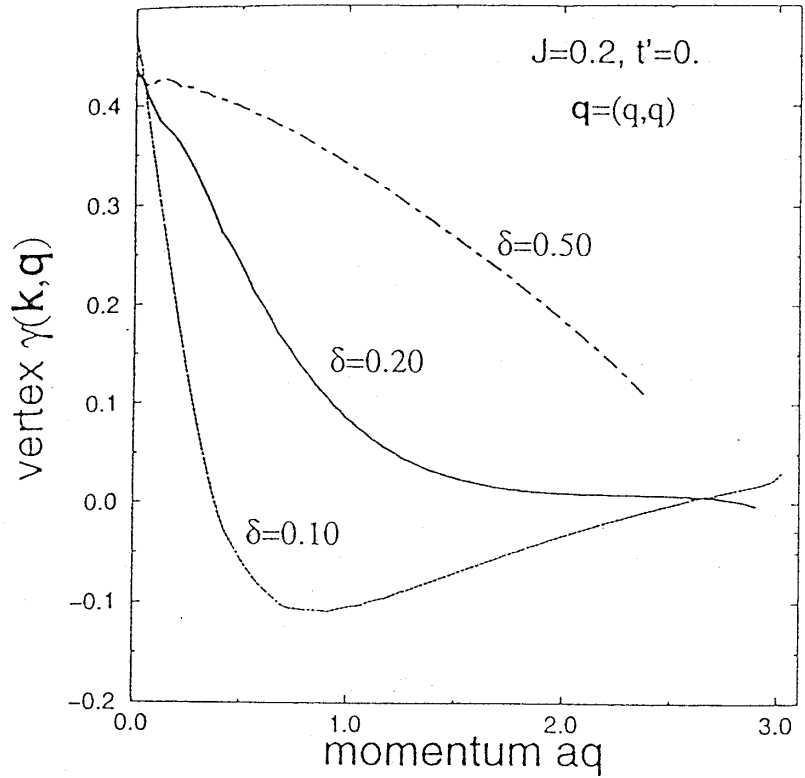}}
\caption{Zero-frequency vertex function $\protect\gamma (\mathbf{k}_{F},%
\mathbf{q})$ of the $t-J$ model as a function of the momentum $aq$ with $%
\mathbf{q}=(q,q)$ for three different doping $\protect\delta $ - from
\protect\cite{Kulic3}. }
\label{VertexFig}
\end{figure}

We stress that $\gamma _{c}(\mathbf{k},q)$ describes a specific screening of
the charge potential due to of strong correlations. In the presence of
perturbation (external source $u$) there is change of the band width, as
well as of the local chemical potential, which comes from the suppression of
doubly occupancy. The central result of the X-method is that for momenta $%
\mathbf{k}$\ laying at (and near) the Fermi surface $\gamma _{c0}(\mathbf{k},%
\mathbf{q},\omega =0)$ has very pronounced\textbf{\ }\textit{forward
scattering peak} at $\mathbf{q}=0$) at low doping concentration $\delta (\ll
1)$, while the backward scattering is substantially suppressed - see Fig.~%
\ref{VertexFig}. The latter means that charge fluctuations are strongly
suppressed (correlated) at small distances. Such a behavior of the vertex
function means that a quasiparticle moving in the strongly correlated medium
digs up a \textit{giant correlation hole} with the radius $\xi _{ch}\approx
a/\delta $, where $a$ is the lattice constant - see Fig.~\ref{CorrHoleFig}.

However, in the highly doped systems with $\delta >0.1$ - which corresponds
to the overdoped HTSC oxides, the effects of strong correlations is
progressively suppressed and the screening mechanism due to strong
correlations is less effective. We stress that when $J<t$ then the last term
in Eq.(\ref{Eq51}) for $\gamma _{c0}(\mathbf{k}_{F},\mathbf{q})$ is
unimportant. On the other hand both terms, the second (due to band
narrowing) and the third (due band shifting) one, are in conjunction
responsible for the development of the forward scattering peak at lower
doping. If we omit in Eq.(\ref{Eq51}) the band shifting term (the third one)
we get very weak forward scattering peak, while omitting the band narrowing
term (the first one) $\gamma _{c0}(\mathbf{k}_{F},\mathbf{q})$ is
practically constant in a broad region of $\mathbf{q}$.

Finally, since the real physics is characterized by $N=2$ one can put the
question - what is the \textit{reliability} of the results for the
quasiparticle properties obtained by the $1/N$ expansion (and $N\rightarrow
\infty $) within the X-method? First, the exact diagonalization of the
charge correlation function $N(\mathbf{k},\omega )$ in the $t-J$ model \cite
{Tohyama} shows clearly that the low-energy charge scattering processes at
large momenta $\mid \mathbf{k}\mid \approx 2k_{F}$ \ are strongly suppressed
compared to the small transferred momenta ($\mid \mathbf{k}\mid \ll 2k_{F}$%
). These calculations confirm unambiguously\textbf{\ }the results obtained
by the X-method in \cite{Kulic1}, \cite{Kulic2}, \cite{Kulic3} on the
suppression of the backward scattering in the vertex. Second, very recent
Monte Carlo (numerical) calculations in the Hubbard model with finite U \cite
{Scalapino-Hanke} show clear development of the forward scattering peak in $%
\gamma _{c}(\mathbf{k}_{F},\mathbf{q})$ by increasing U, thus confirming the
theoretical predictions in \cite{Kulic1}, \cite{Kulic2}, \cite{Kulic3}.

\section{Renormalization of the EPI by strong correlations}

In preceding Sections arguments we argued that because $\lambda _{tr}\ll
\lambda $ the standard Migdal-Eliashberg EPI theory must be corrected in
order to take into account screening properties of strongly correlated
system. This analysis is done in the framework of the X-method in a series
of papers \cite{Kulic1}, \cite{Kulic2}, \cite{Kulic3} which we briefly
discuss below. The renormalization of the EPI by strong correlations has
been studied also by the $SB$-method \cite{Auerbach}, \cite{Kim}, \cite
{Grilli}, \cite{Grilli2} by the $1/N$ expansion in the partition function,
or by using the mean-field approach \cite{Entel1} where also the non-Migdal
correction due to the EPI is considered. We stress that at present there are
no systematic and controllable calculations within the $SB$-method for the
EPI. From that point of view the $X$-method is of indispensable value.

\subsection{The forward scattering peak in the EPI}

The minimal model Hamiltonian for the HTSC oxides contains besides the $t-J $
terms also the EPI, i.e. $\hat{H}=\hat{H}_{tJ}+\hat{H}_{EP}$ where $\hat{H}%
_{EP}=\hat{H}_{EP}^{ion}+\hat{H}_{EP}^{cov}$
\[
\hat{H}=-\sum_{i,j,\sigma }t_{ij}\hat{X}_{i}^{\sigma 0}\hat{X}_{j}^{0\sigma
}\;+\sum_{i,j,}J_{ij}(\mathbf{S}_{i}\cdot \mathbf{S}_{j}-\frac{1}{4}%
n_{i}n_{j}\;)+
\]
\begin{equation}
+\sum_{i,\sigma }\epsilon _{a,i}^{0}\hat{X}_{i}^{\sigma \sigma }+\hat{H}%
_{ph}+\hat{H}_{EP}+\hat{V}_{LC},  \label{Eq54}
\end{equation}
where the ionic contribution to the EPI is
\begin{equation}
\hat{H}_{EP}^{ion}=\sum_{i,\sigma }\hat{\Phi}_{i}(\hat{X}_{i}^{\sigma \sigma
}-\langle \hat{X}_{i}^{\sigma \sigma }\rangle )+\hat{H}_{EP}^{cov}.
\label{Eq55}
\end{equation}
Here, $\hat{\Phi}_{i}$ (given by Eq.(\ref{Eq27})) describes the change of
the atomic energy $\epsilon _{a,i}^{0}$ due to the long-range Madelung
energy, where $L$ and $\kappa $ enumerate unit lattice vectors and atoms in
the unit cell, respectively. $Z_{L\kappa }$ is the effective charge of an
ion at the site $L\kappa $. Note, in Eq.(\ref{Eq55}) we do not assume small
displacement $\mathbf{\hat{u}}_{i}$ and the following analysis holds in
principle also for an \textit{anharmonic} EPI. The term proportional to $%
\langle \hat{X}_{i}^{\sigma \sigma }\rangle $ in Eq.(\ref{Eq55}) is
introduced in order to have $\langle \hat{\Phi}_{i}\rangle =0$ in the
equilibrium state. Note, that there is also covalent contribution to the $%
EPI $ in Eq.(\ref{Eq54}) due to the change of the hopping ($t$) and exchange
energy ($J$) by the ion displacements
\[
\hat{H}_{EP}^{cov}=-\sum_{i,j,\sigma }\frac{\partial t_{ij}}{\partial (%
\mathbf{R}_{i}^{0}-\mathbf{R}_{j}^{0})}(\mathbf{\hat{u}}_{i}-\mathbf{\hat{u}}%
_{j})\hat{X}_{i}^{\sigma 0}\hat{X}_{j}^{0\sigma }\;+
\]
\begin{equation}
+\sum_{i,j,}\frac{\partial J_{ij}}{\partial (\mathbf{R}_{i}^{0}-\mathbf{R}%
_{j}^{0})}(\mathbf{\hat{u}}_{i}-\mathbf{\hat{u}}_{j})\mathbf{S}_{i}\cdot
\mathbf{S}_{j}.  \label{Eq56}
\end{equation}
The treatment of the first term is similar to the Madelung term in Eq.(\ref
{Eq55}) although the equation \ for the \textit{four-point vertex function} $%
\gamma _{c}(1,2;3,4)$ is different than Eq.(\ref{Eq56}). We stress, that the
X-method has advantage also in the treatment of the covalent term, because
it peaks up straightforwardly all important contributions in $\gamma _{c}$,
due to strong correlations. On the other side the corresponding treatment by
the $SB$ method is complicated and not well defined, giving sometimes wrong
results. For instance, in \cite{Kim} several terms in the vertex equation
are omitted leading to incorrect results for the covalent EPI coupling. We
stress, that the covalent part contributes approximately $20-30$ $K$ to the
critical temperature in HTSC oxides as the band structure calculations in
\cite{Weber}, \cite{Weber-Mattheiss} have shown (partly discussed in Section
3.). Its renormalization by strong correlations will be studied elsewhere.
The second term in $\hat{H}_{EP}^{cov}$ is due to the change of the exchange
energy by phonon vibrations. Since it is second order with respect to $t_{ij}
$ it is much smaller than the first covalent term and accordingly
contributes very little to the total EPI.

After technically lengthy calculations, which are performed in \cite{Kulic1}%
, \cite{Kulic2} the expression for the ionic part of the EPI
(frequency-dependent) part of the self-energy reads
\begin{equation}
\Sigma _{EP}^{(dyn)}(1,2)=-V_{EP}(\bar{1}-\bar{2})\gamma _{c}(1,\bar{3};\bar{%
1})g_{0}(\bar{3}-\bar{4})\gamma _{c}(\bar{4},2;\bar{2}),  \label{Eq57}
\end{equation}
where analogously to Eq.(\ref{Eq29}) one has $V_{EP}(1-2)=\varepsilon
_{e}^{-1}(1-\bar{1})V_{EP}^{0}(\bar{1}-\bar{2})\varepsilon _{e}^{-1}(\bar{2}%
-2)$. The propagator of the bare EPI $V_{EP}^{0}(1-2)=-\langle T\hat{\Phi}(1)%
\hat{\Phi}(2)\rangle $ comprises in principle also the anharmonic
contribution. From Eq.(\ref{Eq57}) it is seen that in strongly correlated
systems the ionic part of the EPI is proportional to the square of the%
\textit{\ three-point charge vertex} $\gamma _{c}(1,2;3)$ (due to
correlations). The self-energy is given by
\begin{equation}
\Sigma _{EP}^{(dyn)}(\mathbf{k},\omega )=\int_{0}^{\infty }d\Omega \langle
\alpha ^{2}F(\mathbf{k,k}^{\prime },\Omega )\rangle _{\mathbf{k}^{\prime
}}R(\omega ,\Omega ),  \label{Eq58}
\end{equation}
where $R(\omega ,\Omega )$ is given in \cite{KulicReview}, \cite{Kulic1},
\cite{Kulic2}. The (momentum-dependent) Eliashberg spectral function is
defined by
\[
\alpha ^{2}F(\mathbf{k,k}^{\prime },\omega )=N_{sc}(0)\sum_{\nu }\mid
g_{eff}(\mathbf{k,k-k}^{\prime },\nu )\mid ^{2}\times
\]
\begin{equation}
\times \delta (\omega -\omega _{\nu }(\mathbf{k-k}^{\prime }))\gamma
_{c}^{2}(\mathbf{k,k-k}^{\prime }).  \label{Eq59}
\end{equation}
$n_{B}(\Omega )$ is the Bose distribution function and $\psi $ is di-gamma
function, while $g_{eff}(\mathbf{k,p},\nu )$ is the EPI coupling constant
for the $\nu $-the mode, where the renormalization by long-range Coulomb
interaction is included, i.e. $g_{eff}(\mathbf{k,p},\nu )=g(\mathbf{k,p},\nu
)/\varepsilon _{e}(\mathbf{p})$. $N_{sc}(0)$ is the density of states
renormalized by strong correlations where $N_{sc}(0)=N_{0}(0)/q_{0}$ and $%
q_{0}=\delta /2$ in the $t-t^{\prime }$ model ($J=0$). In the $t-J$ model $%
N_{sc}(0)$ has another form which does not diverge for $\delta \rightarrow 0$
but one has $N_{sc}(0)(\sim 1/J_{0})>N_{0}(0)$, where the bare density of
states $N_{0}(0)$ is calculated, for instance by the $LDA$ scheme.

\subsection{Pairing and transport EPI coupling constants}

Depending on the symmetry of the superconducting order parameter $\Delta (%
\mathbf{k},\omega )$ ($s-$, $d-wave$ pairing) various averages (over the
Fermi surface) of $\alpha ^{2}F(\mathbf{k,k}^{\prime },\omega )$ enter the
Eliashberg equations. Assuming that the superconducting order parameter
transforms according to the representation $\Gamma _{i}$ ($i=1,3,5$) of the
point group $C_{4v}$ of the square lattice (in the $CuO_{2}$ planes) the
appropriate symmetry-projected spectral function is given by

\[
\alpha ^{2}F_{i}(\mathbf{\tilde{k},\tilde{k}}^{\prime },\omega )=\frac{%
N_{sc}(0)}{8}\sum_{\nu ,j}\mid g_{eff}(\mathbf{\tilde{k},\tilde{k}-}T_{j}%
\mathbf{\tilde{k}}^{\prime },\nu )\mid ^{2}\times
\]
\begin{equation}
\times \delta (\omega -\omega _{\nu }(\mathbf{\tilde{k}-}T_{j}\mathbf{\tilde{%
k}}^{\prime }))\mid \gamma _{c}(\mathbf{\tilde{k},\tilde{k}-}T_{j}\mathbf{%
\tilde{k}}^{\prime })\mid ^{2}D_{i}(j).  \label{Eq60}
\end{equation}
$\ \mathbf{\tilde{k}}$ and $\mathbf{\tilde{k}}^{\prime }$ are momenta on the
Fermi line in the irreducible Brillouin zone which is $1/8$ of the total
Brillouin zone. $T_{j}$ , $j=1,..8,$ denotes the eight point-group
transformations forming the symmetry group of the square lattice. This group
has five irreducible representations which we distinguish by the label $%
i=1,2,...5$. In the following the representations $i=1$ \ and $i=3$ , which
correspond to the $s-$ and $d-wave$ symmetry of the full rotation group,
respectively, will be of importance. $D_{i}(j)$ is the representation matrix
of the $j-th$ transformation for the representation $i$. By assuming that
the superconducting order parameter $\Delta (\mathbf{k},\omega )$ does not
vary much in the irreducible Brillouin zone one can average over $\mathbf{%
\tilde{k}}$ and $\mathbf{\tilde{k}}^{\prime }$ in the Brillouin zone. For
each symmetry one obtains the corresponding spectral function $\alpha
^{2}F_{i}(\omega )$
\begin{equation}
\alpha ^{2}F_{i}(\omega )=\langle \langle \alpha ^{2}F_{i}(\mathbf{\tilde{k},%
\tilde{k}}^{\prime },\omega )\rangle _{\mathbf{\tilde{k}}}\rangle _{\mathbf{%
\tilde{k}}^{\prime }}  \label{Eq61}
\end{equation}
which (in the first approximation determines) the transition temperature for
the order parameter with the symmetry $\Gamma _{i}$. In the case $i=3$ the
electron-phonon spectral function $\alpha ^{2}F_{3}(\omega )$ in the $d$%
\textit{-channel} is responsible for $d-wave$ superconductivity represented
by the irreducible representation $\Gamma _{3}$ (or sometimes labelled as $%
B_{1g}$).

Performing similar calculations (as above) for the phonon-limited
resistivity one finds that the latter is related to the \textit{transport
spectral function} $\alpha ^{2}F_{tr}(\omega )$ which is given by
\begin{equation}
\alpha ^{2}F_{tr}(\omega )=\frac{\langle \langle \alpha ^{2}F(\mathbf{k,k}%
^{\prime },\omega )[\mathbf{v}(\mathbf{k})-\mathbf{v}(\mathbf{k}^{\prime
})]^{2}\rangle _{\mathbf{k}}\rangle _{\mathbf{k}^{\prime }}}{2\langle
\langle \mathbf{v}^{2}(\mathbf{k})\rangle _{\mathbf{k}}\rangle _{\mathbf{k}%
^{\prime }}}.  \label{Eq62}
\end{equation}
$\mathbf{v}(\mathbf{k})$ is the Fermi velocity. The effect of strong
correlations on the EPI was discussed in \cite{Kulic1} and more extensively
in \cite{Kulic2} within the model where the phonon frequencies $\omega (%
\mathbf{\tilde{k}-\tilde{k}}^{\prime })$ and $g_{eff}(\mathbf{k,p},\lambda )$
are weakly momentum dependent - due to the long-range screening (RPA). In
order to illustrate the effect of strong correlations on $\alpha
^{2}F_{i}(\omega )$ we consider the latter functions at zero frequency ($%
\omega =0$) which are then reduced to the (so called) ''enhancement''
functions
\begin{equation}
\Lambda _{i}=\frac{1}{8}\frac{N_{sc}(0)}{N_{0}(0)}\sum_{j=1}^{8}\langle
\langle \mid \gamma _{c}(\mathbf{\tilde{k},\tilde{k}-}T_{j}\mathbf{\tilde{k}}%
^{\prime })\mid ^{2}\rangle _{\mathbf{\tilde{k}}}\rangle _{\mathbf{\tilde{k}}%
^{\prime }}D_{i}(j)  \label{Eq63}
\end{equation}
Note, in the case $J=0$ one has $N_{sc}(0)/N_{0}(0)=q_{0}^{-1}$, where $%
q_{0} $ is related to the doping concentration, i.e. $q_{0}=\delta /2$.
Similarly, the correlation effects in the resistivity $\rho (T)(\sim $ $%
\Lambda _{tr})$ renormalize the transport coupling constant $\Lambda _{tr})$
\begin{equation}
\Lambda _{tr}=\frac{N_{sc}(0)}{N_{0}(0)}\frac{\langle \langle \mid \gamma
_{c}(\mathbf{\tilde{k},\tilde{k}-}T_{j}\mathbf{\tilde{k}}^{\prime })\mid
^{2}[\mathbf{v}(\mathbf{k})-\mathbf{v}(\mathbf{k}^{\prime })]^{2}\rangle _{%
\mathbf{k}}\rangle _{\mathbf{k}^{\prime }}}{2\langle \langle \mathbf{v}^{2}(%
\mathbf{k})\rangle _{\mathbf{k}}\rangle _{\mathbf{k}^{\prime }}}
\label{Eq64}
\end{equation}
Note, that for quasiparticles with the isotropic band the absence of
correlations implies that $\Lambda _{1}=\Lambda _{tr}=1$, $\Lambda _{i}=0$
for $i>1$.

\begin{figure}[tbp]
\resizebox{1.0\textwidth}{!} {
\includegraphics*[width=10cm]{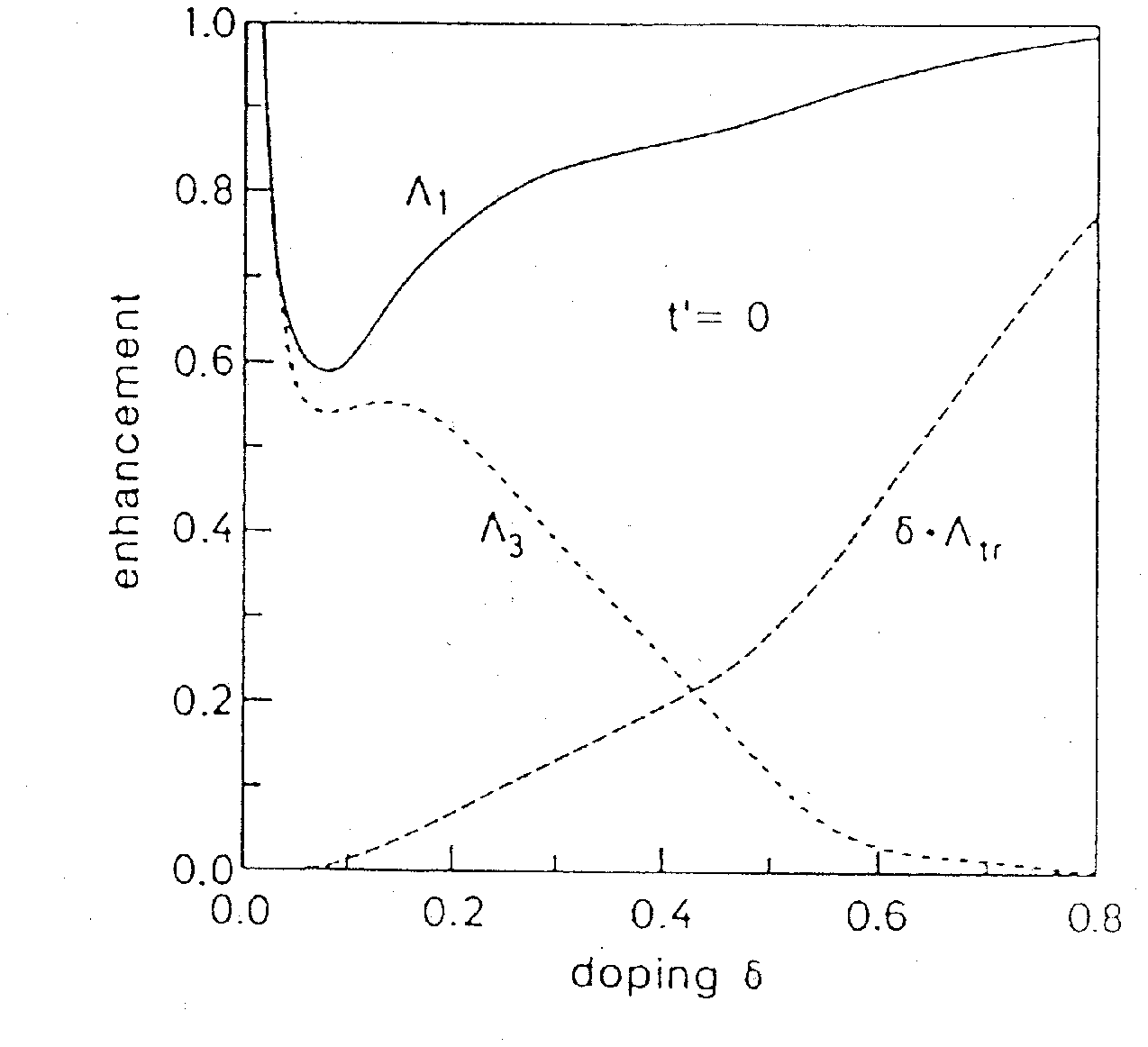}
\includegraphics*[width=10cm]{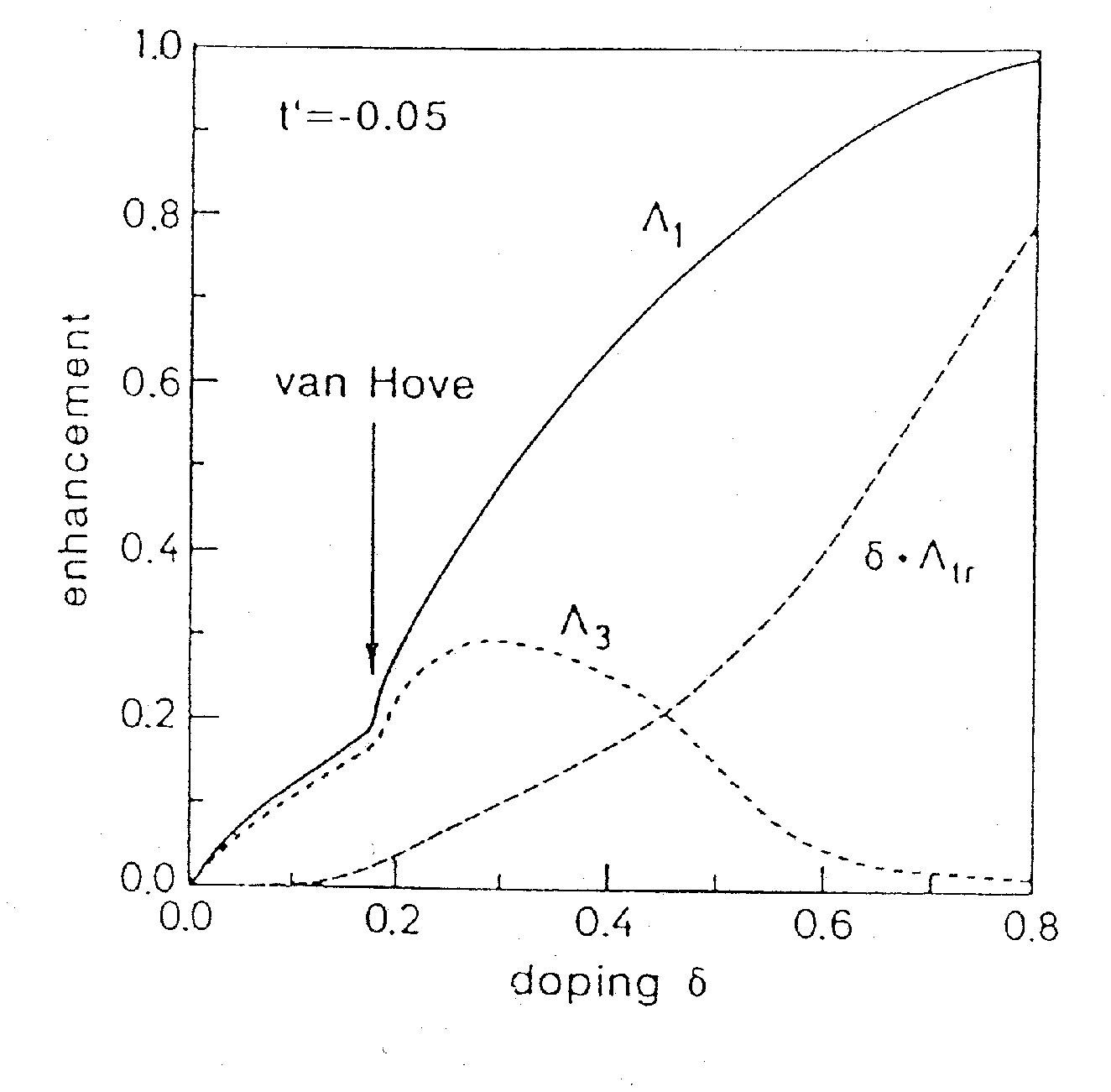}}
\caption{\textbf{(a)} - Enhancements $\Lambda _{1}$ and $\Lambda _{3}$ and $%
\protect\delta \cdot \Lambda _{tr}$ as a function of doping $\protect\delta $
for $\ t^{\prime }=0$ and $J=0$ - from \protect\cite{Kulic2}. \textbf{(b)}
Enhancements $\Lambda _{1}$ and $\Lambda _{3}$ and $\protect\delta \cdot
\Lambda _{tr}$ as a function of doping $\protect\delta $ for $t^{\prime
}=-0.05$ and $J=0$ - from \protect\cite{Kulic2}.}
\label{CouplingFig}
\end{figure}

The averages in $\Lambda _{1},\Lambda _{3}$ and $\Lambda _{tr}$ were
performed numerically in \cite{Kulic2} by using the realistic anisotropic
band dispersion\ in the $t-t^{\prime }-J$ model and the corresponding charge
vertex. The results for $\Lambda _{1},\Lambda _{3}$ and $\Lambda _{tr}$ are
shown in Fig.~\ref{CouplingFig} as functions of doping concentration in the $%
t$ and $t-t^{\prime }$ and $t-t^{\prime }-J$ models, respectively. The three
curves are multiplied with a common factor so that $\Lambda _{1}$ approaches
$1$ in the empty-band limit $\delta \rightarrow 1$, when strong correlations
are absent. Note, that T$_{c}$ in the weak coupling limit and in the $i-th$
channel scales like
\begin{equation}
T_{c}^{(i)}\approx \langle \omega \rangle \exp (-1/(\Lambda _{i}-\mu
_{i}^{\ast }),  \label{Tds}
\end{equation}
where $\mu _{i}^{\ast }$ is the Coulomb pseudopotential in the i-th channel
and $\langle \omega \rangle $ averaged phonon frequency.

Several interesting results, which are seen in Fig.~\ref{CouplingFig},
should be stressed.

\textit{First}, in the empty-band limit $\delta \rightarrow 1$ the $d-wave$
coupling constant $\Lambda _{3}$ is much smaller than the $s-wave$ coupling
constant $\Lambda _{1}$, i.e. $\Lambda _{3}\ll \Lambda _{1}$. Furthermore,
the totally symmetric function $\Lambda _{1}$ decreases with decreasing
doping.

\textit{Second}, in both models $\Lambda _{1}$ and $\Lambda _{3}$ meet each
other (note $\Lambda _{1}>$ $\Lambda _{3}$ for all $\delta $) at some small
doping $\delta \approx 0.1-0.2$ where $\Lambda _{1}\approx $ $\Lambda _{3}$
but still $\Lambda _{1}>$ $\Lambda _{3}$. By taking into account a residual
Coulomb repulsion of quasiparticles with $\mu _{d}^{\ast }\ll \mu _{s}^{\ast
}$ one gets that the $s-wave$ superconductivity (which is governed by the
coupling constant $\Lambda _{1}$) is suppressed, while the $d-wave$
superconductivity (governed by $\Lambda _{3}$) is only weakly affected. In
that case the $d-wave$ superconductivity due to the EPI becomes more stable
than the $s-wave$ superconductivity at sufficiently small doping $\delta $,
i.e. $T_{c}^{(d)}>T_{c}^{(s)}$. Experimentally, this occurs in underdoped,
optimally doped an overdoped HTSC oxides \cite{Tsuei-recent}. This
transition between $s$- \ and $d-wave$ superconductivity is triggered by
electronic correlations because in the calculations it is assumed that the
bare EPI coupling is momentum independent, i.e. the bare coupling constant
contains the $s-wave$ symmetry only.

\textit{Third}, the calculations are performed in the \textit{adiabatic
approximation}, where $\Lambda _{1}$ is less and $\Lambda _{tr}$ much more
suppressed by strong correlations. In the \textit{nonadiabatic regime} $%
\omega >\mathbf{p\cdot v}_{F}(\mathbf{p})$ i.e. for $\omega <\mathbf{p\cdot v%
}_{F}(\mathbf{p})$ the vertex function grows by decreasing $q$ finally
reaching $\gamma _{c}(\mathbf{k}_{F},\mathbf{p=0,}\omega \mathbf{)}=1$. Due
to the latter effect the enhancement function $\gamma _{c}^{2}(\mathbf{k}%
_{F},\mathbf{p,}\omega \mathbf{)}/q_{0}$ may be substantially larger
compared to the adiabatic. This means that different phonons will be
differently affected by strong correlations. For a given frequency the
coupling to phonons with momenta $p<$ $p_{c}=\omega /v_{F}$ will be \textit{%
enhanced,} while the coupling to those with $p>p_{c}=\omega /v_{F}$ is
substantially reduced due to the suppression of the backward scattering by
strong correlations.

\textit{Fourth}, the \textit{transport coupling constant} $\Lambda _{tr}$
(not properly normalized in Fig.~\ref{CouplingFig} - see correction in \cite
{Kulic3}) is reduced in the presence of strong correlations, especially for
lower doping where $\Lambda _{tr}<\Lambda /3$. This is very important result
because it resolves the experimental puzzle that $\lambda _{tr}$ (which
enters resistivity $\rho (T)\sim \lambda _{tr}T$) is much smaller than the
coupling constant $\lambda $ (which enters the self-energy $\Sigma $ and $%
T_{c}$), i.e. why $\lambda _{tr}<<\lambda $. The answer lies in strong
correlations which causes the forward scattering peak in charge scattering
processes - the \textit{FSP theory}.

As we already said, Monte Carlo (numerical) calculations if the
Hubbard model at finite $U$ - performed by Scalapino Group
\cite{Scalapino-Hanke}, show that the forward scattering peak in
the EPI coupling constant (and the charge vertex) develops by
increasing U. The latter effect is more pronounced at lower
doping. The similar (to Monte Carlo) results were obtained quite
recently in \cite{Pietro-SB} in the framework of the
R\"{u}ckenstein-Kotliar (four slave-boson) model. These numerical
results prove the \textit{correctness of the EPI theory} based on
the X-method.

We stress that contrary to the X-method, where the systematic
$1/N$ calculations of the EPI self-energy is uniquely done, this
is still a problem for the $SB$ (Barnes) method where the $1/N$
expansion of the partition function $Z(T,\mu )$ is usually
performed \cite{Kim}. The existing expression (in the literature)
for the vertex function in the SB method is different than that in
the X-method \cite{Kulic2}, \cite{Kulic2}. It seems that such a
not well-controlled procedure omits a class of diagrams giving
inadequate behavior of the coupling constant $\lambda $ as a
function of doping. Additionally, the vertex function in the $SB$
approach is peaked not at $q=0$ but at some finite $q_{\max }$,
where $q_{\max }\rightarrow 0$ only for doping $\delta \rightarrow
0$ - see \cite{Kulic5}.

\section{FSP theory and novel effects}

There are a number of effects which are predicted by the FSP theory. We have
already explained the effects of the forward scattering peak on the EPI. In
Section 2. the shift-puzzle in ARPES was also explained by the FSP theory
(model). We discuss briefly some other predictions of the FSP theory
containing parts not comprised in \cite{KulicReview}.

\subsection{Nonmagnetic impurities and robustness of d-wave pairing}

In the presence of strong correlations\ the impurity potential is also
renormalized and the effective potential in the Born approximation is given
by $u^{2}(\mathbf{q})=\gamma _{c}^{2}(k_{F},\mathbf{q})u_{0}^{2}(\mathbf{q})$%
, where $u_{0}(\mathbf{q})$ is the single impurity scattering potential in
the absence of strong correlations \cite{KuOudo}. Since the charge vertex $%
\gamma _{c}(\mathbf{p}_{F},\mathbf{q})$ is peaked at $\mathbf{q=0}$ the
potential $u(\mathbf{q})$ is also peaked at $\mathbf{q=0}$. This means that
the scattering amplitude contains not only the s-channel (as usually assumed
in studying impurity effects in HTSC oxides), but also the \textit{d-channel}%
, etc. Based on this property the FSP theory succeeded in explaining some
experimental facts, such as: $(\mathbf{i})$ the suppression of the residual
resistivity $\rho _{i}$ \cite{Kulic1}, \cite{Kulic2}. It is observed in the
optimally doped $YBCO$, where the resistivity $\rho (T)$ at $T=0$ $K$ has a
rather small value $<10$ $\mu \Omega cm$.; $(\mathbf{ii})$ the robustness of
$d-wave$ pairing \cite{KuOudo}. The previous theories \cite{Ferenbacher},
which assume $u(\mathbf{q})=const$, i.e. the s-wave scattering channel only,
predict that $T_{c}(\rho _{i,c})=0$ at much smaller residual resistivity $%
\rho _{i,c}^{(s)}\sim 50$ $\mu \Omega cm$, while the experimental range is $%
200$ $\mu \Omega cm<\rho _{i,c}^{\exp }<1500$ $\mu \Omega cm$ \cite{Tolpygo}%
. The latter experimental fact means that d-wave pairing in HTSC is much
more robust than the standard theory predicts, and it is one of the smoking
gun experiments in testing the concept of the forward scattering peak in the
charge scattering potential. It is worth of mentioning that in a number of
papers the pair-breaking effect of non-magnetic impurities in HTSC was
analyzed in terms of the impurity concentration $n_{i}$, i.e. the dependence
$T_{c}(n_{i})$. However, $n_{i}$ is not the parameter which governs this
pair-breaking effect. The more appropriate parameter for discussing the
robustness of d-wave pairing is the impurity scattering amplitude $\Gamma
(\theta ,\theta ^{\prime })$, which can be related to the measured residual
resistivity $\rho _{i}$ which leads to the dependence of $T_{c}(\rho _{i})$.
The robustness of d-wave pairing in HTSC can be revealed only by studying
the experimental curve $T_{c}(\rho _{i})$, what has been first recognized
experimentally in \cite{Tolpygo} and theoretically in \cite{Kulic1}, \cite
{KuOudo}.

The theory of the robustness of d-wave pairing in HTSC was elaborated first
in \cite{KuOudo}, where the FSP theory \cite{Kulic1}, \cite{Kulic2} is
applied to this problem. We shall not go into details - which are given in
\cite{KuOudo}, \cite{KulicReview}, but we give here a general formula for
the $T_{c}(\rho _{i})$ dependence in anisotropic (including unconventional)
superconductors, only. We assume that in anisotropic superconductivity the
superconducting order parameter has the form $\Delta (\theta )=\Delta
_{0}Y(\theta )$ and generally one has $\langle Y(\theta )\rangle \neq 0$ ($%
\langle Y^{\ast }(\theta )\rangle Y(\theta )\rangle =1$) where the momentum
dependent impurity scattering amplitude is $\Gamma (\theta ,\theta ^{\prime
})=\Gamma _{s}(\theta ,\theta ^{\prime })+\Gamma _{d}Y_{d}(\theta
)Y_{d}(\theta ^{\prime })+....$.
\begin{equation}
\ln \frac{T_{c}}{T_{c0}}=\Psi (\frac{1}{2})-\Psi (\frac{1}{2}+(1-\beta
)x)-\langle Y(\theta )\rangle ^{2}[\Psi (\frac{1}{2})-\Psi (x+\frac{1}{2})].
\label{Eq65}
\end{equation}

\begin{figure}[tbp]
\resizebox{.9\textwidth}{!} {
\includegraphics*[width=10cm]{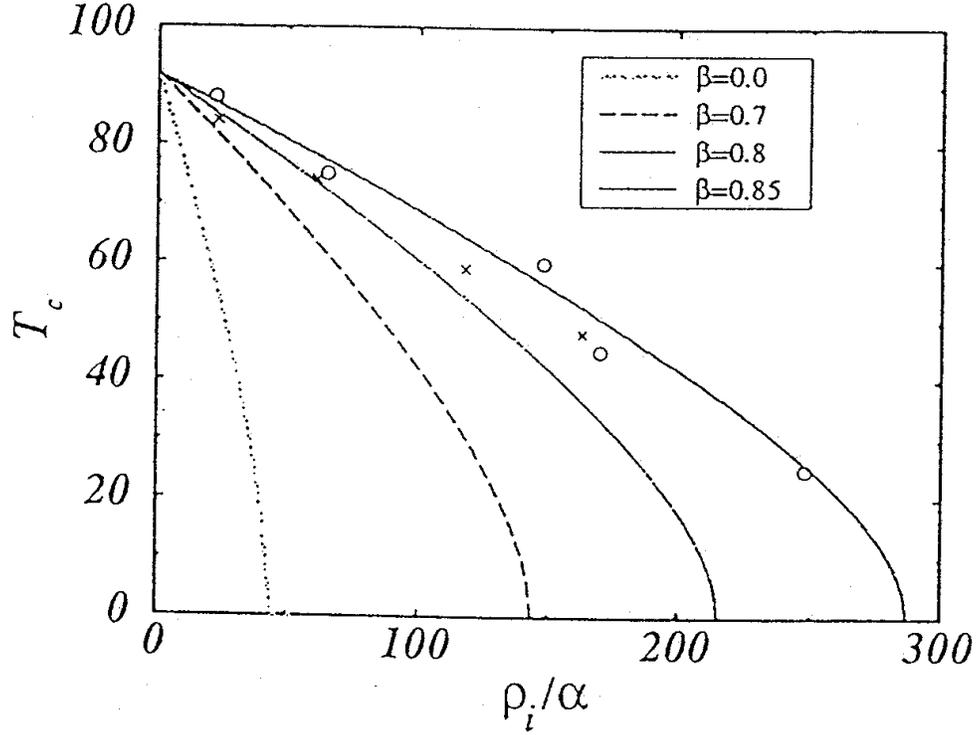}}
\caption{The critical temperature $T_{c}[K]$ of d-wave superconductor as a
function of the experimental parameter $\protect\rho _{i}/\protect\alpha
_{c}[K]$, $\protect\rho _{i} $ is the residual resistivity and $\protect%
\alpha $ is defined in the text. The case $\protect\beta =0$ corresponds to
the prediction of the standard d-wave theory with isotropic scattering
\protect\cite{Ferenbacher}. The experimental data \protect\cite{Tolpygo} are
given by crosses - $YBa_{2}(Cu_{1-x}Zn_{x})_{3}O_{7-\protect\delta }$, and
circles - $Y_{1-y}\Pr_{y}Ba_{2}Cu_{3}O_{7-\protect\delta }$ - from
\protect\cite{KuOudo}.}
\label{ImpurityFig}
\end{figure}

Here, $x=\Gamma _{s}/4\pi T_{c}$, $\beta =\Gamma _{d}/\Gamma _{s}$ and $%
\langle Y(\theta )\rangle $ means an averaging over the Fermi surface. Note,
Eq.(\ref{Eq65}) holds independently of the scattering strength $\Gamma _{s}$,%
$\Gamma _{d}$, i.e. it holds in the Born as well as in the unitary limit.
The residual resistivity $\rho _{i}$ can be related to the transport
scattering rate by $\rho _{i}=4\pi \Gamma _{tr}/\omega _{pl}^{2}$, while the
s-wave amplitude is related to $\Gamma _{tr}$ by $\Gamma _{s}=p\Gamma _{tr}$%
. The parameter $p>1$ can be obtained from the microscopic model (for
instance in the t-J model $p\approx 2-3$) or can be treated as a fitting
parameter - see more in \cite{KulicReview}. In the case of an unconventional
pairing one has $\langle Y(\theta )\rangle =0$ and the last term drops. For
the s-scattering only ($\Gamma _{d}=0$) one has $\beta =0$ and $T_{c}(\rho
_{i})$ should be suppressed very strongly contrary to the experimental
results \cite{Tolpygo} - see Fig.~\ref{ImpurityFig}.

\begin{figure}[tbp]
\resizebox{.7\textwidth}{!} {
\includegraphics*[width=6cm]{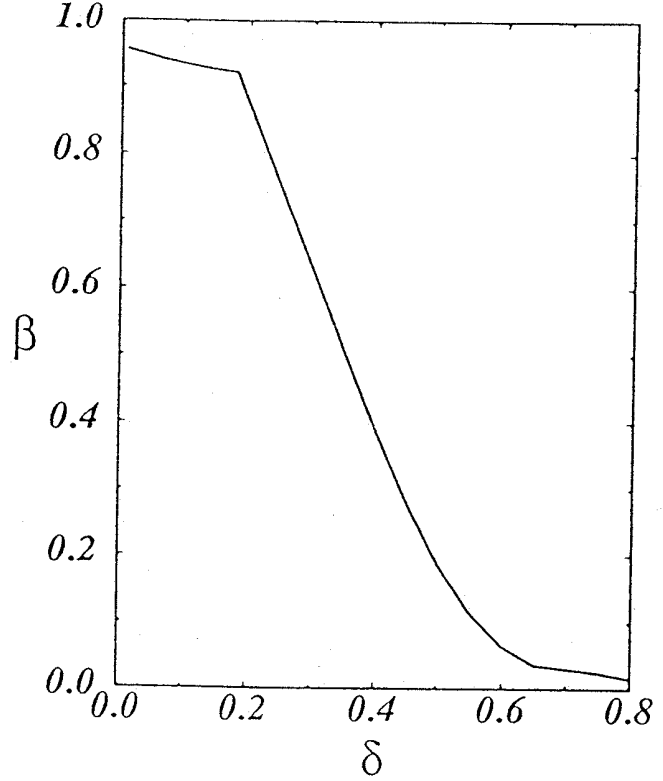}}
\caption{The anisotropy scattering parameter $\protect\beta$ as a function
of doping $\protect\delta$ in the t-J model. From \protect\cite{KuOudo}.}
\label{BetaFig}
\end{figure}

The FSP theory of the impurity scattering in the t-J model\ \cite{KuOudo}
gives that the s-channel and d-channel almost equally contribute to the
impurity scattering amplitude, since $\beta \approx 0.75-0.85$ for doping $%
\delta \approx 0.1-0.2$. The dependence of $\beta (\delta )$ \ is calculated
for the t-J model - see Fig.~\ref{BetaFig}.

Since the d-channel in scattering is not detrimental for d-wave pairing the
FSP theory predicts that $T_{c}(\rho _{i})$ vanishes at much larger $\rho
_{i,c}$, i.e. $\rho _{i,c}^{(FSP)}>>\rho _{i,c}^{(s)}$, what is in the good
agreement with experiments - as shown in Fig.~\ref{ImpurityFig}.

\subsection{Transport properties and superconductivity}

\begin{figure}[tbp]
\resizebox{1.0\textwidth}{!} {
\includegraphics*[width=10cm]{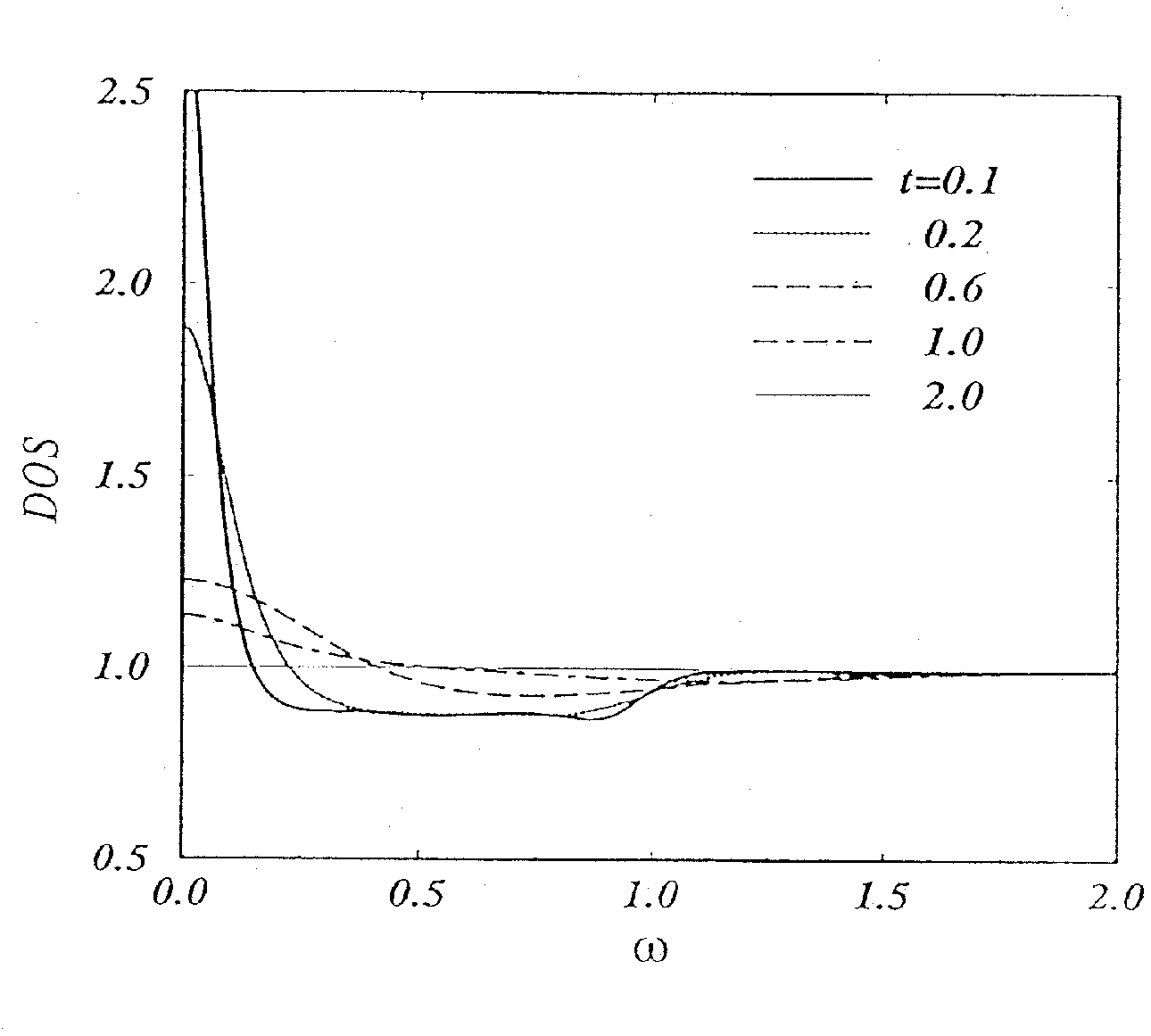}}
\caption{The density of states $N(\protect\omega )$ in the FSP model for the
EPI. with the dimensionless coupling $l(=V_{EP}/\protect\pi \Omega )=0.1$
for various$t(=\protect\pi T/\Omega )$. From \protect\cite{DDKuOudo2}.}
\label{NomegaFSPFig}
\end{figure}

The EPI was studied in the past in the \textit{extreme limit} of
the
forward scattering peak in the Einstein model with the phonon frequency $%
\Omega$ \cite{DDKuOudo2}, where in leading order the spectral function is
singular, i.e. $\alpha ^{2}F(\mathbf{k},\mathbf{k}^{\prime },\omega )\sim
\delta (\mathbf{k}-\mathbf{k}^{\prime })\delta (\omega -\Omega )$. Numerical
calculations of the Eliashberg equations in the normal state \cite{DDKuOudo2}
give very interesting behavior of the \textit{density of states} $N(\omega )$%
, where a strong renormalization of $N(\omega )$ is present, but which is
absent in the standard theory of the isotropic EPI. \textit{First}, $%
N(\omega =0)>N_{bare}(\omega =0)$, where $N_{bare}(\omega =0)$ is the
density of states in the absence of the EPI - see Fig.~\ref{NomegaFSPFig}.

\begin{figure}[tbp]
\resizebox{.9\textwidth}{!} {
\includegraphics*[width=10cm]{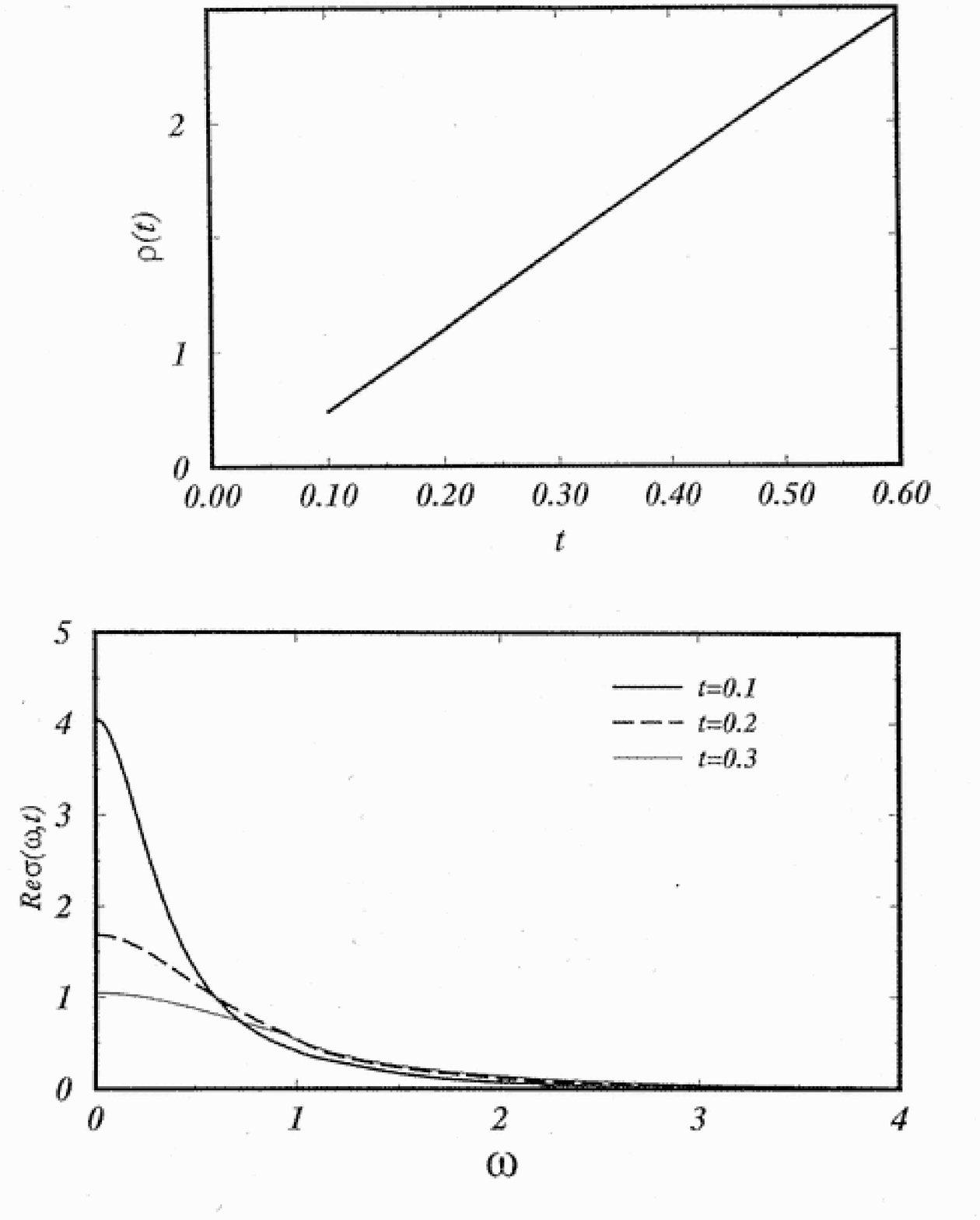}}
\caption{$\protect\rho(T)$ (upper part) and $\protect\sigma _{1}(\protect%
\omega )$ in the FSP model for the EPI. with the dimensionless coupling $%
l(=V_{EP}/\protect\pi \Omega)=0.1$ for various$t(=\protect\pi T/\Omega)$.
From \protect\cite{DDKuOudo2}.}
\label{TransportFSPFig}
\end{figure}

There is a "pseudogap"-like feature in the region $(\Omega /5)<\omega \leq
\Omega $ where $N(\omega )<N_{bare}(\omega )$. The "pseudogap" feature
disappears at $T$ comparable with the phonon energy $\Omega $. Note, that
the usual isotropic EPI does not renormalize the density of states in the
normal state, i.e. $N(\omega )=N_{bare}(\omega )$. As a consequence of the
pseudogap behavior of $N(\omega )$ the transport properties are very
peculiar. For instance, the resistivity $\rho (T)$ is linear in $T$ starting
at very low temperatures, i.e. $\rho (T)\sim T$ for $(\Omega /30)\leq T$ and
extends up to several $\Omega $ - as it is seen in Fig.~\ref{TransportFSPFig}%
. The dynamical conductivity $\sigma _{1}(\omega )$ shows the (extended)
Drude-like behavior with the Drude width $\Gamma _{tr}\sim T$, for $\omega
<T $ - see Fig.~\ref{TransportFSPFig}. The above numbered properties are in
a qualitative agreement with experimental results in HTSC oxides, as
discussed in Section 2.

In this extreme forward scattering peak limit one can calculate T$_{c}$. In
leading order w.r.t. ($\Omega /T_{c}<<1$) one has

\begin{equation}
T_{c0}\approx N(0)V_{EP}=\lambda N(0)/4,  \label{Tco-FSP}
\end{equation}
where $\lambda =N(0)V_{EP}$. In that case the maximal superconducting gap is
given by $\Delta _{0}=2T_{c}$ which is reached on the Fermi surface, while
away from it the gap decreases, i.e.
\begin{equation}
\Delta _{k}=\Delta _{0}\sqrt{1-(\xi _{k}/\Delta _{0})^{2}}.  \label{FSP-gap}
\end{equation}
The expression for T$_{c}$ tells us that it can be large even for $\lambda
<0.1$, since in HTSC oxides the bare density of states is $N_{bare}(0)\sim
1states/eV$. It is apparent that in this order there is no isotope effect,
i.e. $\alpha =0$. We stress that such an extreme limit is never realized in
nature, but for the self-energy it is a good starting point, since the
effects of the finite width ($k_{c}$) of $\alpha ^{2}F(\mathbf{k},\mathbf{k}%
^{\prime },\omega )$, whenever $k_{c}\ll k_{F}$, change mainly the
quantitative picture - see \cite{DDKuOudo2}. In case when $k_{c}v_{F}\ll
\Omega $ the reduction of T$_{c}$ is given by

\begin{equation}
T_{c}=T_{c0}(1-\frac{7\zeta (3)k_{c}v_{F}}{4\pi ^{2}T_{c0}}).
\label{Tc-corr}
\end{equation}

Very interesting calculations in the more realistic FSP model with the
finite width $k_{c}$, but $k_{c}\ll k_{F}$, were done in \cite{VarelogFEP},
where the FSP theory for the EPI and the SFI theory (based on
spin-fluctuation mechanism of pairing) were compared. For instance, the FSP
theory can explain the appreciable increase of the anisotropy ratio $R\equiv
\Delta (\pi ,0)/\Delta (\pi /2,\pi /2)$ when $T\rightarrow T_{c}$, while the
SFI is unable. Furthermore, the FSP theory of the EPI can explain the
pronounced orthorhombic ($a\neq b$) effect in $YBCO$ on the gap ratio $%
\Delta _{a}/\Delta _{b}$, penetration depth anisotropy $\lambda
_{a}^{2}/\lambda _{b}^{2}$ and supercurrent ratio in the c-axis $Pb-YBCO$
junction. On the other hand, the SFI theory is ineffective, since it
predicts at least one order of magnitude smaller effects - \cite{VarelogFEP}%
, \cite{KulicReview}.

\subsection{Nonadiabatic corrections of $T_{c}$}

\begin{figure}[tbp]
\resizebox{.9\textwidth}{!} {
\includegraphics*[width=10cm]{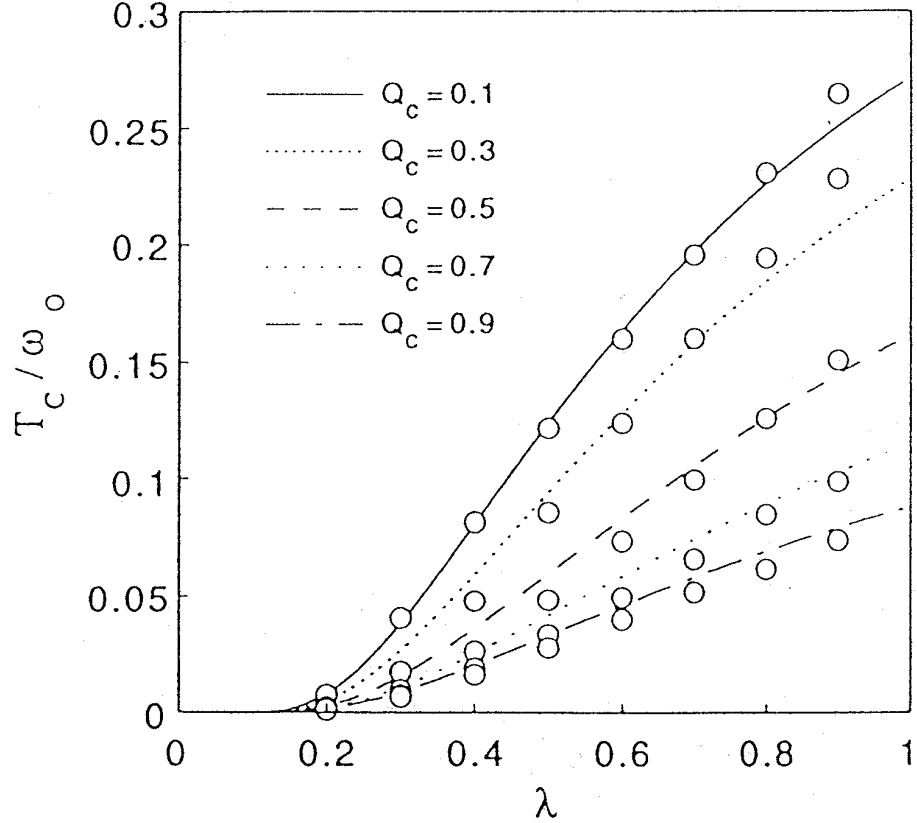}}
\caption{The approximative analytic (solid lines) and numerical (circles)
solution of $T_{c}(\protect\lambda )$ in the first nonadiabatic
approximation for various cutoff $Q_{c}$ and for $m=(\protect\omega
_{D}/E_{F})=0.2$. From \protect\cite{Grimaldi}.}
\label{NonMigdalFig}
\end{figure}

HTSC oxides are characterized not only by strong correlations but also by
relatively small Fermi energy $E_{F}$, which is not much larger than the
characteristic (maximal) phonon frequency $\omega _{ph}^{\max }$, i.e. $%
E_{F}\simeq 0.1-0.3$ $eV$, $\omega _{ph}^{\max }\simeq 80$ $meV$. The
situation is even more pronounced in \textit{fullerene compounds} $%
A_{3}C_{60}$, with $T_{c}=20-35$ $K$, where $E_{F}\simeq 0.2$ $eV$ and $%
\omega _{ph}^{\max }\simeq 0.16$ $eV$. \ This fact implies a possible
breakdown of the Migdal's theorem \cite{Migdal}, \cite{AllenMit}, which
asserts that the relevant vertex corrections due to the $E-P$ interaction
are small if $(\omega _{D}/E_{F})\ll 1$. In that respect a comparison of the
intercalated graphite $KC_{8}$ and the fullerene $A_{3}C_{60}$ compounds,
given in \cite{Grimaldi}, is very instructive, because both compounds have a
number of similar properties. However, the main difference in these systems
lies in the ratio $\omega _{D}/E_{F}$, since $(\omega _{D}/E_{F})\ll 1$ in $%
KC_{8}$, while it is rather large $(\omega _{D}/E_{F})\sim 1$ in $%
A_{3}C_{60} $. Due to the appreciable magnitude of $\omega _{D}/E_{F}$ in
the fullerene compounds and in HTSC oxides it is necessary to correct the
Migdal-Eliashberg theory by vertex corrections due to the EPI. It is
well-known that these vertex corrections lower T$_{c}$ in systems with
isotropic EPI. However, the vertex corrections in systems with the forward
scattering peak and with the cut-off $q_{c}<<k_{F}$ the increases of T$_{c}$
appreciable. The calculations by the Pietronero group \cite{Grimaldi} gave
two important results: (\textbf{1}) there is a drastic increase of T$_{c}$
by lowering $Q_{c}=q_{c}/2k_{F}$, for instance $T_{c}(Q_{c}=0.1)\approx
4T_{c}(Q_{c}=1)$; (\textbf{2}) Even small values of $\lambda <1$ can give
large T$_{c}$. The latter results open a new possibility in reaching high T$%
_{c}$ in systems with appreciable ratio $\omega _{D}/E_{F}$ and with the
forward scattering peak. The difference between the Migdal-Eliashberg and
non-Migdal theories can be explained qualitatively in the framework of an
approximative McMillan formula for T$_{c}$ (for not too large $\lambda $)
which reads
\begin{equation}
T_{c}\approx \langle \omega \rangle e^{-1/[\tilde{\lambda}-\mu ^{\ast }]}.
\label{Eq66}
\end{equation}
The \textit{Migdal-Eliashberg theory} predicts
\begin{equation}
\tilde{\lambda}\approx \frac{\lambda }{1+\lambda },  \label{Eq67}
\end{equation}
while the non-Migdal theory \cite{Grimaldi} gives

\begin{equation}
\tilde{\lambda}\approx \lambda (1+\lambda ).  \label{Eq68}
\end{equation}
For instance $T_{c}\sim 100$ $K$ in HTSC oxides can be explained by the
Migdal-Eliashberg theory for $\lambda \sim 2$, while in the non-Migdal
theory much smaller coupling constant is needed, i.e. $\lambda \sim 0.5$ as
it is seen in Fig.~\ref{NonMigdalFig}. The pioneering approach done in \cite
{Grimaldi} deserves more attention in the future.

\subsection{Pseudogap behavior in the FSP model for the EPI}

In this review we did not discuss a number of interesting topics
such as the possible existence of stripes, the properties of the
pseudogap state, etc.. There is a believe that the understanding
of these properties might give some hints for pairing mechanism in
HTSC oxides. Especially, the pseudogap (PG) problem is a very
intriguing one and is not surprising at all, that a number of
theoretical approaches were proposed for explaining the PG. We are
not going to discuss it but only quote some of them. The
\textit{first} one is based on the assumption that the PG phase
represents pre-formed pairs \cite{Randeria}, and the true critical
temperature T$_{c}$ is smaller than the mean-field one
T$_{c}^{MF}$. In the region T$_{c}$<T<T$_{c}^{MF}$ pre-formed
pairs exist giving rise to the dip in the density of states
$N(\omega )$. This approach is physically plausible having in mind
that HTSC oxides are characterized by the short coherence length
and quasi-two dimensionality. From the experimental side there are
some supports. For instance, the specific heat measurements
\cite{Junod} point to the non-mean field character of the
superconducting phase transition, particularly for the underdoped
systems. As we have already mentioned in the Introduction the
ARPES measurements show that the PG has a d-wave like form
\begin{equation}
\Delta _{pg}(\mathbf{k})\approx \Delta _{pg,0}(\cos k_{x}-\cos k_{y})
\label{og}
\end{equation}
(like the superconducting gap) and $\Delta _{pg,0}$ increases by lowering
doping. The \textit{second} approach assumes that the PG is due to a
competing order, but usually without the long-range order, such as due to
''spin-density wave'' alias for strong antiferromagnetic fluctuations \cite
{Sadovskii}, \cite{Gabovich}. There are other approaches which are based on
the RVB and orbital current model, d-wave order, etc., but we shall not
discussed it here.

However, the FSP theory, which predicts the long-range force due to the
renormalization of the EPI by strong correlations, opens an additional
possibility for the PG. As we discussed in Section 6.2, due to the forward
scattering peak the critical temperature has a non-BCS dependence, i.e. $%
T_{c}^{MF}=V_{EP}/4$. However, this is the mean-field value, which is
inevitably reduced by phase and internal Cooper pair fluctuations present in
systems with long-range attractive forces, i.e. with the forward scattering
peak.

The interesting problem of fluctuations in systems with long-range
attractive forces was recently studied in \cite{Yang}. It was shown there,
that such a long-ranged superconductor exhibits a class of fluctuations in
which the internal structure of the Cooper pair is soft. This leads to a
''pseudogap'' behavior in which the actual transition temperature T$_{c}$ is
greatly depressed from its mean-field value T$_{c}^{MF}$. We stress that
these fluctuations are not the standard phase fluctuations in
superconductors. Since the problem is very interesting and deserve much more
attention in the future we discuss it here briefly. In the following the
weak coupling limit is assumed, where the pairing Hamiltonian has the form

\[
H=\sum_{\sigma }\int d\mathbf{x}\psi _{\sigma }^{\dagger }(\mathbf{x})\xi (%
\hat{\mathbf{p}})\psi _{\sigma }(\mathbf{x})
\]
\begin{equation}
-\int d\mathbf{x}d\mathbf{\mathbf{x}}^{\prime }V(\mathbf{x-\mathbf{x}}%
^{\prime })\psi _{\uparrow }^{\dagger }(\mathbf{x})\psi _{\downarrow
}^{\dagger }(\mathbf{x}^{\prime })\psi _{\downarrow }(\mathbf{x}^{\prime
})\psi _{\uparrow }(\mathbf{x}).  \label{H-long}
\end{equation}

In the MFA the order parameter $\Delta (\mathbf{x},\mathbf{\mathbf{x}}%
^{\prime })$ is given by
\begin{equation}
\Delta (\mathbf{x},\mathbf{\mathbf{x}}^{\prime })=V(\mathbf{x-\mathbf{x}}%
^{\prime })\langle \mathbf{\psi }_{\downarrow }(\mathbf{x}^{\prime })\mathbf{%
\psi }_{\uparrow }(\mathbf{x})\rangle .  \label{O-P}
\end{equation}
$\Delta (\mathbf{x},\mathbf{\mathbf{x}}^{\prime })$ depends in fact on the
internal coordinate $\mathbf{r}=\mathbf{x}-\mathbf{\mathbf{x}}^{\prime }$
and the center of mass $\mathbf{R}=(\mathbf{x}+\mathbf{\mathbf{x}}^{\prime }%
\mathbf{\mathbf{)/2}}$, i.e. $\Delta (\mathbf{x},\mathbf{\mathbf{x}}^{\prime
})=\Delta (\mathbf{r},\mathbf{\mathbf{R}})$. In usual superconductors with
the short-range pairing potential $V_{sr}(\mathbf{x-\mathbf{x}}^{\prime
})\approx V_{0}\delta (\mathbf{x-\mathbf{x}}^{\prime })$ one has $\Delta (%
\mathbf{r},\mathbf{\mathbf{R}})=\Delta (\mathbf{\mathbf{R}})$ and therefore
there are practically the spatial ($\mathbf{\mathbf{R}}$) fluctuations of
the order parameter, only. In the case of long-range pairing potential there
are additional fluctuations of the internal ($\mathbf{r}$) degrees of
freedom. In the following we sketch the analysis given in \cite{Yang}.

When the range of the pairing potential is large, i.e. $r_{c}>\xi $ (the
superconducting coherence length), fluctuations of the internal Cooper
wave-function are important since they give rise to a tremendous reduction
of the mean-field quantities. In order to make the physics of internal
wave-function fluctuations we study much simpler Hamiltonian the so called
reduced BCS Hamiltonian,
\begin{equation}
H=\sum_{\mathbf{k}\sigma }\xi _{\mathbf{k}}c_{\mathbf{k}\sigma }^{\dagger
}c_{\mathbf{k}\sigma }-\sum_{\mathbf{k},\mathbf{k}^{\prime }}V_{\mathbf{k}-%
\mathbf{k}^{\prime }}c_{\mathbf{k}\uparrow }^{\dagger }c_{-\mathbf{k}%
\downarrow }^{\dagger }c_{-\mathbf{k}^{\prime }\downarrow }c_{\mathbf{k}%
^{\prime }\downarrow }.  \label{Hred}
\end{equation}
Since we shall study excitations around the ground state we assume that
there are no unpaired electrons which allows us to study the problem in the
pseudo-spin Hamiltonian \cite{Ander-pseudo}
\[
H=\sum_{\mathbf{k}\sigma }2\xi _{\mathbf{k}}S_{\mathbf{k}\sigma }^{z}-\frac{1%
}{2}\sum_{\mathbf{k},\mathbf{k}^{\prime }}V_{\mathbf{k}-\mathbf{k}^{\prime
}}(S_{\mathbf{k}}^{+}S_{\mathbf{k}^{\prime }}^{-}+S_{\mathbf{k}^{\prime
}}^{+}S_{\mathbf{k}}^{-})
\]
\begin{equation}
=\sum_{\mathbf{k}\sigma }2\xi _{\mathbf{k}}S_{\mathbf{k}\sigma }^{z}-\sum_{%
\mathbf{k},\mathbf{k}^{\prime }}V_{\mathbf{k}-\mathbf{k}^{\prime }}(S_{%
\mathbf{k}}^{x}S_{\mathbf{k}^{\prime }}^{x}+S_{\mathbf{k}^{\prime }}^{y}S_{%
\mathbf{k}}^{y}),  \label{Hpseudo}
\end{equation}
where the pseudo-spin 1/2 operators $S_{\mathbf{k}\sigma }^{z}$, $S_{\mathbf{%
k}\sigma }^{+}=(S_{\mathbf{k}\sigma }^{-})^{\dagger }$ are given by
\[
S_{\mathbf{k}\sigma }^{z}=\frac {1}{2}(c_{\mathbf{k}\uparrow }^{\dagger }c_{%
\mathbf{k}\uparrow }-c_{-\mathbf{k}\downarrow }^{\dagger }c_{-\mathbf{k}%
\downarrow }-1),
\]
\begin{equation}
S_{\mathbf{k}\sigma }^{+}=c_{\mathbf{k}\uparrow }^{\dagger }c_{-\mathbf{k}%
\downarrow }^{\dagger }.  \label{ps-spin}
\end{equation}
We see that Eq.(\ref{Hpseudo}) belongs to the class of the Heisenberg
ferromagnetic ($V_{\mathbf{k}-\mathbf{k}^{\prime }}>0$) Hamiltonian
formulated on the lattice in the Brillouin zone. The mean-field
approximation (MFA) for this Hamiltonian is given by
\begin{equation}
H_{MFA}=-\sum_{k}\mathbf{h}_{\mathbf{k}}\mathbf{S}_{\mathbf{k}}  \label{Hmfa}
\end{equation}
with the mean-field $\mathbf{h}_{\mathbf{k}}$ given by
\begin{equation}
\mathbf{h}_{\mathbf{k}}=-2\xi _{\mathbf{k}}\mathbf{z+}\sum_{\mathbf{k}%
^{\prime }}V_{\mathbf{k}-\mathbf{k}^{\prime }}\langle S_{\mathbf{k}^{\prime
}}^{x}\mathbf{x+}S_{\mathbf{k}^{\prime }}^{y}\mathbf{y\rangle ,}  \label{mf}
\end{equation}
where $\mathbf{x,y}$ and $\mathbf{z}$ are unit vectors. Since x- and y-axis
are equivalent one can searches $\mathbf{h}_{\mathbf{k}}$ in the form $%
\mathbf{h}_{\mathbf{k}}=-2\xi _{\mathbf{k}}\mathbf{z+}2\Delta _{\mathbf{k}}%
\mathbf{x}$, where the order parameter $\Delta _{\mathbf{k}}$ is the
solution of the equation
\begin{equation}
\Delta _{\mathbf{k}}=\sum_{\mathbf{k}^{\prime }}V_{\mathbf{k}-\mathbf{k}%
^{\prime }}\langle S_{\mathbf{k}^{\prime }}^{x}\rangle =\sum_{\mathbf{k}%
^{\prime }}V_{\mathbf{k}-\mathbf{k}^{\prime }}\frac{V_{\mathbf{k}-\mathbf{k}%
^{\prime }}\Delta _{\mathbf{k}^{\prime }}}{2E_{\mathbf{k}}}\tanh \frac{\beta
E_{\mathbf{k}}}{2},  \label{gap-eq}
\end{equation}
with $E_{\mathbf{k}}=\sqrt{\xi _{\mathbf{k}}^{2}+\Delta _{\mathbf{k}}^{2}}$.

In the case of \textit{short-range BCS-like forces }$V_{BCS}(\mathbf{x-%
\mathbf{x}}^{\prime })\approx V_{0}\delta (\mathbf{x-\mathbf{x}}^{\prime })$
one has $V_{\mathbf{k}-\mathbf{k}^{\prime }}=V_{0}$ for all momenta. This
''long-range force'' in the momentum space it is the ''long-range force''
means that the MFA is good approximation with the standard BCS solution of
Eq.(\ref{gap-eq}).

For the \textit{long-range attractive forces} the function $V_{\mathbf{k}-%
\mathbf{k}^{\prime }}$ is peaked at $\mid \mathbf{k}-\mathbf{k}^{\prime
}\mid =0$, for instance in the extreme forward scattering peak case (see
Section 6.2) one has $V_{\mathbf{k}-\mathbf{k}^{\prime }}=V_{0}\delta (%
\mathbf{k-\mathbf{k}}^{\prime })$. In the following we analyze s-wave
pairing only where the solution of Eq.(\ref{gap-eq}) gives $%
T_{c}^{MF}=V_{0}/4$ and $\Delta _{0}=2T_{c}^{MF}$. (Note, that in
the BCS case one has $\Delta _{0}=1.76T_{c}^{MF}$.). The coherence
length is defined by $\xi =v_{F}/\pi \Delta _{0}$. The important
fact is that in the case of long-ranged superconductors the
Heisenberg like Hamiltonian in the momentum space is short-ranged
giving rise to low-laying spin-wave spectrum. The latter spectrum
are in fact the low-energy bound states (excitons) which loosely
correspond to the low-energy collective modes (in the true
many-body theory based on Eq.(\ref{H-long})). This problem is
studied in \cite{Yang} for the long-range (but finite) potential
$V(r)=V_{0}\exp \{-r^{2}/2r_{c}^{2}\}$ \ (its Fourier transform is
$V_{k}=(2\pi r_{c}^{2})V_{0}\exp \{-k^{2}r_{c}^{2}/2\}$) where it
was found a large number $N_{cm}\sim \pi k_{F}r_{c}/6\xi $ (for
$r_{c}\gg \xi $) of the excitonic like collective modes $\omega
_{mn}^{exc}$ at zero momentum. These excitonic modes lie between
the ground state and the two particle continuum for $\omega
>2\Delta _{0}$. Note, that since we assume that $\Delta _{0}\ll
E_{F}$ the system is far from the Bose-Einstein condensation
limit.

The above analysis is useful for physical understanding, but the fully
many-body fluctuation problem, which is based on the Hamiltonian in Eq.(\ref
{H-long}), is studied in \cite{Yang} where the Ginzburg-Landau (G-L)
equation is derived for the long-ranged superconductor. Due to the
fluctuations of the internal wave-function the G-L free-energy functional $%
F\{\Delta (\mathbf{R},\mathbf{k})\}$ for the order parameter $\Delta (%
\mathbf{R},\mathbf{k})=\int d\mathbf{r}\Delta (\mathbf{R}-\mathbf{r/2},%
\mathbf{R}+\mathbf{r/2})\exp \{-i\mathbf{kr}\}$ has much more complicated
form

\[
F\{\Delta (\mathbf{R},\mathbf{k})\}=\sum_{k}\int d\mathbf{R}\{A_{\mathbf{k}%
}\mid \Delta (\mathbf{R},\mathbf{k})\mid ^{2}+B_{\mathbf{k}}\mid \Delta (%
\mathbf{R},\mathbf{k})\mid ^{2}
\]
\begin{equation}
+\frac{1}{2M}\mid \partial _{\mathbf{k}}\Delta (\mathbf{R},\mathbf{k})\mid
^{2}+\frac{1}{2m_{\mathbf{k}}}\mid \partial _{\mathbf{R}}\Delta (\mathbf{R},%
\mathbf{k})\mid ^{2}\},  \label{F-GL}
\end{equation}
where $M=r_{c}^{2}V_{0}$ and
\[
A_{\mathbf{k}}=\frac{1}{V_{0}}-\frac{\tanh (\beta \xi _{\mathbf{k}}/2)}{2\xi
_{\mathbf{k}}}
\]
\[
\frac{1}{2m_{\mathbf{k}}}=\frac{\beta ^{2}v_{F}^{2}\sinh (\beta \xi _{%
\mathbf{k}}/2)}{32\xi _{\mathbf{k}}\cosh ^{3}(\beta \xi _{\mathbf{k}}/2)}
\]
\begin{equation}
B_{\mathbf{k}}=\frac{\tanh (\beta \xi _{\mathbf{k}}/2)}{8\xi _{\mathbf{k}%
}^{3}}-\frac{\beta }{16\xi _{\mathbf{k}}^{2}\cosh ^{3}(\beta \xi _{\mathbf{k}%
}/2)}.  \label{Coeff}
\end{equation}

The term due to the partial derivative $\partial_{\mathbf{k}}$ is a direct
consequence of the long-ranged pairing potential, and it describes of
fluctuations of the internal Cooper wave-function. The effect of these
fluctuations, described by the free-energy functional in Eq.(\ref{F-GL}), is
studied in the Hartree-Fock approximation in the limit $r_{c}\gg \xi $,
where it is found the large reduction of the mean-field critical temperature
\begin{equation}
T_{c}\sim \frac{T_{c}^{MF}}{(r_{c}/\xi )}.  \label{Tc-fl}
\end{equation}

The latter result means that $T_{c}$ in the long-ranged superconductors is
\textit{controlled by thermal fluctuations of collective modes} which is in
contrast with the short-range (BCS-like) superconductivity. In the
temperature interval $T_{c}<T<T_{c}^{MF}$ the system is in the pseudogap
regime where the electrons are paired but there is no long-range phase
coherence. The latter sets in only at $T<T_{c}$. We shall not further
discuss this interesting approach but only stress that it can be generalized
by including the repulsive interaction due to spin fluctuations, what shall
be discussed elsewhere.

In conclusion, the forward scattering peak in the EPI gives rise to the
long-ranged superconductivity in which the soft excitonic modes of the
internal Cooper wave function reduce $T_{c}$ strongly. In the region $%
T_{c}<T<T_{c}^{MF}$ the pseudogap (PG) phase is realized. In this approach
the PG has the same symmetry as the superconducting gap.

\section{Electron-phonon interaction vs spin-fluctuations}

\subsection{Interaction via spin fluctuations (SFI) and pairing}

\begin{figure}[tbp]
\resizebox{.9\textwidth}{!} {
\includegraphics*[width=10cm]{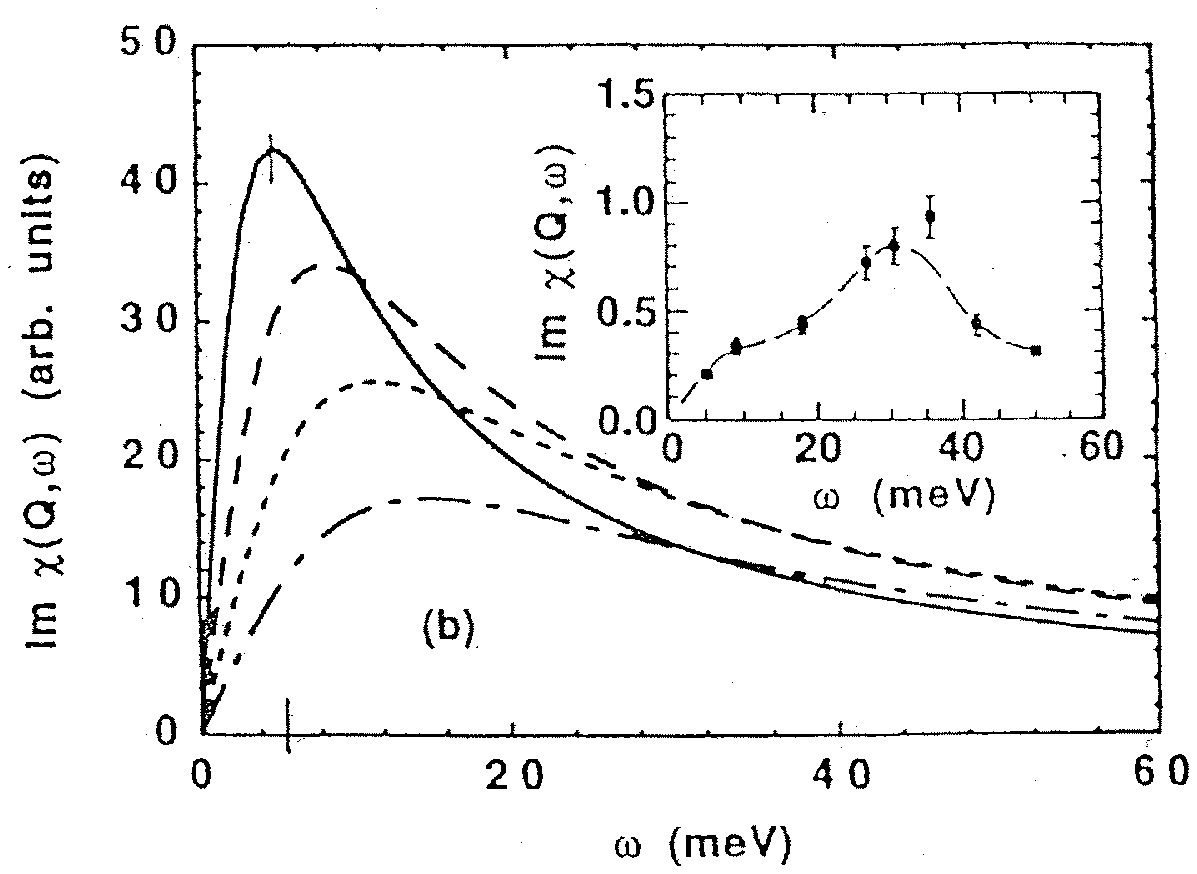}}
\caption{Spectral function $\mathrm{Im}\protect\chi ({},\protect\omega )$
for the \textit{MMP model} of spin-mediated interactions $\mathrm{Im}\protect%
\chi ({},\protect\omega )$ in $YBa_{2}Cu_{3}O_{7-\protect\delta}$. The
spectral function is calculated at ${}=(\protect\pi ,\protect\pi )$ and $T=0$
$K$ (solid line), $100$ $K$ (long-dashed line), $200$ $K$ (short dashed
line), and $300$ $K$ (dot-dashed line). Inset: experimental data of $%
YBa_{2}Cu_{3}O_{6.6}$ at $T=100$ $K$ - the line is to guide the eye. From
\protect\cite{Levin1}, \protect\cite{Norman}.}
\label{ImchiPinFig}
\end{figure}

At present one of the possible candidates in explaining experimental results
in HTSC oxides appears to be the theory based on the spin fluctuation\
pairing mechanism - the SFI theory. The latter is usually described by the
single band Hubbard model, or on the phenomenological level by the
postulated form of the self-energy (written below)\cite{Pines}, \cite{Pines3}%
, \cite{MMP}, \cite{Levin1}, \cite{Norman}, \cite{Schuttler}. In the
approach of Pines-school to the SFI the effective potential $V_{eff}(\mathbf{%
k},\omega )$ (see Eq.(\ref{Eq3}) in Sections 2.) depends on the imaginary
part of the spin susceptibility $\mathrm{Im}\chi (\mathbf{k-k}^{\prime
},\omega )$ ($\omega $ real). According to this school, the shape and the
magnitude of $\mathrm{Im}\chi (\mathbf{q},\omega )$, which is peaked at $%
\mathbf{Q}=(\pi ,\pi )$, plays an important role in obtaining T$_{c}$ in
this mechanism. There are two phenomenological approaches, which can be
theoretically justified in a very weak coupling limit $g_{sf}\ll 1$ only,
where $\mathrm{Im}\chi (\mathbf{q},\omega )$ is inferred from different
experiments:

$\mathbf{(1)}$ From $NMR$ experiments at very low $\omega $ - the \textit{%
MMP model} \cite{Pines}, \cite{Pines3}, \cite{MMP}, where $Im \chi _{MMP}$
is modelled by

\begin{equation}
Im \chi _{MMP}(\mathbf{q},\omega +i0^{+})=\frac{\omega }{\omega _{sf}}\frac{%
\chi _{Q}}{[1+\xi _{M}^{2}\mid \mathbf{q-Q}\mid
^{2}+(\omega/\omega_{sf})^{2}]^{2}}\Theta (\omega _{c}^{MMP}-\mid \omega
\mid ),  \label{Eq69}
\end{equation}
with the frequency cutoff $\omega _{c}^{MMP}=400$ $meV$. They fit the $NMR$
experiments by assuming very large value for $\chi _{Q}\approx (30-40)\chi
_{0}\sim 100$ $eV^{-1}$. From Fig.~\ref{ImchiPinFig} it is seen that the
imaginary susceptibility is peaked at low frequency $\omega _{peak}\approx
5-10$ $meV$.

\begin{figure}[tbp]
\resizebox{.9\textwidth}{!} {
\includegraphics*[width=10cm]{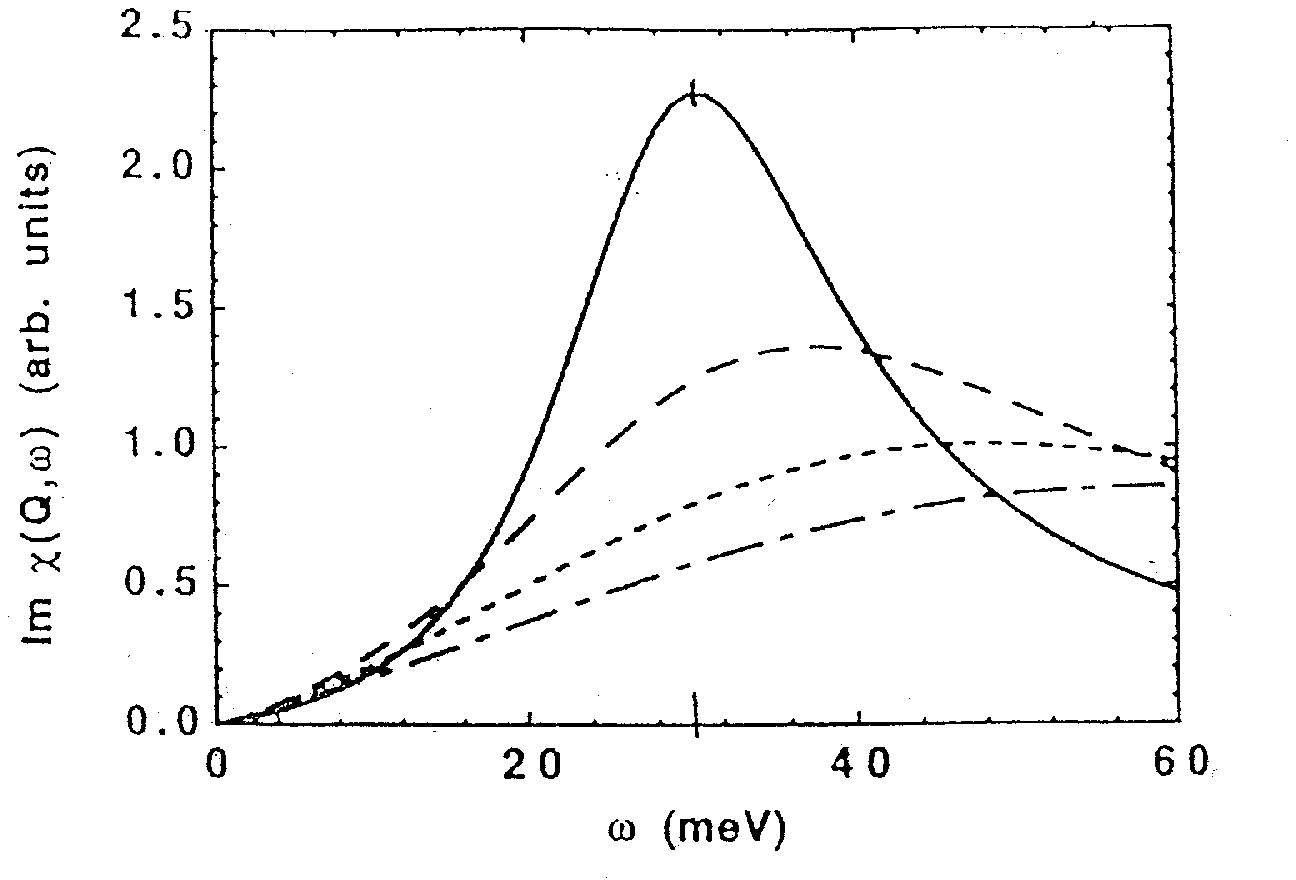}}
\caption{Spectral function $\mathrm{Im}\protect\chi ({},\protect\omega )$
for the \textit{RULN model} of spin-mediated interactions $\protect\chi ({},%
\protect\omega )$ in $YBa_{2}Cu_{3}O_{7-\protect\delta }$. The spectral
function is calculated at ${}=(\protect\pi ,\protect\pi )$ and $T=0$ $K$
(solid line), $100$ $K$ (long-dashed line), $200$ $K$ (short dashed line),
and $300$ $K$ (dot-dashed line). Inset: experimental data of $%
YBa_{2}Cu_{3}O_{6.6}$ at $T=100$ $K$ - the line is to guide the eye. From
\protect\cite{Levin1}, \protect\cite{Norman}.}
\label{ImchiNeutFig}
\end{figure}

$\mathbf{(2)}$ From the neutron scattering experiments \cite{Levin1}, \cite
{Norman}) - the \textit{RULN model}, where $Im \chi _{RULN}$ is modelled by

\[
Im\chi _{RULN}(\mathbf{q},\omega +i0^{+})=C[\frac{1}{1+J_{0}[\cos q_{x}+\cos
q_{y}]}]^{2}\times
\]
\begin{equation}
\times \frac{3(T+5)\omega }{1.5\omega ^{2}-60\mid \omega \mid +900+3(T+5)^{2}%
}\Theta (\omega _{c}^{RULN}-\mid \omega \mid ),  \label{Eq70}
\end{equation}
where $\omega _{c}^{RULN}=100$ $meV$, $J_{0}=0.3$, $C=0.19$ $eV^{-1}$ with $T
$ and $\omega $ measured in $meV$. From Fig.~\ref{ImchiNeutFig} it is seen
that $Im \chi _{RULN}$ is peaked around 30 meV, which is much larger than in
the MMP model.

\begin{figure}[tbp]
\resizebox{.8\textwidth}{!} {
\includegraphics*[width=10cm]{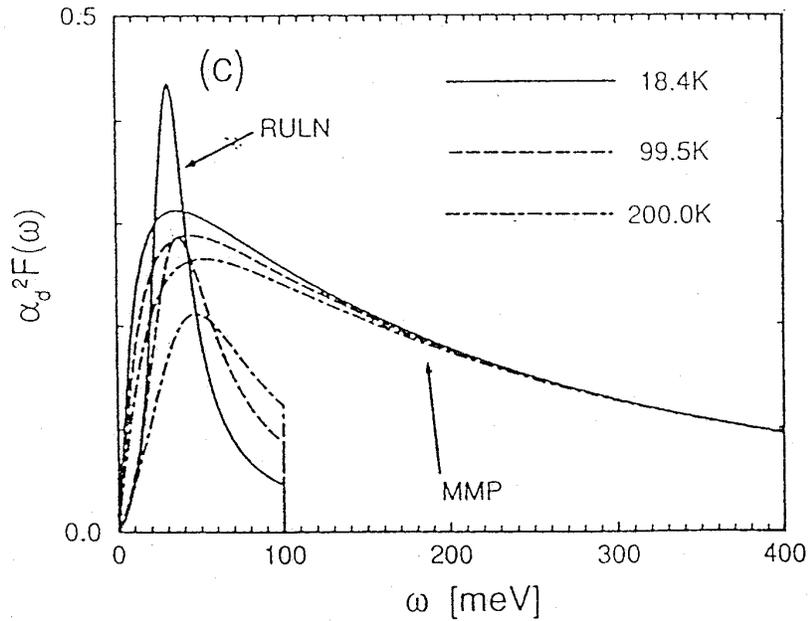}}
\caption{The spectral function $\protect\alpha _{d}^{2}F(\protect\omega )$
in the $d-wave$ pairing channel for the $MMP$ and $RULN$ model at different
temperature - from \protect\cite{Schuttler}.}
\label{sfdSpecFig}
\end{figure}

By knowing $Im \chi $ one can calculate the effective pairing potential $%
V_{eff}(\mathbf{k},\omega )$ from Eq.(\ref{Eq3}) and the spectral function
for the d-wave pairing $\alpha _{d}^{2}F(\omega )$

\begin{equation}
\alpha _{d}^{2}F(\omega )=-\frac{\langle \langle Y_{d}(\mathbf{k})Y_{d}(%
\mathbf{k}^{\prime })V_{SF}(\mathbf{k-k}^{\prime },\omega +i0^{+})\rangle
\rangle }{\langle Y_{d}^{2}(\mathbf{k})\rangle }.  \label{Eq71}
\end{equation}
Here, $Y_{d}(\mathbf{k})=\cos k_{x}-\cos k_{y}$ is the $d-wave$ pairing
function ($\Delta (\mathbf{k},\omega )\approx \Delta (\omega )Y_{d}(\mathbf{k%
})$). The bracket means an average over the Fermi surface. The spectral
function $\alpha _{d}^{2}F(\omega )$ for two models is shown in Fig.~\ref
{sfdSpecFig}, where it is seen that $\alpha _{d}^{2}F(\omega )^{RULN}$ is
much narrower function than $\alpha _{d}^{2}F(\omega )^{MMP}$. The latter is
peaked almost at the same $\omega $ as $\alpha _{d}^{2}F(\omega )^{RULN}$,
while $\alpha _{d}^{2}F(\omega )^{MMP}$ is much broader than $\alpha
_{d}^{2}F(\omega )^{RULN}$.

Due to different shapes of the susceptibility and of $\alpha
_{d}^{2}F(\omega )$ in these two approaches the calculated (from Eliashberg
equations) critical temperatures are also very different. Since the MMP
spectral function is much broader than the RULN one it turns out that $%
T_{c}^{(MMP)}$ can reach $100$ $K$ for rather large value of $g_{SF}\sim 0.64
$ $eV$, while $T_{c}^{(RULN)}$ saturates already at $50$ $K$ even for $%
g_{SF}\gg 1$. From the physical pint of view the RULN model is more
plausible than the MMP one, since the former is based on the neutron
scattering measurements which comprise much larger frequencies than the NMR
measurements. Note, that a valid model for HTSC oxides must be able to
explain the high values of T$_{c}$ (which needs $\lambda_{sf} (= 2\int
(\alpha _{d}^{2}F(\omega )/\omega)d\omega )\sim 2$) and the resistivity $%
\rho (T)$ (and its slope $\rho ^{\prime }$ with small $\lambda _{tr}\sim 0.6$%
). It turns out that $T_{c}^{(MMP)}$ can fit $T_{c}\approx 100$ $K$ on the
expense of large coupling $g_{sf}^{(MMP)}\sim 0.64$ $eV$ and $%
\lambda_{sf}^{(MMP)}\sim 2.5$. However, the value $g_{sf}^{(MMP)}\sim 0.64$ $%
eV$ gives much larger value for $\rho (T)$ and $\rho ^{\prime }$ than the
experiments do. On the other hand if one fits $\rho (T)$ and $\rho ^{\prime }
$ with the MMP model one gets very small $T_{c}<7$ $K$, thus making the MMP
model ineffective in HTSC oxides.

In the physically more plausible \textit{RULN model} $T_{c}^{(RULN)}$
saturates at $50$ $K$ even for $g_{sf}^{(RULN)}\gg 1$ $eV$. If one chooses
an appropriate value for $g_{sf}^{(neut)}$ to fit $\rho (T)$ and $\rho
^{\prime }$ one gets $T_{c}^{(RULN)}\approx 7$ $K$. This analysis gives a
convincing evidence that the existing SFI theories are ineffective in HTSC
oxides.

We stress again, that the large effective coupling constant, assumed in the
SFI theories, $g_{sf}\sim 0.64$ $eV$ , is difficult to justify theoretically
(if at all). By analyzing theoretically the possible strength of the
coupling constant $g_{sf}$, in both the weak ($N(0)U\ll 1$) and strong ($%
N(0)U\gg 1$) coupling limit, one obtains that $g_{sf}<0.2$ $eV$ and $\lambda
_{sf}<0.2$ (note $\lambda _{sf}\sim g_{sf}^{2}$), which means that $\lambda
_{sf}\ll \lambda _{sf}^{(MMP)}$ and $T_{c}^{(sf)}\ll T_{c}^{(MMP)}$ \cite
{KulicReview}. This analysis is supported by the recent theoretical results
where $g_{sf}<0.2$ $eV$ is extracted from the calculation : \textbf{1.} of
the width of the magnetic resonance peak at $41$ $meV$ \cite{Kivelson2};
\textbf{2.} of the small magnetic moment ($\mu <0.1$ $\mu _{B}$) in the
antiferromagnetic order, which coexists with superconductivity in $%
La_{2-x}Sr_{x}CuO_{4}$ \cite{KulicKulic}.

\subsection{Are the EPI and SFI compatible in d-wave pairing?}

The phenomenological SFI theories (MMP, RULN, FLEX approximation \cite
{Monthoux}] became popular because they can produce d-wave pairing, due to
the repulsive character of spin-fluctuations (in the momentum space.) which
are peaked in the backward ($q=Q$) scattering. However, as we have argued in
previous Sections, a number of experiments point to a large EPI with $%
\lambda >1$. On the other hand if one assumes that the EPI is momentum
independent (isotropic), like in the standard Migdal-Eliashberg theory, then
it is strongly pair-breaking for d-wave pairing. So, if one assumes (for the
moment) that superconductivity in HTSC oxides is due to the SFI with $%
\lambda_{sf}\approx 2$, then in that case $T_{c}^{(sf)}$ would be
drastically reduced (to almost zero) by the isotropic EPI even for moderate $%
\lambda_{EP}\sim 1$. The latter was shown in \cite{LichtKul} where the
Eliashberg equations are solved for the SFI treated in the FLEX
approximation \cite{Monthoux} and the EPI in the Einstein model with various
momentum dependent $V_{EP}(\vec{k},\omega)$. This result means, that if the
SFI would be the basic pairing interaction in the presence of the isotropic
and momentum independent EPI, then in order to reach $T_{c}\sim 100$ $K$ the
bare critical temperature should be $T_{c}^{(sf)}\sim (600-700)$ $K$, which
needs unrealistically large $\lambda _{sf}$. This is not only highly
improbable, but would give enormous large resistivity and its slope, in
contrast to experiments. The similar results were obtained in \cite
{MaksimovReview}.

The calculations in \cite{LichtKul} show that the SFI interaction is
dominant in (d-wave) pairing if some strong constraints are realized, such
as: (\textbf{1}) very large SFI coupling constant $\lambda_{sf}\approx 2$; (%
\textbf{2}) a strong forward scattering peak in the EPI with small EPI
coupling $\lambda \ll 1$. Both these conditions are incompatible with
experiments and theoretical analysis - see also Section 2.1. The way out
from this controversy is that the EPI with the forward scattering peak is
inevitably dominant interaction in the quasiparticle scattering and pairing
in HTSC oxides. As we already discussed in Section 5. the forward scattering
peak in the EPI gives rise to the large coupling constant in the d-wave
channel, which is of the order of the one in the s-wave channel in the range
of doping around the optimal one, i.e. $\lambda_{EP,d}\approx \lambda_{EP,s}$%
. This means that the residual Coulomb repulsion (by including also the SFI
with the backward scattering peak (BSP)) with $\lambda_{c}< \lambda_{EP,d}$
\textit{triggers} d-wave pairing.

\section{Is there high-temperature superconductivity in the Hubbard and t-J
model?}

\subsection{Hubbard model}

There are a number of papers dealing with numerical calculations, such as
Monte Carlo, exact Lanczos diagonalization, in the 2-dimensional (2D)
single-band and three-band Hubbard model. One can say that the single-band
Hubbard model does very well in describing the magnetic properties of HTSC
oxides. Concerning the existence of superconductivity the situation is not
definitely resolved. So far the calculations are done on finite clusters and
rather high temperatures $T>0.1t$ \cite{Dagotto}, \cite{Kampf} which show no
tendency to superconductivity. It is worth of mentioning that most of these
calculations deal with the pairing susceptibility - see Eq.(\ref{chiSC})
below, defined in terms of the bare electron operators $c_{mathbf{k}\sigma}$
in Eq.(\ref{SCdelta}). Since superconducting pairing is realized on
quasiparticles with the weight $z<1$ there is a last hope that the accuracy
of the present numerical calculations is not sufficient to pick up the
suppressed pairing susceptibility. In that respect a very important approach
to the problem of superconductivity (in any microscopic model), which is
formulated without using any order parameter, was given by the Scalapino
group \cite{D-Ds}. The method of calculations is based on the two most
important hallmarks of superconductivity: \textbf{(i)} \textit{ideal
diamagnetism} (the Meissner effect) and \textbf{(ii)} \textit{ideal
conductivity}. In that respect they study the superfluid density $D_{s}$
(proportional to $\lambda ^{-2}$, $\lambda $ is the penetration depth) and
the Drude weight $D$ in the single-band n.n. (nearest neighbors) Hubbard
model. The dynamical conductivity along the x-axis is given by
\begin{equation}
\sigma _{xx}(\omega )=-\frac{e^{2}}{i}\frac{\langle -T_{x}\rangle -\Lambda
_{xx}(\mathbf{q}=0,\omega )}{\omega +i\delta }.  \label{D-Ds1}
\end{equation}
Here, $\langle -T_{x}\rangle =\langle -T\rangle /2$ where T is the kinetic
energy in the n.n. tight-binding model - see Eq.(\ref{Eq9}), where the
current-current response function $\Lambda _{xx}(\mathbf{q},\omega )$ is
obtained from
\begin{equation}
\Lambda _{xx}(\mathbf{q},i\omega _{m})=\frac{1}{N}\int_{0}^{\beta }d\tau
e^{i\omega _{m}\tau }\langle j_{x}^{p}(\mathbf{q},\tau )j_{x}^{p}(-\mathbf{q}%
,0)\rangle ,  \label{D-Ds2}
\end{equation}
with $\omega _{m}=2\pi mT$, by the standard analytic continuation $i\omega
_{m}\rightarrow \omega +i\delta $ and
\begin{equation}
j_{x}^{p}(\mathbf{q},\tau )=it\sum_{l,\sigma }e^{-i\mathbf{ql}}(c_{l+x\sigma
}^{\dagger }c_{l\sigma }-c_{l\sigma }^{\dagger }c_{l+x\sigma }).
\label{D-Ds3}
\end{equation}
In the pure Hubbard model $\sigma _{xx}(\omega )$ contains the delta
function contribution
\begin{equation}
\sigma _{xx}(\omega )=D\delta (\omega )+\sigma _{reg}(\omega ),
\label{D-Ds4}
\end{equation}
where the Drude weight $D(\equiv (n/m)^{\ast }$, which measures the ratio of
the density of the mobile charge carriers to their mass, is defined by

\begin{equation}
\frac{D}{\pi e^{2}}=\langle -T_{x}\rangle -\Lambda _{xx}(\mathbf{q}=0,\omega
\rightarrow 0).  \label{D-Ds5}
\end{equation}

The Meissner effect is the current response to a static and transverse gauge
potential $\mathbf{q}\cdot \mathbf{A}(\mathbf{q},\omega =0)=0$. In the small
$\mathbf{q}$ limit one has

\begin{equation}
\langle j_{\alpha }(\mathbf{q\rightarrow 0})\rangle =-\frac{D_{s}}{\pi }%
(\delta _{\alpha \beta }-q_{\alpha }q_{\beta }/q^{2})A_{\beta }
\label{D-Ds6}
\end{equation}
where $D_{s}(\equiv (n_{s}/m)^{\ast }$

\begin{equation}
D_{s}=\langle -T_{x}\rangle -\Lambda _{xx}(q_{x}=0,q_{y}\rightarrow 0,\omega
=0).  \label{D-Ds7}
\end{equation}
Based on the above definitions of D and D$_{s}$ we can study various phases
of the system: (\textbf{1}) D=D$_{s}$=0 - an \textit{isolator}; (\textbf{2})
D$\neq $0 and D$_{s}$=0 - a \textit{nonsuperconducting metal}; (\textbf{3}) D%
$_{s}\neq $0, D$\neq $0 - a \textit{superconducting metal}.

The numerical Monte Carlo calculation in the repulsive Hubbard model
(U=4t>0) \cite{D-Ds} on an 8$\times $8 lattice show that D$_{s}$=0 and D$%
\neq $0 in a broad range of the filling 0.5<n<0.9 and for $T>0.1t$. This
means that there is no tendency to high-temperature superconductivity in the
single-band Hubbard model. This conclusion is supported by the projector-QMC
calculations \cite{Assad} for the quarter filling case n=0.5 and at T=0.
That these results ($D_{s}=0$) are not a finite size effect confirm the
calculations on the attractive Hubbard model (U=-4t<0), also on an 8$\times $%
8 lattice, where the clear tendency to superconductivity is found, since D$%
_{s}\neq $0, D$\neq $0 already at T<0.2t.

The paper \cite{D-Ds} is of great importance numerical studies of
pairing in model systems, not only because it hints on the absence
of superconductivity in the repulsive Hubbard model, but also
because of the following two reasons: (\textbf{1}) It uses the
general and unbiased criterion for superconductivity, which is
independent on the type of the pairing amplitude; (\textbf{2}) It
shows that the attractive interaction is more favorable for
(high-temperature) superconductivity than the repulsion.

\subsection{t-J model}

The SFI phenomenological approaches root on their basic $t-J$ Hamiltonian
Eq.(\ref{Eq36}). On can put a legitimate question - is there
superconductivity in the t-J model? In the past there were various
approaches confronting with this important problem. In spite of a number of
controversial statements it seems that the results converge to the unique
answer - \textit{there is no superconductivity with appreciable T$_{c}$}. If
superconductivity exists $T_{c} $ is very low. As the strong support of this
claim serve the recent calculations based on the high-temperature expansion
in the t-J-V model \cite{Pryadko},

\begin{equation}
\hat{H}=-\sum_{i,j,\sigma }t_{ij}\hat{X}_{i}^{\sigma 0}\hat{X}_{j}^{0\sigma
}\;+\sum_{i,j=n.n.}[J(\mathbf{S}_{i}\cdot \mathbf{S}_{j})+(V-\frac{J}{4}%
)n_{i}n_{j}\;)].  \label{Hab}
\end{equation}
Here, the V-term mimic the screened Coulomb interaction which is always
present in metals, where one expects that $V>J$. In \cite{Pryadko} it was
calculated the uniform susceptibility for the superconducting pairing
\begin{equation}
\chi _{SC}\equiv \frac{1}{N}\int_{0}^{\beta }d\tau \langle \langle \hat{T}%
_{\tau }e^{H\tau }O_{SC}e^{-H\tau }O_{SC}^{\dagger }\rangle \rangle
\label{chiSC}
\end{equation}

\begin{equation}
O_{SC}=\frac{1}{2}\sum_{\mathbf{r}}(\Delta _{\mathbf{r,r+x}}\pm \Delta _{%
\mathbf{r,r+y}})  \label{OSC}
\end{equation}

\begin{equation}
\Delta _{ij}\equiv c_{i\uparrow }c_{j\downarrow }+c_{j\uparrow
}c_{i\downarrow }.  \label{SCdelta}
\end{equation}
where + sign holds for the s-Sc and the - sign for d-Sc.

\begin{figure}[tbp]
\resizebox{.9\textwidth}{!}
{\includegraphics*[width=8cm]{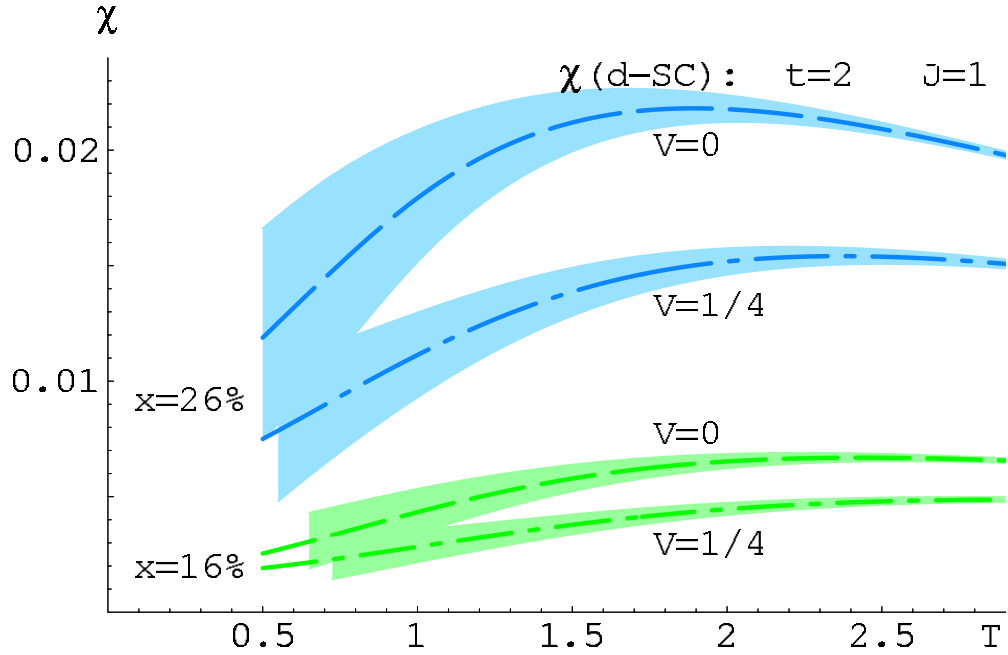}} \caption{Superconducting
(d-wave) susceptibility $\protect\chi (d-Sc)(T)$ for $t=2J$.
Pairing correlations are already weak for $V=0$, dashes, and
decreases further by increasing $V$-n.n. repulsion (dot-dash).
From \protect\cite{Pryadko}} \label{NoScFig}
\end{figure}

In the physical region of parameters $t>J$ both $\chi _{s-SC}$ and $\chi
_{d-SC}$ are small and further decrease by decreasing T. For rather small $%
V=J/4$, which is even much smaller than expected, the
superconducting susceptibilities are drastically decreased as it
is seen in Fig.~\ref {NoScFig}. This means that in the more
realistic models for HTSC oxides, such as the t-J-V, \textit{there
is no tendency to high-temperature superconductivity}. If
superconductivity exists at all its T$_{c}$ must be very low.

\section{Summary and conclusions}

A number of experiments, such as optics ($IR$ and Raman),
transport, tunnelling, ARPES, neutron scattering, give convincing
evidence that the electron-phonon interaction (EPI) in HTSC oxides
is sufficiently strong and contributes to pairing. These
experiments give also evidence for the presence of strong
correlations which modify the EPI not only quantitatively but also
qualitatively. The most spectacular result of the $EP $I theory in
strongly correlated systems is the appearance of the
\textit{forward scattering peak} in the EPI, as well as in other
charge scattering processes such as the residual Coulomb
interaction and scattering on non-magnetic impurities - the FSP
\textit{\ theory } \cite{Kulic1}, \cite{Kulic2}, \cite {Kulic3},
\cite{Kulic4} \cite{KulicReview}. The forward scattering peak is
especially pronounced at lower doping $\delta $. This fundamental
result allows us to resolve a number of experimental facts which
can not be explained by the old theory based on th isotropic
Migdal-Eliashberg equations for the EPI. The most important
predictions of the FSP theory of the EPI and other charge
scattering processes are: (\textbf{1}) the
transport coupling constant $\lambda _{tr}$ (entering the resistivity, $%
\varrho \sim \lambda _{tr}T$) is much smaller than the pairing one $\lambda $%
, i.e. $\lambda _{tr}<\lambda /3$; (\textbf{2}) the strength of pairing in
HTSC oxides is basically due to the EPI, while the residual Coulomb
repulsion (including spin--fluctuations) \textit{triggers} d-wave pairing; (%
\textbf{3}) d-wave pairing is very robust in the presence of
non-magnetic impurities; (\textbf{4}) the nodal kink in the
quasiparticle spectrum is unshifted in the superconducting state,
while the anti-nodal singularity is shifted.

We stress the following two facts coming from the theoretical analysis: (%
\textbf{i}) the forward scattering peak in the EPI of strongly correlated
systems is a general phenomenon by \textit{affecting electronic coupling to
all phonons}; (\textbf{ii}) the existence of the forward scattering peak in
the EPI is confirmed numerically by the Monte Carlo calculations for the
Hubbard-Holstein model with finite U \cite{Scalapino-Hanke}, by exact
diagonalization \cite{Tohyama}, as well as by some other methods \cite
{Pietro-SB}.

Tunnelling experiments and ARPES measurements of the real part of
the self-energy give evidence that the EPI coupling constant
$\lambda
>1$. At present there are no reliable microscopic
calculations of $\lambda $ in HTSC oxides, which properly include (\textbf{a}%
) the ionic-metallic coupling (due to the long-range Madelung energy) and
covalent coupling and (\textbf{b}) strong electronic correlations.

In the last several years a large number of published papers were devoted to
the study of spin-fluctuation (SFI) interaction as a mechanism of pairing in
HTSC oxides. In spite of many efforts and well financed theoretical projects
(headed by the greatest authorities in the field), which have opened some
new research directions in the theory of electron magnetism, there is no
theoretical evidence for the effectiveness of the non-phononic mechanism of
pairing. Until now superconductivity could not been proved in the repulsive
single-band Hubbard model as well as in its derivative the $t-J$ model. Just
opposite, quite recent numerical calculations \cite{Pryadko} show in a
convincing way, that there is no high-temperature superconductivity in the
t-J model. If it exists its T$_{c}$ is extremely low. Finally, the numerical
calculations in the Hubbard model \cite{D-Ds} show that the repulsive
Hubbard interaction is unfavorable for high-temperature superconductivity,
contrary to the attractive interaction which favors it.

The explanation of the high critical temperature in HTSC oxides should be
searched in the electron-phonon interaction which is renormalized by strong
electronic correlations.

To conclude, \textit{one can not avoid unavoidable}.

\begin{theacknowledgments}

I am thankful to Professors Ferdinando Mancini and Adolfo Avella,
the organizers of the ''\textit{VIII Training Course in the
Physics of Correlated Electron Systems and High-Tc
Superconductors}'' held in October 2004 in Vietri sul Mare
(Salerno) Italy, for giving me the chance to inform the young and
talented scientists on the fundamental problems in HTSC physics.

With the great honor, I express my deep gratitude and respect to
Vitalii Lazarevich Ginzburg for supporting me for many years, for
his deep understanding of superconductivity, sharing it generously
with us - his students, collaborators and friends.

I am very thankful to Oleg V. Dolgov and Evgenii G. Maksimov for
numerous elucidating discussions on superconductivity theory, on
optical properties of HTSC oxides and on many-body theory, as well
as for their permanent support. Discussions with Ivan
Bo\v{z}ovi\'{c}, A. V. Boris and N. N. Kovaleva on optics, with Z.
-X. Shen and D. J. Scalapino on ARPES measurements and theory are
acknowledged. I appreciate very much support of Ivan
Bo\v{z}ovi\'{c}, Ulrich Eckern and Igor M. Kuli\'{c}.

\end{theacknowledgments}

\section{Appendix: Derivation of the t-J model}

\subsection{Hubbard model for finite U in terms of Hubbard operators}

For simplicity we study the nearest neighbor (n.n.) Hubbard model \cite
{Izyumov}

\begin{equation}
H=-t\sum_{m\neq n\sigma }\hat{c}_{m\sigma }^{\dagger }\hat{c}_{n\sigma
}+U\sum_{m}\hat{n}_{m\uparrow }\hat{n}_{m\downarrow },  \label{h1}
\end{equation}
where the operator $\hat{c}_{m\sigma }^{\dagger }$ creates an electron at
the m-th site with the spin projection $\sigma $.

The Hilbert space at the given lattice site contains four states \{$\mid
\alpha >\Longrightarrow \mid 0>,\mid 2>,\mid \uparrow >,\mid \downarrow >$%
\}. Let us introduce the Hubbard projection operators $X^{\alpha \beta }$ ; $%
\alpha ,\beta =0,2,\sigma $ (where $\sigma =\uparrow (+),\sigma =\downarrow
(-)$)

\begin{equation}
X^{\alpha \beta }=\mid \alpha ><\beta \mid .  \label{h2}
\end{equation}
They fulfill the projection properties
\begin{equation}
X^{\alpha \beta }X^{\gamma \delta }=\delta _{\beta \gamma }X^{\alpha \delta
},  \label{h3}
\end{equation}
and rather ''ugly'' (anti)commutation algebra

\begin{equation}
X_{i}^{\alpha \beta }X_{j}^{\gamma \delta }\pm X_{j}^{\gamma \delta
}X_{i}^{\alpha \beta }=\delta _{ij}(\delta _{\beta \gamma }X_{i}^{\alpha
\delta }\pm \delta _{\delta \alpha }X^{\gamma \beta }.  \label{h4}
\end{equation}
The completeness relation of the Hilbert space reads

\begin{equation}
X_{i}^{00}+X_{i}^{22}+\sum_{\sigma }X_{i}^{\sigma \sigma }=1.  \label{h5}
\end{equation}
The Hubbard operators describe the composite object. There is a connection
between $\hat{c}_{i\sigma }$ and $X^{\alpha \beta }$ \ (if $\sigma =\uparrow
$ $\Longrightarrow $ $\bar{\sigma}=\downarrow $)
\begin{equation}
\hat{c}_{i\sigma }=X_{i}^{0\sigma }+\sigma X_{i}^{\bar{\sigma}2};\hat{c}%
_{i\sigma }^{\dagger }=X_{i}^{\sigma 0}+\sigma X_{i}^{2\bar{\sigma}}
\label{h6}
\end{equation}
\begin{equation}
n_{i}=1-X_{i}^{00}+X_{i}^{22}
\end{equation}
\[
S_{i}^{+}=\hat{c}_{i\uparrow }^{\dagger }\hat{c}_{i\downarrow
}=X_{i}^{+-}=(S_{i}^{-})^{\dagger }=(X_{i}^{-+})^{\dagger }
\]
\begin{equation}
S_{i}^{z}=\frac{1}{2}(\hat{c}_{i\uparrow }^{\dagger }\hat{c}_{i\uparrow }-%
\hat{c}_{i\downarrow }^{\dagger }\hat{c}_{i\downarrow })=\frac{1}{2}%
(X_{i}^{++}-X_{i}^{--}),  \label{h8}
\end{equation}
and vice versa
\begin{equation}
X^{\sigma 0}=\hat{c}_{\sigma }^{\dagger }(1-\hat{n}_{\bar{\sigma}})
\end{equation}
\begin{equation}
X^{\sigma \sigma }=\hat{n}_{\sigma }(1-\hat{n}_{\bar{\sigma}});\ \ X^{\sigma
\bar{\sigma}}=\hat{c}_{\sigma }^{\dagger }\hat{c}_{\bar{\sigma}}  \label{h10}
\end{equation}

\begin{equation}
X^{00}=(1-\hat{n}_{\uparrow })(1-\hat{n}_{\downarrow })  \label{h11}
\end{equation}
\
\begin{equation}
X^{2\sigma }=\sigma \hat{c}_{\bar{\sigma}}^{\dagger }\hat{n}_{\sigma };\ \
X^{20}=\sigma \hat{c}_{\bar{\sigma}}^{\dagger }\hat{c}_{\sigma }
\end{equation}
\begin{equation}
X^{22}=n_{\uparrow }n_{\downarrow }  \label{h13}
\end{equation}
The Hubbard Hamiltonian $H=H_{1}+H_{12}+H_{0}$ in terms of $X^{\alpha \beta
} $ is given by
\begin{equation}
H_{1}=-t\sum_{ij\sigma }(X_{i}^{\sigma 0}X_{j}^{0\sigma }+X_{i}^{2\sigma
}X_{j}^{\sigma 2})  \label{h14}
\end{equation}
\begin{equation}
H_{12}=-t\sum_{ij\sigma }\sigma (X_{i}^{\sigma 0}X_{j}^{\bar{\sigma}%
2}+X_{i}^{2\bar{\sigma}}X_{j}^{0\sigma })  \label{h15}
\end{equation}
\begin{equation}
H_{0}=U\sum_{i}X_{i}^{22}  \label{h16}
\end{equation}
The first term in $H_{1}$ describes the\ motion of single electron in the
lower (L) Hubbard band, while the second term describes the motion of the
doubly occupied electrons from j-th to the i-th side in the upper (U)
Hubbard band. The term $H_{0}$ is the repulsive energy of two electrons on
the i-th site. Finally, $H_{12}$ connects the two (lower and upper) bands.

\subsection{Effective Hamiltonian for $U>>t$}

There are various ways to obtain the effective Hamiltonian in the case $U>>t$%
. Because of its generality and simplicity we use here the canonical
transformation method \cite{Izyumov}, where the operator $S$ mixes lower and
upper band. Under the action of $S$ the Hamiltonian is transformed into $%
H_{eff}=e^{S}He^{-S}$
\begin{equation}
H_{eff}=H+[S,H]+\frac{1}{2}[S,[S,H]]+..  \label{h17}
\end{equation}
with $S$ in the form
\begin{equation}
S=\kappa \sum_{ij\sigma }(X_{i}^{\sigma 0}X_{j}^{\bar{\sigma}2}-X_{i}^{2\bar{%
\sigma}}X_{j}^{0\sigma }).  \label{h18}
\end{equation}
Now, we choose $\kappa$ so that all first-order in t processes between the
L- and U-band disappear from $H_{eff}$, i.e. one has
\begin{equation}
H_{12}+[S,H_{2}]=0.  \label{h19}
\end{equation}
The solution of Eq.(\ref{h19}) is $\kappa =-t/U$, and $H_{eff}$ reads
\[
H_{eff}=-t\sum_{ij\sigma }X_{i}^{\sigma 0}X_{j}^{0\sigma }+H_{3s}
\]
\begin{equation}
+J\sum_{ij\sigma }(\mathbf{S}_{i}\mathbf{S}_{j}-\frac{1}{4}\hat{n}_{i}\hat{n}%
_{j})+H_{2},  \label{h20}
\end{equation}
where $J=2t^{2}/U$ is the exchange energy.

The term $H_{2}$ describes motion of ''doublons'' in the U-band
\begin{equation}
H_{2}=U\sum_{i}X_{i}^{22}-t\sum_{ij\sigma }X_{i}^{2\sigma }X_{j}^{\sigma 2},
\label{h21}
\end{equation}
while $H_{3s}$ describes the three-sites hopping.

\begin{equation}
H_{3s}=\frac{J}{2}\sum_{ijl\sigma }(X_{i}^{\bar{\sigma}0}X_{l}^{\sigma \bar{%
\sigma}}X_{j}^{0\sigma }-X_{i}^{\sigma 0}X_{l}^{\bar{\sigma}\bar{\sigma}%
}X_{j}^{0\sigma }).  \label{h22}
\end{equation}
Usually this term is neglected in the t-J model.

However, it may have tremendous effect on superconductivity by strongly
suppressing it \cite{Alligia-privat}. By projecting $H_{eff}$ onto the lower
Hubbard band one gets the famous t-J model Hamiltonian $H_{tJ}=PH_{eff}P$%
\[
H_{tJ}=-t\sum_{ij\sigma }X_{i}^{\sigma 0}X_{j}^{0\sigma }++J\sum_{ij\sigma }(%
\mathbf{S}_{i}\mathbf{S}_{j}-\frac{1}{4}\hat{n}_{i}\hat{n}_{j})
\]
\begin{equation}
=-t\sum_{ij\sigma }X_{i}^{\sigma 0}X_{j}^{0\sigma }+\frac{J}{2}%
\sum_{ij\sigma }(X_{i}^{\sigma \bar{\sigma}}X_{j}^{\bar{\sigma}\sigma
}-X_{i}^{\sigma \sigma }X_{j}^{\bar{\sigma}\bar{\sigma}}).  \label{h23}
\end{equation}
Before we are going to discuss some representations for non-canonical
operators $X_{i}^{\sigma 0}$ in terms of bosons and fermions let us stress
that the so called ''spin'' operators $S^{\pm },S^{z}$ do not describe
correctly the electron spin. Although they satisfy the correct
spin-commutation relations
\[
\lbrack S_{i}^{+},S_{j}^{-}]=2\delta _{ij}S_{i}^{z}
\]
\begin{equation}
\lbrack S_{i}^{z},S_{j}^{\pm }]=\pm \delta _{ij}S_{i}^{\pm },  \label{h24}
\end{equation}
they describe a particle with spins $S=0,1/2$ at the lattice site, since $%
\mathbf{S}_{i}^{2}$ fulfills
\begin{equation}
\mathbf{S}_{i}^{2}=\frac{3}{4}\hat{n}_{i}\neq \frac{3}{4}.  \label{h25}
\end{equation}

Since $X^{\alpha \beta }$ obey the non-canonical ( ''ugly'') algebra the
question is how to treat the Hamiltonian $H_{tJ}$ ? In Section 4.-5. we have
shown that one can study directly with these operators by using the
functional technique and 1/N expansion for the self-energy. However, there
are very popular approaches which represent $X^{\alpha \beta }$ in terms of
bosons and fermions with canonical commutation relations.

\paragraph{Slave boson (SB) method}

Here, one introduces the fermion with spin (spinon) $F_{i\sigma }$ and the
boson without spin (holon) $B_{i}$ operators, where $X^{0\sigma }=F_{\sigma
}B^{\dagger }$. The \textit{constraint} on the SB Hilbert space
(completeness), at the given lattice site, is given by $B^{\dagger
}B+\sum_{\sigma }F_{\sigma }^{\dagger }F_{\sigma }=1$ and the t-J
Hamiltonian reads
\begin{equation}
H_{tJ}=-t\sum_{ij\sigma }F_{i\sigma }^{\dagger }F_{j\sigma
}B_{i}B_{j}^{\dagger }+\frac{J}{2}\sum_{ij\sigma \sigma ^{\prime
}}F_{i\sigma }^{\dagger }F_{j\sigma }F_{j\sigma ^{\prime }}^{\dagger
}F_{i\sigma ^{\prime }}.  \label{h26}
\end{equation}
We stress that the \textit{constraint} strongly limits the SB Hilbert space
of bosons and fermions (at the given lattice site) which effectively means
their strong interaction. In that respect any uncontrollable decoupling in
the SB method (as in some RVB approaches) leads to a spin-charge
(spinon-holon) separation, which is not realized in 2D and 3D systems. In
order to correct this one introduces the so called gauge fields which keep
the spin and charge together. We already discussed the difficulties of the
SB method in studying the electron-phonon interaction.

\paragraph{Slave fermion (SF) method}

In the SF method the boson has spin and fermion not, i.e. $X^{0\sigma
}=B_{\sigma }^{\dagger }F$ with the constraint on the Hilbert space $%
F^{\dagger }F+\sum_{\sigma }B_{\sigma }^{\dagger }B_{\sigma }=1$.

\paragraph{Spin fermion method}

Here, the real fermion $\hat{c}^{\dagger }$ with the ''spin'' $\mathbf{S}$
is represented via the auxiliary fermion $F^{\dagger }$ and spin $\mathbf{s}$
by $\hat{c}_{\uparrow }^{\dagger }\hat{c}_{\uparrow }+\hat{c}_{\downarrow
}^{\dagger }\hat{c}_{\downarrow }=1-F^{\dagger }F$ and $\mathbf{S=s}%
(1-F^{\dagger }F)$. The t-J Hamiltonian is rather complicated
\[
H_{tJ}=2t\sum_{ij}F_{i}^{\dagger }F_{j}(\mathbf{s}_{i}\mathbf{s}_{j}+\frac{1%
}{4})
\]
\begin{equation}
+J\sum_{ij}(1-F_{i}^{\dagger }F_{i})(\mathbf{s}_{i}\mathbf{s}_{j}-\frac{1}{4}%
)(1-F_{j}^{\dagger }F_{j}).  \label{h27}
\end{equation}
Usually this method is used for analyzing motion of single hole in the
half-filled system where the antiferromagnetic order is realized.

\end{document}